\documentclass[
11pt, 
oneside, 
UKenglish, 
doublespacing, 
nolistspacing, 
parskip, 
headsepline, 
]{thesis} 

\usepackage[T2A]{fontenc}
\usepackage{graphicx}
\usepackage{subfigure}
\usepackage{setspace}
\usepackage{amsmath}
\usepackage{amssymb}
\usepackage{bm}
\usepackage{verbatim}
\usepackage{cite}

\def\half{\frac{1}{2}}
\def\third{\frac{1}{3}}
\def\quarter{\frac{1}{4}}
\def\0{\varnothing} 
\def\p{\parallel} 
\def\x{\mathsf{x}}

\def\T{\mathsf{T}} 
\def\BI{\gamma_{\mathrm{BI}}} 

\def\R{\mathcal{R}}
\def\K{\mathcal{K}}

\def\bp{\mathcal{P}}
\def\Rfour{{}^{(4)}\!R}

\def\xty{\left({x}\leftrightarrow{y}\right)}
\def\gud{\mathrm{gd}}
\def\sech{\mathrm{sech}}
\def\area{\lozenge}
\newcommand{\three}[1]{{}^{(3)}\!{#1}}
\newcommand{\four}[1]{{}^{(4)}\!{#1}}
\newcommand{\partdif}[2]{\frac{\partial {#1}}{\partial {#2}}}
\newcommand{\funcdif}[2]{\frac{\delta {#1}}{\delta {#2}}}
\newcommand{\totdif}[2]{\frac{\mathrm{d} {#1}}{\mathrm{d} {#2}}}

\newcommand{\figref}[1]{Fig.~\ref{#1}}

\newcommand{\secref}[1]{section~\ref{#1}}
\newcommand{\subsecref}[1]{subsection~\ref{#1}}
\newcommand{\chapref}[1]{chapter~\ref{#1}}
\newcommand{\appref}[1]{appendix~\ref{#1}}
\newcommand{\sgn}[1]{\operatorname{sgn} (#1)}

\usepackage[autostyle=true]{csquotes} 

\thesistitle{Deformed general relativity} 
\supervisor{Prof. Mairi \textsc{Sakellariadou}} 
\examiner{David \textsc{Wands} and Elizabeth \textsc{Winstanley}} 
\degree{Doctor of Philosophy} 
\author{Rhiannon \textsc{Cuttell}} 
\addresses{King's College London, The Strand, London, WC2R 2LS, United Kingdom} 

\subject{Theoretical Physics} 
\keywords{deformed general relativity, gravity, loop quantum gravity, cosmology, loop quantum cosmology, general relativity, scalar-tensor} 
\university{\href{https://www.kcl.ac.uk}{King's College London}} 
\department{\href{https://www.kcl.ac.uk/nms/depts/physics}{Department of Physics}} 
\group{\href{https://www.kcl.ac.uk/nms/depts/physics/research/tppc}{Theoretical Particle Physics and Cosmology}} 
\faculty{\href{https://www.kcl.ac.uk/nms}{School of Natural and Mathematical Sciences}} 

\AtBeginDocument{
\hypersetup{pdftitle=\ttitle} 
\hypersetup{pdfauthor=\authorname} 
\hypersetup{pdfkeywords=\keywordnames} 
}

\begin{document}

\frontmatter 

\pagestyle{plain} 


\begin{titlepage}
\begin{center}

\vspace*{.03\textheight}
{\scshape\LARGE \univname\par}\vspace{1.2cm} 
\textsc{\Large Doctoral Thesis}\\[0.5cm] 

\HRule \\[0.4cm] 
{\huge \bfseries \ttitle\par}\vspace{0.4cm} 
\HRule \\[1.5cm] 
 
\begin{minipage}[t]{0.4\textwidth}
\begin{flushleft} \large
\emph{Author:}\\
\texorpdfstring{\href{mailto:rhiannon.cuttell@kcl.ac.uk}{\authorname}}{\authorname} 
\end{flushleft}
\end{minipage}
\begin{minipage}[t]{0.4\textwidth}
\begin{flushright} \large
\emph{Supervisor:} \\
\texorpdfstring{\href{mailto:mairi.sakellariadou@kcl.ac.uk}{\supname}}{\supname} 
\end{flushright}
\end{minipage}\\[1cm]
 
\vfill

\large \textit{A thesis submitted in fulfillment of the requirements\\ for the degree of \degreename}\\[0.2cm] 
\textit{in the}\\[0.3cm]
\groupname\\\deptname\\[0.5cm] 
 
\vfill

{\large 1st April 2019}\\[0.6cm] 

\vfill

\includegraphics[height=3cm]{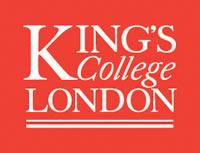} 

\end{center}
\end{titlepage}


\begin{declaration}
\addchaptertocentry{\authorshipname} 
\noindent I, \authorname, declare that this thesis titled, \enquote{\ttitle} and the work presented in it are my own. I confirm that:

\begin{itemize} 
\item This work was done wholly or mainly while in candidature for a research degree at this University.
\item Where any part of this thesis has previously been submitted for a degree or any other qualification at this University or any other institution, this has been clearly stated.
\item Where I have consulted the published work of others, this is always clearly attributed.
\item Where I have quoted from the work of others, the source is always given. With the exception of such quotations, this thesis is entirely my own work.
\item I have acknowledged all main sources of help.
\item Where the thesis is based on work done by myself jointly with others, I have made clear exactly what was done by others and what I have contributed myself.
\end{itemize}
 
\noindent Signed:\qquad\qquad\includegraphics[height=28pt, trim={1pt 1pt 1pt 1pt}, clip]{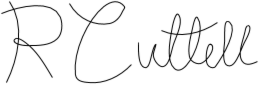}\\
\rule[0.5em]{25em}{0.5pt} 
 
\noindent Date:{\quad\qquad\qquad 1st April 2019}\\
\rule[0.5em]{25em}{0.5pt} 
\end{declaration}

\cleardoublepage


\vspace*{0.2\textheight}

\noindent\enquote{\itshape I really don't know what I'm doing... I don't. It's terrible...}\bigbreak

\hfill Leonardo DiCaprio


\begin{abstract}
\addchaptertocentry{\abstractname} 
In this thesis, I investigate how to construct a self-consistent model of deformed general relativity using canonical methods and metric variables.  The specific deformation of general covariance is predicted by some studies into loop quantum cosmology.

I firstly find the minimally-deformed model for a scalar-tensor theory, thereby establishing a classical reference point, and investigate the cosmological effects of a non-minimal coupled scalar field.
By treating the deformation perturbatively, I derive the deformed gravitational action which includes the nearest order of curvature corrections.
Then working more generally, I derive the deformed scalar-tensor constraint to all orders and I find that the momenta and spatial derivatives from gravity and matter must combine in a very specific form.  It suggests that the deformation should be equally affected by matter field derivatives as it is by gravitational curvature.
Finally, I derive the deformed gravitational action to all orders, and find how intrinsic and extrinsic curvatures differently affect the deformation.  The deformation seems to be required to satisfy a non-linear equation usually found in fluid mechanics.
\end{abstract}


\begin{acknowledgements}
\addchaptertocentry{\acknowledgementname} 

Thanks to my supervisor, Mairi, who patiently facilitated and enabled this, and helped guide me away from dead ends.

Thanks to Martin, who helped me find and correct a serious error in my methodology before it was too late.

Thanks to those in the physics department who have helped so much with navigating through difficult situations, especially Jean, Julia and Rowena.

Thanks to Marc, Brinda, Agnes, Ruth and Gwyn, the medical and mental health professionals that helped me keep afloat.

Thanks to my parents Jeff and Liz, who gave enough encouragement for me to be doing this and on whom I have depended too much.

Thanks to Joanna, for being the perfect big sister I don't deserve, and thanks to Jim for being there for her in turn.

Thanks to Naomi, for always being there with love, support and a goofy joke. I couldn't imagine managing to finish this without you. 
(\;${}^{\varphi}\,\omega{}^{\varphi}$)\includegraphics[height=10pt,trim={1pt 1pt 1pt 1pt},clip]{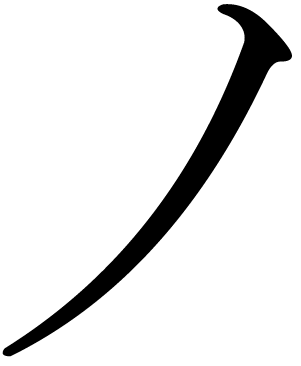}$\,\sim$\includegraphics[height=11pt,trim={1pt 1pt 1pt 1pt},clip]{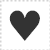}

Thanks to Xin Xin, Jack, Pebbles, Muffin and Tuxedo Kamen for just being you.
\end{acknowledgements}

\begin{spacing}{1}
\tableofcontents 
\listoffigures 
\end{spacing}

\begin{symbols}{ll} 

$g_{ab}$ & space-time metric \\
$q_{ab}$ & spatial metric \\
$q$ & determinant of $q_{ab}$ \\
$\partial_q$ & logarithmic partial derivative with respect to $q$ \\
$Q_{abcd}$ & symmetric combination of two spatial metrics\\
$p^{ab}$ & momentum canonically conjugate to $q_{ab}$ \\
$v_{ab}$ & normal derivative of $q_{ab}$ \\
$\bp$ & traceless part of $p^{ab}$ squared\\
$w$ & traceless part of $v_{ab}$ squared\\
$K_{ab}$ & extrinsic curvature tensor \\
$\K$ & standard extrinsic curvature contraction \\
$\mathcal{L}_{Y}X$ & Lie derivative of $X$ with respect to $Y^a$ \\
$\mathcal{H}$ & Hubble expansion rate \\
$R$ & Ricci curvature scalar\\
$N$ & lapse function \\
$N^a$ & shift vector \\
$n^a$ & future-pointing vector normal to the spatial manifold \\
$S$ & action \\
$L$ & Lagrangian \\
$H$ & total Hamiltonian \\
$C$ & Hamiltonian constraint \\
$D_a$ & diffeomorphism constraint \\
$\partial_a$ & partial derivative with respect to the coordinate $x^a$ \\
$\nabla_a$ & covariant derivative with respect to the coordinate $x^a$ \\
$w_\rho$ & cosmological equation of state \\

\addlinespace 

$\beta$ & deformation function \\
$\psi$ & scalar field \\
$\partial_\psi$ & partial derivative with respect to $\psi$ \\
$\pi$ & momentum canonically conjugate to $\psi$ \\
$\nu$ & normal derivative of $\psi$ \\
$\varphi$ & minimally coupled scalar field \\
$\omega$ & gravitational coupling \\
$\phi_I$ & constraints \\
$\delta^{ab}_{cd}$ & symmetric combination of two Kronecker delta functions \\
$\pi_{\circ}$ & ratio of a circle's circumference to its diameter \\

\end{symbols}


\dedicatory{For Naomi} 


\mainmatter 

\pagestyle{thesis} 



\chapter{Introduction}
\label{sec:intro}





In this thesis I investigate deformed general relativity, which is a semi-classical model attempting to capture the leading effects of a correction to general relativity predicted in some studies of loop quantum gravity.  It uses the methods of canonical gravity but with space-time covariance deformed by a phase-space function.  
By assuming a general deformation, I find the general models which are consistent with it, demonstrating multiple routes which can be taken to find them.






Before going into more depth on this, I must first discuss the motivations for this investigation.


\section{The need for a theory of quantum gravity}
\label{sec:intro_quantum_gravity}

It is known that matter fields are quantised due to the remarkable agreement of experimental results with quantum field theory \cite{Levine:1997zb, Bennett:2006fi, Hanneke:2008tm}.  There have been some attempts to allow for classical gravity to couple to quantum fields at a fundamental level \cite{Ruffini:1969qy, Kibble:1979jn}, and some interesting phenomena have been discovered from considering effective models of quantum fields on a curved space-time \cite{Hawking:1974sw, Unruh:1976db, Bunch:1978yq}.  However, it is generally expected that gravity must be quantised too
\cite{kieferqg, bojowald2010canonical}.
\begin{displayquote}
    The gravitational field, like all other fields, therefore must be quantized, or else the logical structure of quantum field theory must be profoundly altered, or both.
    \cite[B.~DeWitt]{dewittquantization}
\end{displayquote}

Besides gravity being known to couple to quantum fields, there are known limitations to the current common understanding.  General relativity predicts its own demise due to singularities arising in the equations describing black holes and the very early universe \cite{hawkingpenrose}.  They are known to exist due to robust experimental observations supporting the existence of black holes \cite{TheLIGOScientific:2016wfe} and supporting an early universe which closely matches what is predicted of a hot big bang \cite{Planck2018cosmo}.
These phenomena exist at the intersection of general relativity and quantum mechanics since they involve both massive systems and small scales.  It seems they cannot be fully understood without a framework which consistently bridges the gap.

As a precedent for the singularity problem, classical mechanics could not sufficiently account for experimental results showing that atoms contained small, massive nuclei orbited by electrons (the Nagaoka-Rutherford model).  This is due to accelerating point charges (electric field singularities) being known to emit radiation as per the Landau formula, and therefore an electron orbit should radiatively decay, causing atoms to be unstable.
However, the development of quantum mechanics resolved this by introducing discrete and stationary orbitals in the Bohr model.  The hope is that quantising gravity will similarly cure it of some of its pathologies.

One might not want to jettison all that is good about general relativity in pursuit of a quantised theory.
The key underlying idea, equivalence of all frames, is considered a philosophically and aesthetically satisfying aspect.  Conversely, the requirement in the orthodox interpretation of quantum mechanics for an external observer is considered troubling, hence why Einstein spent much of the latter part of his career challenging it \cite{Einstein:1935rr}.



One crucial sticking point in reconciling general relativity and quantum mechanics is the problem of time \cite{Rovelli:1989jn, Isham:1992ms}.  In quantum mechanics time is a fixed external parameter, in general relativity it is internal to the system and is not uniquely defined.  These are seemingly incommensurable differences, and to bridge the gap requires significant compromise.

The solution in canonical gravity for reconciling the two is to split space-time at the formal level, but include symmetry requirements so that the full general covariance is kept implicitly \cite{Arnowitt:1962hi, Gourgoulhon:2007ue, bojowald2010canonical}.  One is left with a description of a spatial slice evolving through time rather than one of a static and eternal bulk.
These methods are often required for numerically simulating general relativity due to the necessity of specifying a time coordinate when setting up an evolution simulation.

\begin{sloppypar}
This introduces on each spatial manifold a conserved quantity or `constraint' given by ${\phi_I\to0}$ for each dimension of time and space, analogous to a generalisation of the conservation of energy and momentum. 
These constraints form an algebra which contains important information about the geometric nature of space-time, and is of the form ${\left\{\phi_I,\phi_J\right\}=f_{IJ}^K\phi_K}$ \cite{dirac1964, bojowald2010canonical}.
This is a Lie algebroid which describes the relationships between the constraints and generates transformations between different choices of coordinates \cite{hojman_geometrodynamics_1976, Bojowald:2016hgh}.
\end{sloppypar}

The important $\{C,C\}$ part of this algebra ensures that the spatial manifold evolving through time is equivalent to a stack of spatial manifolds embedded in a geometric space-time manifold.
\begin{displayquote}
    In this more general case of gravitation in interaction with other fields, [the equation\footnote{the equation referenced in the quote as the same as \eqref{eq:con-alg_CC}}] not only guarantees the embeddability of the 3-geometries in a space-time but also ensures that these additional fields evolve consistently within this space-time. \cite[C.~Teitelboim]{teitelboim_how_1973}
\end{displayquote}
This part of the algebra is what I am going to consider to be deformed, but where does this hypothesis come from?


\section{Loop quantum gravity}
\label{sec:intro_loop}




Though there are several candidates for a theory of quantum gravity, I am going to only consider loop quantum gravity \cite{Thiemann:2002nj, Rovelli:2014ssa}.  There are other somewhat related theories which also deal directly with quantising gravity, such as: causal dynamical triangulations \cite{Ambjorn:2006jf}; causal set theory \cite{Henson:2006kf}; group field theory \cite{Oriti:2006se}; and asymptotically safe gravity \cite{Percacci:2007sz}.
The main alternative candidate is string theory and its variants, which prioritises bringing gravity into the established framework for quantum particles in order to create a unified theory \cite{Polchinski:1994mb, Taylor:2006ye}.

Loop quantum gravity focuses on maintaining some key concepts from general relativity such as background independence and local dynamics throughout the process of combining gravity and quantum mechanics.  It describes space-time as not being a continuous manifold, but instead being a network of nodes connected by ordered links with quantum numbers for geometrical quantities such as volume.  Such a network is not merely embedded in space but \emph{is space itself}.  As such, due to the quantisation of geometry, one cannot shrink the length of a link between nodes to being infinitesimal as in the classical case.

If general relativity is truly the classical limit of loop quantum gravity, then there should be a semi-classical limit where the dynamics are well approximated by general relativity with minor quantum corrections.  These should become larger at small scales and in regions of high curvature.


A closely related theory is loop quantum cosmology, which uses concepts and techniques from loop quantum gravity and applies them directly at the cosmological level by using midi-superspace models \cite{bojowald_quantum_2012, Ashtekar:2011ni}.  That is, by quantising a universe which already has certain symmetries assumed such as isotropy to simplify the process.  There has been some progress towards proving that loop quantum gravity can be symmetry-reduced to loop quantum cosmology, but as yet this has not been shown definitively \cite{Gielen:2013kla, Alesci:2016gub}.

For models of loop quantum cosmology to be self-consistent and anomaly-free while including some of the interesting effects from the discrete geometry, it seems that the algebra of constraints must be deformed.  Specifically, some of the structure functions become more dependent on the phase space variables through a deformation function ${f_{IJ}^K(q)\to{}\beta(q,p)f_{IJ}^K(q)}$ \cite{Bojowald:2008gz, Bojowald:2008bt, Perez:2010pm, Mielczarek:2011ph, Cailleteau2012a, Mielczarek:2012pf, Cailleteau2013}.  Deforming rather than breaking the algebra in principle maintains general covariance but the transformations between different choices of coordinates become highly non-linear \cite{tibrewala_inhomogeneities_2013}.  It becomes less clear to what extent one can still interpret space-time geometrically, at least in terms of classical notions of geometry.

However, there is ambiguity in the correct choice of variables used for loop quantum gravity.  The results cited in the previous paragraph are for real variables for which there has been significant difficulty including matter and local degrees of freedom \cite{Bojowald:2016itl}.  The main alternative, self-dual variables, have had some positive results for including those degrees of freedom without deforming the constraint algebra \cite{BenAchour:2016brs}, but might not have the desired quality of resolving curvature singularities \cite{Brahma:2016tsq}.

Interesting predictions coming from loop quantum gravity include: a bouncing universe \cite{Ashtekar2006}; black hole singularity resolution and transition to white holes \cite{Rovelli:2013osa}; and signature change of the effective metric \cite{Mielczarek:2012pf}.
Some of these predictions are closely associated with a deformation of classical symmetries in regions of high energy density.


\section{Why study deformed general relativity?}
\label{sec:intro_deformed}

Deformed general relativity builds directly from the idea that the constraint algebra is deformed \cite{bojowald_deformed_2012}. It is constructed by taking the deformed constraint algebra, and finding a corresponding model which includes local degrees of freedom \emph{a priori}.
This can be done because, if one starts from an algebra and makes some reasonable assumptions, one can deduce the general form of all the constraints \cite{kuchar_geometrodynamics_1974, hojman_geometrodynamics_1976}.  This should provide a more intuitive understanding of how the deformation affects dynamics and may provide a guide for how to include the problematic degrees of freedom when working with real variables in loop quantum gravity.

The constraint algebra is important because, as said previously, it closely relates to the structure of space-time \cite{teitelboim_how_1973}.  
Quantum geometry will behave differently to classical geometry, and deformed general relativity attempts to capture some of the effects in a semi-classical model which is more amenable to phenomenological investigations.

Phenomenological models which are comparable to deformed general relativity, such as deformed special relativity \cite{AmelinoCamelia:2000mn} and rainbow gravity \cite{Magueijo:2002xx}, struggle to go beyond describing individual particles coupled to an energy-dependent metric.  They can suffer from a breakdown of causality \cite{hossenfelder_box-problem_2009}, or find it difficult to describe multi-particle states \cite{hossenfelder_multi-particle_2007}.  Deformed general relativity does not suffer from these problems by construction.


\section{Overview of this thesis}
\label{sec:intro_overview}



The main focus of this thesis is to investigate how to construct a self-consistent  model of deformed general relativity using canonical methods and metric variables.
I review important concepts and methodology in \chapref{sec:methodology}.
In \chapref{sec:2ndst}, I find the minimally-deformed model for a scalar-tensor theory, establishing a classical reference point.
Then in \chapref{sec:pert}, I derive the deformed gravitational action which includes the lowest non-trivial order of perturbative curvature corrections coming from the deformation.
In \chapref{sec:allst}, I derive the deformed scalar-tensor constraint to all orders and I find that the momenta and space derivatives must combine in a specific form.
Finally, in \chapref{sec:allact}, I find the deformed gravitational action to all orders, and find how intrinsic and extrinsic curvatures differently affect the deformation.
I identify some of the cosmological consequences for the significant results of each chapter.

There are several research questions which I attempt to answer in this thesis.
How are the form of the deformation function and the form of the model related?
In particular, what is the deformed scalar-tensor Hamiltonian and what is the deformed gravitational Lagrangian, using either perturbative or non-perturbative methods? How do they relate to the classical limit and to each other?
How can matter fields be incorporated in deformed models?
How does the deformation function depend on curvature, and is it different for intrinsic and extrinsic curvatures?

The research chapters \ref{sec:2ndst} and \ref{sec:pert} are adapted from the previously published papers \cite{cuttell2018} and \cite{cuttell2014}, respectively.  The other research chapters, \ref{sec:allst} and \ref{sec:allact}, were recently submitted for publication \cite{cuttellconstraint, cuttellaction}


\section{Wider impact}
\label{sec:intro_impact}

This study is directly motivated by the prediction of a deformed constraint algebra appearing in loop quantum cosmology\cite{Bojowald:2008gz, Bojowald:2008bt, Perez:2010pm, Mielczarek:2011ph, Cailleteau2012a, Mielczarek:2012pf, Cailleteau2013}.  As such it should provide insight into the lingering questions of how matter and local degrees of freedom need to be incorporated into the motivating theory in the presence of a deformation, and how spatial and time derivatives are differently affected.

There are also potentially wider implications for this study.
For example, it has been shown that taking the deformed constraint algebra to the flat-space limit gives a deformed version of the Poincar\'{e} algebra, which leads to a modified dispersion relation \cite{Bojowald:2012ux, Brahma:2016tsq}.  This might indicate something such as a variable speed of light or an observer-independent energy scale.  In this respect it is similar to the phenomenological models of deformed special relativity \cite{AmelinoCamelia:2000mn} and rainbow gravity \cite{Magueijo:2002xx}.

The deformation might indicate a non-commutative character to geometry \cite{Amelino-Camelia:2016gfx, Bojowald:2017kef} although apparently not a multifractional one \cite{Calcagni:2016ivi}.  It might represent a variable dimensionality of space-time and a running of the spectral dimension \cite{Mielczarek:2016zfz}.
The deformation function may change sign, as suggested in the motivating studies \cite{Mielczarek:2012pf}.  This makes the hyperbolic equations become elliptical and implies a phase transition from classical Lorentzian space-time to an effectively Euclidean quantum regime \cite{Bojowald:2016hgh, Bojowald:2016vlj}.  It therefore may be a potential mechanism for the Hartle-Hawking no-boundary proposal \cite{Hartle1983}.

\chapter{Methodology}
\label{sec:methodology}

In this thesis I am primarily building on preceding work done by others \cite{kuchar_geometrodynamics_1974, hojman_geometrodynamics_1976, bojowald_deformed_2012} and elaborating on previously published material \cite{cuttell2014, cuttell2018}.

\section{Space-time decomposition}
\label{sec:methodology_decomposition}

Quantum mechanics naturally works in the canonical or Hamiltonian framework.
The canonical framework takes variables defined at a certain time and evolves them through time.  That evolution defines a canonical momentum for each variable.
To make general relativity more amenable to quantum mechanics, one must likewise make a distinction between the time dimension and the spatial dimensions.
So I foliate the bulk space-time manifold $\mathcal{M}$ into a stack of labelled spatial hypersurfaces, $\Sigma_t$. I assume it is globally hyperbolic, so topologically $\mathcal{M}=\Sigma\times\mathbb{R}$ \cite{bojowald2010canonical, Arnowitt:1962hi, Gourgoulhon:2007ue}.

\begin{sloppypar}
A future-pointing vector normal to the spatial hypersurface $\Sigma_t$ is defined such that ${g_{ab}n^{a}n^{b}=-1}$.  The spatial slices $\Sigma_t$ are themselves Riemannian manifolds with an induced metric
${q_{ab}=g_{ab}+n_{a}n_{b}}$, such that ${q_{ab}n^{b}=0}$.  The spatial metric has an inverse defined as 
${q^{ab}=g^{ab}+n^{a}n^{b}}$,
so that 
${q_{a}^{b}:=q_{ac}q^{bc}=\delta_{a}^{b}+n_{a}n^{b}}$
acts as a spatial\footnote{by `spatial', I mean tangential to the spatial manifold} projection tensor.
\end{sloppypar}

If the spatial foliation, and therefore the spatial coordinates, are arbitrary, the time-evolution vector field $t^{a}$ cannot be uniquely determined by the time function $t$. One can project it into its normal and spatial components, defining the lapse function $N=-n_{a}t^{a}$, and the spatial shift vector $N^{a}=q^{a}_{b}t^{b}$.  Therefore, $t^{a}=Nn^{a}+N^{a}$.

Since the coordinates are arbitrary, it is convenient to take the normal to the spatial surface as the time-like direction for defining velocities rather than using the time-vector itself.  Therefore,
\begin{equation}
\begin{split}
    v_{ab} : = \mathcal{L}_n q_{ab} 
    = \frac{1}{N} \big( \dot{q}_{ab} - 2 \nabla_{(a} N_{b)} \big),
\quad
    \nu_I := \mathcal{L}_n \psi_I 
    = \frac{1}{N} \big( \dot{\psi}_{I} - N^a \partial_a \psi_I \big),
\end{split}
        \label{eq:methodology_normalderivatives}
\end{equation}
where ${\dot{X}:=\mathcal{L}_tX}$,
and the extrinsic curvature of the spatial slice is related to this by ${K_{ab}=\half{}v_{ab}}$.


\section{Canonical formalism}

I take a general first-order action for a model with dynamical fields $\psi_I$, and non-dynamical fields $\lambda_I$,
\begin{equation}
    S = \int \mathrm{d}^4 x \, L \left( \psi_I, \partial_a \psi_I, \lambda_I \right),
\end{equation}
where $\displaystyle\partial_a\psi_I:=\partdif{\psi_I}{x^a}=:\psi_{I,a}$.
Varying the action with respect to each field, fixing the variation at the boundaries, and imposing the principle of least action,
\begin{equation}
    \funcdif{ S }{ \psi_I } \approx 0,
\quad
    \funcdif{ S }{ \lambda_I } \approx 0,
\end{equation}
gives the Euler-Lagrange equations of motion,
\begin{subequations}
\begin{gather}
    0 \approx \partdif{ L }{ \psi_I } - \partial_a \left( \partdif{ L }{ \left( \partial_a \psi_{I} \right) } \right),
        \label{eq:euler-lagrange_equation}
\\
    0 \approx \partdif{ L }{ \lambda_I }.
        \label{eq:euler-lagrange-multiplier}
\end{gather}
    \label{eq:euler-lagrange}%
\end{subequations}
The approximation symbol is used to indicate something that is true in the dynamical regime, or `on-shell', rather than something that is true kinematically, or `off-shell'.  The non-dynamical fields $\lambda_I$ can be seen to produce constraints on the system given by \eqref{eq:euler-lagrange-multiplier}, they are also known as Lagrange multipliers.

Making a space-time decomposition as in \secref{sec:methodology_decomposition}, one can define the canonical momenta of each field,
\begin{equation}
    \pi_{\psi}^{I} : = \funcdif{ S }{ \dot{ \psi }_I } = \partdif{ L }{ \dot{ \psi }_I },
\quad
    \pi_{\lambda}^{I} : = \funcdif{ S }{ \dot{ \lambda }_I } = \partdif{ L }{ \dot{ \lambda }_I }.
\end{equation}
Since $L$ does not depend on $\dot{\lambda}_I$, one can see that $\pi_{\lambda}^{I}\approx0$ are primary constraints on the system.
If the matrix $\displaystyle\partdif{^2L}{\dot{\psi}_I\partial\dot{\psi}_J}$ is non-degenerate, then the above equation can be inverted to find $\dot{\psi}_I=\dot{\psi}_I(\psi_J,\pi_\psi^J,\lambda_J)$, and so one can replace the time derivatives in the action.
Making a Legendre transform to find the Hamiltonian associated to this action,
\begin{equation}
    H = \int \mathrm{d} t \mathrm{d}^3 x \left( 
        \sum_I \dot{\psi}_I \pi_\psi^I 
        + \sum_I \mu^\lambda_I \pi_\lambda^I
    \right)
    - S,
\end{equation}
where $\mu^\lambda_I$ is a coefficient which acts like a Lagrange multiplier.
The Poisson bracket of a quantity with the Hamiltonian equals the time derivative of that quantity on-shell,
\begin{equation}
    \dot{F} \approx \left\{ F, H \right\}
    = \int \mathrm{d}^3 x \left\{ 
        \sum_I \funcdif{ F }{ \psi_I (x) } \funcdif{ H }{ \pi_\psi^I (x) }
        + \sum_I \funcdif{ F }{ \lambda_I (x) } \funcdif{ H }{ \pi_\lambda^I (x) }
    \right\} 
    - \left( F \leftrightarrow H \right),
    \label{eq:methodology_timederivative}
\end{equation}
and if $F\approx0$ should be true at all times, then $\dot{F}\approx0$ must also be true \cite{dirac1964}.  Therefore, evaluating $\{\pi_\lambda^I,H\}$ either gives back a function of the primary constraints $\pi_\lambda^J$, produces a secondary constraint $\phi_I(\psi_J,\pi_\psi^J,\lambda_J)\approx0$, or gives a specific form for the coefficients of the constraints $\mu_I$.  The equations \eqref{eq:euler-lagrange-multiplier} appear here as secondary constraints.  

I repeat the process with $\{\phi_I,H\}$ until I have found all the constraints on the system, at which point there is no need to differentiate between primary and secondary constraints, and I have found the generalised Hamiltonian,
\begin{equation}
    H^\star = \int \mathrm{d} t \mathrm{d}^3 x \left( 
        \sum_I \dot{\psi}_I \pi_\psi^I
        + \sum_I \mu_I \phi_I
    \right)
    - S \approx H.
\end{equation}
The set of constraints has a Poisson bracket structure
\begin{equation}
    \{ \phi_I, \phi_J \} = f_{IJ}^K \phi_K + \alpha_{IJ},
\quad
    \alpha_{IJ} \notin \{ \phi_K \},
    \label{eq:con-alg_general}
\end{equation}
and if $\alpha_{IJ}\neq0$ then some of $\phi_I$ are what are called `second-class' constraints, in which case some of the coefficients $\mu_I$ are uniquely determined.  If $\alpha_{IJ}=0$ then all of $\phi_I$ are `first-class', in which case the constraints not only restrict the values of the dynamical fields, but also generate gauge transformations \cite{dirac1964, bojowald2010canonical}.  This is because, in general the evolution \eqref{eq:methodology_timederivative} will depend on $\mu_I$.  For an undetermined $\mu_I$ to influence the mathematics but not the physical observables, a change of its value must correspond to a gauge transformation generated by the relevant first-class constraint.

For classical general relativity, the action does not depend on $\dot{N}$ or $\dot{N}^a$ (up to boundary terms) and is only linearly dependent on $N$ and $N^a$.%
\footnote{Or rather, it is only linearly dependent on $N$ and $N^a$ when velocities are represented by normal derivatives \eqref{eq:methodology_normalderivatives}.}
As such, there are primary constraints given by $\pi_N$ and $\pi^N_a$, which generate secondary constraints known as the Hamiltonian constraint and diffeomorphism constraint respectively,
\begin{equation}
    C : = \funcdif{ H }{ N } = \left\{ H, \pi_N \right\},
\quad\quad
    D_a : = \funcdif{ H }{ N^a } = \left\{ H, \pi^N_a \right\},
\end{equation}
which are all first-class constraints.  This means that $N$ and $N^a$ are gauge functions which do not affect the observables, and therefore the spatial slicing does not affect the dynamics.  The theory is background independent and the constraints generate gauge transformations%
\footnote{The square brackets indicates the constraint is `smeared' over the spatial surface using the function in the brackets, e.g. $C[N]=\int\mathrm{d}^3xN(x)C(x)$.},
\begin{equation}
    \{ F, C [ N ] \} = N \mathcal{L}_n F,
\quad
    \{ F, D_a [ N^a ] \} = \mathcal{L}_N F.
    \label{eq:methodology_gauge_transformations}
\end{equation}
The Hamiltonian can be rewritten as a sum of the constraints up to a boundary term,
\begin{equation}
    H = \int \mathrm{d} t \mathrm{d}^3 x \left(
        N C + N^a D_a + \mu_N \pi_N + \mu_N^a \pi^N_a
    \right).
\end{equation}

\begin{sloppypar}
Considering the Poisson bracket structure of these constraints, given by \eqref{eq:con-alg_general} with ${\phi_I\in\{C,D_a\}}$, one finds that they form a Lie algebroid%
\footnote{`Algebroid' refers to the fact that some of the structure coefficients $f^K_{IJ}$ are phase space functions}\cite{Bojowald:2016hgh},\end{sloppypar}
\vspace{-2\baselineskip}
\begin{subequations}
\begin{align}
    \big\{ D_a [N_1^a], D_b [N_2^b] \big\} & = D_a \big[ \mathcal{L}_{N_2} N_1^a \big],
        \label{eq:con-alg_DD}
        \\
    \big\{ C [N_1], D_a [N_2^a] \big\} & = C \big[ \mathcal{L}_{N_2} N_1 \big],
        \label{eq:con-alg_CD}
        \\
    \big\{ C [N_1], C [N_2] \big\} & = D_a \big[ \, q^{ab} \left( N_1 \partial_b N_2 - \partial_b N_1 N_2 \right) \big].
        \label{eq:con-alg_CC}
\end{align}
    \label{eq:con-alg}%
\end{subequations}
where $(N_1,N_1^a)$ and $(N_2,N_2^a)$ each represent the lapse and shift of two different hypersurface transformations.
As interpreted in ref.~\cite{teitelboim_how_1973}, \eqref{eq:con-alg_DD} shows that $D_a$ is the generator of spatial morphisms, \eqref{eq:con-alg_CD} shows that $C$ is a scalar density of weight one (as defined in \appref{sec:diff}) and \eqref{eq:con-alg_CC} specifies the form of $C$ such that it ensures the embeddability of the spatial slices in space-time geometry.


\section{Choice of variables}
\label{sec:methodology_variables}

Classical canonical general relativity can be formulated equivalently using different variables.  There is geometrodynamics, which uses the spatial metric and its canonical momentum $(q_{ab}, p^{cd})$, the latter of which is directly related to extrinsic curvature,
\begin{equation}
    p^{ab} = \frac{ \omega }{ 2 } \sqrt{ q } \left( K^{ab} - K q^{ab} \right),
\end{equation}
where $q:=\det{q_{ab}}$ and $\omega$ is the gravitational coupling.
An alternative is connection dynamics, which uses the Ashtekar-Barbero connection and densitised triads $(A^I_a, E^b_J)$, where capital letters signify internal indices rather than coordinate indices \cite{Ashtekar:1986yd, Barbero:1994ap}. This can be related to geometrodynamics by using the equations \cite{bojowald2010canonical},
\begin{subequations}
\begin{gather}
    q \, \delta_{IJ} = q_{ab} E^a_I E^b_J ,
\\
    A_a^I = \Gamma_a^I + \BI K_a^I ,
\\
    \Gamma_a^I = \frac{ 1 }{ 2 \sqrt{ q } } q_{bc} \epsilon^{IJK} E^b_J \nabla_a \left( \frac{ E^c_{K} }{ \sqrt{ q } } \right) ,
\\
    K_a^I = \frac{ 1 }{ \sqrt{ q } } \delta^{IJ} K_{ab} E^{b}_J ,
\end{gather}
\end{subequations}
where $\BI$ is the Barbero-Immirzi parameter and $\epsilon^{IJK}$ is the covariant Levi-Civita tensor.  The exact value of $\BI$ should not affect the dynamics \cite{Immirzi:1996dr}.

The other alternative I mention here is loop dynamics, which uses holonomies of the connection and gravitational flux $(h_{\ell}[A],F^I_{\ell}[E])$.  Classically, $h_{\ell}[A]$ is given by the path-ordered exponential of the connection integrated along a curve ${\ell}$ and $F^I_{\ell}[E]$ is the flux of the densitised triad through a surface that the curve ${\ell}$ intersects. If ${\ell}$ is taken to be infinitesimal, one can easily relate loop dynamics and connection dynamics because then $h_{\ell} = 1 + A(\dot{\ell}) + \mathcal{O}(|{\ell}|^2)$ \cite[p.~21]{Rovelli:2014ssa}.

When each set of variables is quantised, they are no longer equivalent, for example the value of $\BI$ does now affect the dynamics \cite{Ashtekar:1997yu, Brahma:2016tsq}. For complex $\BI$, care has to be taken to make sure the classical limit is real general relativity, rather than complex general relativity.  Significantly, quantising loop variables (loop quantum gravity) discretises geometry, and so $\ell$ cannot be taken to be infinitesimal \cite[p.~105]{Rovelli:2014ssa}.

In this work, I choose to use metric variables to build a semi-classical model of gravity. This is because the comparison to other modified gravity models should be clearer, and there is no ambiguity arising from $\BI$.

\section{Higher order models of gravity}
\label{sec:methodology_higher}

In four dimensions, the Einstein-Hilbert action for general relativity is given by 
\begin{equation}
    S = 
    \frac{ \omega }{ 2 } \int \mathrm{d}^4 x 
    \sqrt{ - g } \, \four{R}.
\end{equation}    
where $\omega=1/8\pi_{\circ}{G}$ is the gravitational coupling and $g:=\det{g_{ab}}$.
The integrand is the four dimensional Ricci curvature scalar which is contracted from the Riemann curvature tensor $\four{R}:=\four{R}^{a}_{\;\;bac}g^{bc}$. For any Riemannian manifold, this is defined using the commutator of two covariant derivatives of an arbitrary vector, 
\begin{equation}
    \nabla_{c} \nabla_{d} A^{a} - \nabla_{d} \nabla_{c} A^{a} = R^{a}_{\;\;bcd} A^{b}.
\end{equation}

There are many reasons why theoretical physicists seek to find models of gravity which go beyond the Einstein-Hilbert action.  For instance, mysteries known as dark matter \cite{Garrett:2010hd} and dark energy \cite{Li:2011sd} may originate with gravity behaving differently than expected rather than being due to unknown dark substances \cite{Clifton:2011jh}.  The indication that there was a period of inflationary expansion in the early universe has also caused a search for relevant models \cite{Starobinsky:1980te, Linde:1981mu}.
Moreover, the classical equations of gravity predict their own demise in extraordinary circumstances such as in a black hole or at a hot big bang.  A theory of gravity that solves these problems to which classical general relativity is the low-curvature, large-scale limit may have a semi-classical regime where corrections appear, at leading orders, similar to these theories of modified gravity \cite{Starobinsky:1980te, Machado:2007ea, Klinkhamer:2008ff}.

One way of attempting to find alternative models of gravity is by constructing actions from higher order combinations of the Riemann tensor, so you instead have the general action
\begin{equation}
    S = \frac{ \omega }{ 2 } 
    \int \mathrm{d}^4 x \sqrt{ - g }
    F \left( \four{R}^{a}_{\;\;bcd} \right).
\end{equation}
To bring this in line with the space-time split, I replace the determinant, 
$ g = - N^2 q $.
The Riemann tensor must be decomposed by projecting it along its normal and tangential components relative to the spatial slice,
\begin{subequations}
\begin{align}
    q_a^e q_b^f q_c^g q_d^h \, \four{R}_{efgh} & =
    \quarter v_{ac} v_{bd} - \quarter v_{ad} v_{bc}
    + {}^{(3)}\!R_{abcd},
\label{eq:riemann_identities_gauss} \\
    q_a^e q_b^f q_c^g n^h \, \four{R}_{efgh} & =
    \half \nabla_a v_{bc} - \half \nabla_b v_{ac},
\label{eq:riemann_identities_codazzi} \\
    q_a^e n^f q_b^g n^h \, \four{R}_{efgh} & =
    - \half \mathcal{L}_n v_{ab} + \quarter q^{bc} v_{ac} v_{bd} + \frac{1}{N} \nabla_{(a} \nabla_{b)} N.
\label{eq:riemann_identities_ricci}
\end{align}%
    \label{eq:riemann_identities}%
\end{subequations}
These identities are respectively known as the Gauss equation, the Codazzi equation, and the Ricci equation \cite{bojowald2010canonical, Deruelle2010}.  All other projections vanish due to the tensor's antisymmetry.  As can be seen from \eqref{eq:riemann_identities_ricci}, there are second order time derivatives included in the Riemann tensor.  Including second order time derivatives in an action is problematic because it may introduce the Ostrogradsky instability \cite{Woodard:2006nt}. 
To demonstrate what this means, I take a one dimensional model action,
\begin{equation}
    S = \int \mathrm{d} t L \left( q, \dot{q}, \ddot{q} \right),
        \label{eq:methodology_higher_test_1}
\end{equation}
I cannot find the associated Hamiltonian when there are time derivatives higher than second order, and the Euler-Lagrange equations may involve fourth order time derivatives,
\begin{equation}
    0 \approx 
    \partdif{ L }{ q }
    - \totdif{}{t} \left( \partdif{ L }{ \dot{q} } \right)
    + \totdif{^2}{t^2} \left( \partdif{ L }{ \ddot{q} }\right) ,
        \label{eq:methodology_higher_test_2}
\end{equation}
if 
$\displaystyle{\partdif{^2L}{\ddot{q}^2}\neq0}$. 
So I must introduce an additional variable to absorb the higher order terms.  The Ostrogradsky method \cite{Schmidt:1994iz} is to replace $\dot{q}$ with an independent variable $v$.
\begin{equation}
    S = \int \mathrm{d} t \left\{ L \left( q, v, \dot{v} \right) + \psi \left( v - \dot{q} \right) \right\},
        \label{eq:methodology_higher_test_3}
\end{equation}
however, I instead do this slightly differently for reasons which will be apparent later.  Following the method used in ref.~\cite{Deruelle:2009pu, Deruelle2010} and using variables like in ref.~\cite{Hawking:1984ph}, I instead replace $\ddot{q}$ with an auxiliary variable $a$,
\begin{equation}
    S = \int \mathrm{d} t \left\{
        L \left( q, \dot{q}, a \right) + \psi \left( \ddot{q} - a \right)
    \right\},
        \label{eq:methodology_higher_test_4}
\end{equation}
and integrate by parts to move the second order time derivative to the Lagrange multiplier $\psi$, promoting it to a dynamical variable,
\begin{equation}
    S = \int \mathrm{d} t \left\{
        L \left( q, \dot{q}, a \right) - \dot{q} \dot{\psi} - \psi a
    \right\},
        \label{eq:methodology_higher_test_5}
\end{equation}
which gives the canonical momenta,
\begin{equation}
    p := \funcdif{ S }{ \dot{q} } = \partdif{ L }{ \dot{q} } - \dot{\psi} ,
\quad
    \pi := \funcdif{ S }{ \dot{\psi} } = - \dot{q},
\quad
    \pi_{a} := \funcdif{ S }{ \dot{a} } = 0.
        \label{eq:methodology_higher_test_6}
\end{equation}
So I can invert these definitions to find the velocities in terms of the momenta.  Then make a Legendre transform to find the Hamiltonian,
\begin{equation}
\begin{split}
    H & = \int \mathrm{d} t \left( \dot{q} p + \dot{\psi} \pi + \mu_{a} \pi_{a} \right) - S,
\\ &
    = \int \mathrm{d} t \left\{ 
        - p \pi + \mu_{a} \pi_{a} - L \left( q, \pi, a \right) + \psi a
    \right\},
\end{split}
    \label{eq:methodology_higher_test_7}
\end{equation}
where $\mu_a$ is a Lagrange multiplier.
The equation of motion for $a$ produces the secondary constraint $\displaystyle{\phi=\partdif{L}{a}-\psi\approx0}$.  
Finding ${\{\phi,H\}\approx0}$ produces an equation for $\mu_a$ and therefore $\phi$ is a second-class constraint and $a$ is uniquely determined.  The constraint can be solved for $a\left(q,\psi,\pi\right)$ as long as 
$\displaystyle{\partdif{^2L}{a^2}\neq0}$ 
and this can be substituted into the Hamiltonian without incident, in which case I find,
\begin{equation}
    H = \int \mathrm{d} t \left\{ 
        - p \pi 
        - L \left( q, \psi, \pi \right) 
        + \psi \, a \left( q, \psi, \pi \right)
    \right\}
        \label{eq:methodology_higher_test_8}
\end{equation}
which is only linear in $p$.  This means that the energy is unbounded from below and above, and so the model may be unstable \cite{Schmidt:1994iz}.  For specific models of this kind rather than this simple example, I can find a well behaved Hamiltonian if there are sufficient restrictions on the values that $\psi$ can take \cite{Hawking:1984ph}.  

If I do have a well behaved Hamiltonian, it is clear that the higher order derivative action $L(q,\dot{q},\ddot{q})$ contains an additional degree of freedom, which has been absorbed by $\psi$.

\subsection{Non-minimally coupled scalar from \texorpdfstring{$F\left(\four{R}\right)$}{higher order} gravity}
\label{sec:methodology_FR}

In ref.~\cite{Deruelle:2009pu, Deruelle2010}, it was shown how to find the Hamiltonian form of any $F\left(\four{R}^{a}_{\;\;bcd}\right)$ action.  The Riemann tensor is split into its normal and tangential components \eqref{eq:riemann_identities}, and auxiliary tensors are introduced as in \eqref{eq:methodology_higher_test_4}.  The tensor which is the Lagrange multiplier of \eqref{eq:riemann_identities_ricci} becomes dynamical by integrating by parts.  This turns the action into being first order in time derivatives, and therefore one can find the associated Hamiltonian.  This field contains the additional degrees of freedom allowed by the higher order derivatives.

To include tensor contractions such as $\four{R}^{ab}\,\four{R}_{ab}$ and $\four{R}^{abcd}\,\four{R}_{abcd}$ produces several additional degrees of freedom, and requires considering spatial derivatives of velocity or momenta because of \eqref{eq:riemann_identities_codazzi}.  For the sake of simplicity, in this chapter and throughout the thesis, I will only consider models which are comparable with $F\left(\four{R}\right)$.  So the action is given by,
\begin{equation}
    S = \frac{\omega}{2} \int \mathrm{d}t \mathrm{d}^3x N \sqrt{ q } \left\{ F \left( \rho \right) + \psi \left( \Rfour - \rho \right) \right\}.
        \label{eq:action_rho}
\end{equation}
I decompose the Ricci scalar using \eqref{eq:riemann_identities},
\begin{equation}
    \four{R} = R + q^{ab} \mathcal{L}_n v_{ab} + \quarter v^2
    - \frac{3}{2} v_{ab} v^{ab} - \frac{2}{N} \Delta N,
\quad
    R = \three{R},
        \label{eq:ricci_decomposition}
\end{equation}
where 
${\Delta:=q^{ab}\nabla_a\nabla_b}$.
Then integrate the action \eqref{eq:action_rho} by parts to move the second order time derivative to $\psi$,
\begin{equation}
    S = \frac{\omega}{2} \int \mathrm{d}t \mathrm{d}^3x N \sqrt{q} \left\{ F \left( \rho \right) + \psi \left( R - \mathcal{K} - \frac{2}{N} \Delta N - \rho \right) - \nu v \right\},
\end{equation}
where $q:=\det{q_{ab}}$, $\nu:=\mathcal{L}_n\psi$, and 
${\mathcal{K}:=\left(v^2-v_{ab}v^{ab}\right)/4}$ 
is the standard extrinsic curvature contraction.  The conjugate momenta are,
\begin{subequations}
\begin{align}
\begin{split}
    p^{ab} & := \funcdif{ S }{ \dot{q}_{ab} } = \frac{ 1 }{ N } \funcdif{ S }{ v_{ab} } 
\\ &
    = \frac{ \omega }{ 2 } \sqrt{ q } \left\{ \frac{ \psi }{ 2 } v_{cd} \left( Q^{abcd} - q^{ab} q^{cd} \right) - \nu q^{ab} \right\},
\end{split}
        \\
    \pi & := \funcdif{ S }{ \dot{\psi} } = \frac{ 1 }{ N } \funcdif{ S }{ \nu } = \frac{ - \omega }{ 2 } \sqrt{ q } \, v,
\end{align}
        \label{eq:geo_momenta}%
\end{subequations}
where $Q^{abcd}:=q^{a(c}q^{d)b}$ for convenience.  I can invert these to find,
\begin{equation}
    v_{ab} = \frac{2}{\omega\sqrt{q}} \left( \frac{2}{\psi} p^\T_{ab} - q_{ab} \pi \right),
        \quad
    \nu = \frac{2}{3\omega\sqrt{q}} \left( \psi \pi - p \right).
        \label{eq:geo_velocities}
\end{equation}
where I have separated the trace and the traceless parts of the momentum,
\begin{equation}
    p^{ab} = p_\T^{ab} + \third q^{ab} p .
    \label{eq:momentum_split}
\end{equation}

I Legendre transform the action to find the associated Hamiltonian,
\begin{equation}
\begin{split}
    H & = \int \mathrm{d}^3 x \left( \dot{q}_{ab} p^{ab} + \dot{\psi} \pi + \mu_\rho \pi_\rho + \mu_N \pi_N + \mu^N_a \pi_N^a \right) - S ,
\\
    & = \int \mathrm{d}^3 x \left( N C + N^a D_a + \mu_\rho \pi_\rho + \mu_N \pi_N + \mu^N_a \pi_N^a \right),
    \label{eq:geo_legendre_transform}
\end{split}
\end{equation}
with the corresponding Hamiltonian constraint,
\begin{equation}
    C := \funcdif{H}{N}
    = \frac{2}{\omega\sqrt{q}} \left( \frac{1}{\psi} \bp - \third p \pi + \frac{\psi}{6} \pi^2 \right)
    + \frac{\omega\sqrt{q}}{2} \bigg( \psi \rho - \psi R - F \left( \rho \right) + 2 \Delta \psi \bigg),
        \label{eq:geo_constraint}
\end{equation}
where $\bp:=p^\T_{ab}p_\T^{ab}$.
Finding $\{\pi_\rho,H\}$ gives a secondary constraint,
\begin{equation}
    \phi_\rho = \frac{ \omega }{ 2 } N \sqrt{q} \, \Big( \psi - F^{\,\prime} \left( \rho \right) \Big) \approx 0,
\end{equation}
which is second-class.  It can be solved to find $\rho(\psi)$ as long as $F''\neq0$, in which case we can find the Hamiltonian constraint in terms of only the metric and the scalar field $\psi$. This leaves me with a term depending on $\psi$ which acts like a scalar field potential,
\begin{equation}
    U_\mathrm{geo} \left( \psi \right) 
    = \frac{\omega}{2} \Big( \psi \rho \left( \psi \right) - F \big( \rho \left( \psi \right) \big) \Big)
    = \frac{ \omega }{ 2 } \Big\{ \psi \big( F' \big)^{-1} ( \psi ) - F \Big( \big( F' \big)^{-1} ( \psi ) \Big) \Big\},
        \label{eq:geo_potential}
\end{equation}
which I call the geometric scalar potential.
As I will further elaborate in \secref{sec:2ndst}, this scalar-tensor model I have derived from an $F\left(\four{R}\right)$ model of gravitation is equivalent to letting the gravitational coupling in the Einstein-Hilbert action become dynamical, ${\omega\to\omega\psi}$.

So models of gravity that have an action which is an arbitrary function of the space-time curvature scalar $\four{R}$ can be converted into a scalar-tensor theory in the Hamiltonian formalism.  The structure of general covariance underlying general relativity should be preserved in these models, though they do contain an additional degree of freedom.


\section{Deformed constraint algebra}

As previously mentioned in \secref{sec:intro_loop}, loop quantum cosmology predicts that the symmetries of general relativity should be deformed in a specific way in the semi-classical limit \cite{Bojowald:2008gz, Bojowald:2008bt, Perez:2010pm, Mielczarek:2011ph, Cailleteau2012a, Mielczarek:2012pf, Cailleteau2013}.  This appears from incorporating loop variables in a mini-superspace model, but specifying that all anomalies $\alpha_{IJ}$ in \eqref{eq:con-alg_general} vanish while allowing counter-terms to deform the classical form of the algebra.
This ensures that the constraints are first-class, retaining the gauge invariance of the theory and of the arbitrariness of the lapse and shift.  
If anomalous terms \emph{were} to appear in the constraint algebra, then the gauge invariance would be broken and the constraints could only be solved at all times for specific $N$ or $N^a$.  This means that there would a privileged frame of reference, and therefore no general covariance.

In the referenced studies, it is strongly indicated that the bracket of two Hamiltonian constraints \eqref{eq:con-alg_CC} is deformed by a phase space function $\beta$,
\begin{equation}
    \{ C [N_1] , C [N_2] \} = D_a [ \beta q^{ab} \left( N_1 \partial_b N_2 - \partial_b N_1 N_2 \right) ].
    \label{eq:con-alg_def}
\end{equation}
This has not been shown generally, but has been shown for several models independently.
There are no anomalies in the constraint algebra, so a form of general covariance is preserved.  However, it may be that the interpretation of a spatial manifold evolving with time being equivalent to a foliation of space-time (also known as `embeddability') is no longer valid.

These deformations only appear to be necessary for models when the Barbero-Immirzi parameter $\BI$ is real.  For self-dual models, when $\BI=\pm{i}$, this deformation does not appear necessary \cite{BenAchour:2016brs}.  However, self-dual variables are not desirable in other ways.  They do not seem to resolve curvature singularities as hoped, and obtaining the correct classical limit is non-trivial \cite{Brahma:2016tsq}.  Because of this, even though I use metric variables in this work, considering $\beta\neq1$ and ensuring the correct classical limit means there should be relevance to the models of loop quantum cosmology with real $\BI$.


\section{Derivation of the distribution equation}
\label{eq:methodology_dist-eqn}

From the constraint algebra, I am able to find the specific form of the Hamiltonian constraint $C$ for a given deformation $\beta$.  The diffeomorphism constraint $D_a$ is not affected when the deformation is a weightless scalar%
\footnote{See \appref{sec:diff} for information about weight.}
and so is completely determined as shown in \appref{sec:diff}.  With $D_a$ and $\beta$ as inputs, I can find $C$ by manipulating \eqref{eq:con-alg_def}.

Firstly, I must find the unsmeared form of the deformed algebra.  At this point I do not need to specify my canonical variables, and leave them merely as $\left(q_I,p_I\right)$,
\begin{subequations}
\begin{align}
    0 & = \{ C [N_1] , C [N_2] \} - D_a [ \beta q^{ab} \left( N_1 \partial_b N_2 - \partial_b N_1 N_2 \right) ],
\\
    & = \int \mathrm{d}^3 z \left\{ 
        \sum_I \funcdif{ C [N_1] }{ q_I(z) } \funcdif{ C [N_2] }{ p_I (z) }
        - \left( D^a \beta N_1 \partial_a N_2 \right)_z
    \right\}
    - \left( N_1 \leftrightarrow N_2 \right).
    \label{eq:con-alg_def_expanded}
\end{align}
\end{subequations}
Take the functional derivatives with respect to $N_1(x)$ and $N_2(y)$,
\begin{equation}
    0 = \sum_I \int \mathrm{d}^3 z
        \funcdif{ C (x) }{ q_I(z) } \funcdif{ C (y) }{ p_I (z) }
    - \left( D^a \beta \partial_a \right)_x \delta \left( x, y \right)
    - \xty,
\end{equation}
where $\delta(x,y)$ is the three dimensional Dirac delta distribution\footnote{Defined such that $\funcdif{q_I(x)}{q_I(y)}=\delta(x,y)$. It is non-zero when $x^a=y^a$, behaves as a scalar with respect to its first argument and as a scalar density with respect to its second argument.}.
If I note that I will only consider constraints without spatial derivatives on momenta, this simplifies,
\begin{equation}
    0 = \sum_I \funcdif{ C (x) }{ q_I (y) } \left. \partdif{ C }{ p_I } \right|_y
    - \left( \beta D^a \partial_a \right)_x \delta \left( x, y \right) 
    - \xty.
    \label{eq:dist-eqn_con}
\end{equation}

For when I wish to derive the action instead of the constraint, I can transform the equation by noting that,
\begin{equation}
    \funcdif{ C [N] }{ q_I } = - \funcdif{ L [N] }{ q_I },
\quad
    N v_I = \funcdif{ C [N] }{ p_I },
\end{equation}
where $v_I:=\mathcal{L}_n{q_I}$ and the Lagrangian is here defined such that
$ S = \int \mathrm{d} t \mathrm{d}^3 x N L $.
I substitute these into \eqref{eq:con-alg_def_expanded}, then take the functional derivatives to remove $N_1$ and $N_2$,
\begin{equation}
    0 = \sum_I \funcdif{ L (x) }{ q_I (y) } v_I (y)
    + \left( \beta D^a \partial_a \right)_x \delta \left( x, y \right)
    - \xty.
    \label{eq:dist-eqn_act}
\end{equation}
To find a useful form for this, I need to use a specific form for the diffeomorphism constraint.  Because it depends on momenta, I must replace them using,
\begin{equation}
    p_I := \funcdif{ S }{ \dot{q_I} } = \frac{1}{N} \funcdif{ L [N] }{ v_I },
    \label{eq:momentum_def_full}
\end{equation}
and, as before, if I note that I will only consider actions without spatial derivatives of momenta this simplifies to
\begin{equation}
    p_I = \partdif{ L }{ v_I }.
    \label{eq:momentum_def}
\end{equation}
Therefore, substituting the diffeomorphism constraint found in \appref{sec:diff} and momenta \eqref{eq:momentum_def} into \eqref{eq:dist-eqn_act}, I find the distribution equation which can be used for restricting the form of the deformed action.

So, the key equations I use as a basis for finding the action or constraint for deformed general relativity are \eqref{eq:dist-eqn_con} and \eqref{eq:dist-eqn_act}.


\section{Order of the deformed action and constraint}
\label{sec:methodology_order}

I can determine the relationship between the order of the deformation function and the order of the associated constraint (or action) by comparing orders of momenta (or velocity).

\subsection{Hamiltonian route}
\label{sec:methodology_order_constraint}

As an example, take the distribution equation \eqref{eq:dist-eqn_con} with only a scalar field,
\begin{equation}
    0 = \funcdif{ C (x) }{ \psi (y) } \left. \partdif{ C }{ \pi } \right|_y
    - \left( \beta \pi \partial^a \psi \partial_a \right)_x \delta \left( x, y \right)
    - \xty,
    \label{eq:dist-eqn_con_scalar}
\end{equation}
where I have used the diffeomorphism constraint \eqref{eq:diff_scalar}.
I take a simplified model with two spatial derivatives represented by $\Delta$, only taking even orders of derivatives because of assuming spatial parity.  I take the distribution equation \eqref{eq:dist-eqn_con_scalar} and put it into schematic form,
\begin{equation}
    0 = \partdif{ C }{ \Delta } \partdif{ C }{ \pi } 
    - \beta \, \pi.
    \label{eq:schematic_constraint_all}
\end{equation}
so that I can consider orders of $\pi$ in a way analogous to dimensional analysis.
This equation must be satisfied independently at each order of momenta, so I isolate the coefficient of $\pi^n$,
\begin{equation}
    0 = \sum_{m=1}^{n_C} m \partdif{ C^{(n-m+1)} }{ \Delta } C^{(m)}
    - \beta^{(n-1)},
    \label{eq:schematic_constraint_coefficient}
\end{equation}
where I have expanded the constraint and deformation, 
\begin{equation}
    C=\sum_{m=0}^{n_C}C^{(m)}\pi^m,
\quad
    \beta=\sum_{m=0}^{n_\beta}\beta^{(m)}\pi^m.
\end{equation}
The highest order contribution to \eqref{eq:schematic_constraint_coefficient} comes when ${m=n_C}$ and ${n-m+1=n_C}$, in which case ${n=2n_C-1}$.  This is the highest order at which $\beta$ won't automatically be constrained to vanish, so I find its highest order of momenta to be ${n_\beta=2n_C-2}$.
However, this result does not take into account the fact that the combined order of momenta and spatial derivatives may be restricted.  If this is the case (as is found in \chapref{sec:allst}), then the highest order contribution to the \eqref{eq:schematic_constraint_coefficient} will be when ${n-m+1=n_C-2}$, in which case I find the relation
\begin{equation}
    2 n_C - n_\beta = 4 .
    \label{eq:methodology_order_constraint}
\end{equation}
I see that a deformed second order constraint only requires considering a zeroth order deformation as I do in \chapref{sec:2ndst}, but a fourth order constraint requires considering a fourth order deformation.  I consider the constraint to general order in \chapref{sec:allst}.  Note that this relation suggests there are higher order deformations which allow for constraints given by finite order polynomials.

\subsection{Lagrangian route}
\label{sec:methodology_order_action}

Consider the distribution equation \eqref{eq:dist-eqn_act} with only a scalar field,
\begin{equation}
\begin{split}
    0 & =
    \funcdif{ L (x) }{ \psi (y) } \nu (y)
    + \left( \beta \partdif{ L }{ \nu } \partial^a \psi \partial_a \right)_x \delta ( x, y )
    - \xty,
\end{split}
\end{equation}
where I have used the diffeomorphism constraint \eqref{eq:diff_scalar} and the momentum definition \eqref{eq:momentum_def}.
Let me consider a simplified model to match the derivative orders for the deformation and the derivative orders for the Lagrangian in a way analogous to dimensional analysis.  First order time derivatives are given by $\nu$ and two orders of spatial derivatives are given by $\Delta$.  I can collect terms in the distribution equation of the same order of time derivatives as they are linearly independent.  Schematically, the distribution equation is given by,
\begin{equation}
    0 = \partdif{L}{\Delta} \nu + \partdif{L}{\nu} \beta,
\end{equation}
and expanding the Lagrangian and deformation in powers of $\nu$, 
\begin{equation}
    L = \sum_{m=0}^{n_L} L^{(m)} \nu^m,
\quad
    \beta = \sum_{m=0}^{n_\beta} \beta^{(m)} \nu^m,
\end{equation}
the coefficient of $\nu^n$ is then given by,
\begin{equation}
    0 = \partdif{ L^{(n-1)} }{ \Delta }
    + \sum_{m=0}^{n_\beta} \left( n - m + 1 \right) L^{(n-m+1)} \beta^{(m)}.
\end{equation}
I can relabel and rearrange to find a schematic solution for the highest order of $L$ appearing here,
\begin{equation}
    L^{(n)} =
    \frac{ -1 }{ n \beta^{ (0) } }
    \left\{
        \partdif{ L^{ (n-2) } }{ \Delta }
        + \sum_{m=1}^{n_\beta} \left( n - m \right) \beta^{(m)} L^{(n-m)}
    \right\}.
\end{equation}
I can see that if $n_\beta>0$, then this equation is recursive and $n_L\to\infty$ because there is no natural cut-off, suggesting that $L$ is required to be non-polynomial.  If I wish to truncate the action at some order, then it must be treated as an perturbative approximation.  I consider a perturbative fourth order action in \chapref{sec:pert}, and the completely general action in \chapref{sec:allact}.


\section{Cosmology}
\label{sec:methodology_cosmo}

Since the main motivations for this study centre around cosmological implications of the deformed constraint algebra, I need to lay out how I find the cosmological dynamics of a model.
I restrict to an isotropic and homogeneous space, using the Friedmann-Lema\^{i}tre-Robertson-Walker metric (FLRW),
\begin{equation}
    q_{ab} = a^2(t) \Sigma_{ab},
\quad
    a = \left( \det{q_{ab}} \right)^{1/6}
\quad
    N^a = 0,
        \label{eq:methodology_cosmo_metric}
\end{equation}
where $\Sigma_{ab}$ is time-independent and describes a three dimensional spatial slice with constant curvature $k$. When space is flat, $k=0$, this is given by $\Sigma_{ab}=\delta_{ab}$.
The normal derivative of the spatial metric is given by,
\begin{equation}
    v_{ab} = \frac{2}{N} a \dot{a} \Sigma_{ab},
\quad
    \therefore \;
    \mathcal{K} = \frac{ 6 \dot{a}^2 }{ a^2 N^2 } = : \frac{ 6 }{ N^2 } \mathcal{H}^2,
        \label{eq:methodology_cosmo_derivative}
\end{equation}
where $\mathcal{H}$ is the Hubble expansion rate, and the Ricci curvature scalar is given by,
\begin{equation}
    R = \frac{ 6 k }{ a^2 }.
        \label{eq:methodology_cosmo_curvature}
\end{equation}
When using canonical coordinates, the metric momentum is given by
\begin{equation}
    p^{ab} = \bar{p} \, \Sigma^{ab},
\quad
    \bar{p} = \left( \det{ p^{ab} } \right)^{1/3},
\end{equation}
which changes the metric's commutation relation,
\begin{equation}
    \Big\{ q_{ab} (x), p^{cd} (y) \Big\} = \delta_{ab}^{cd} (x) \delta ( x, y )
\;\to\;
    \Big\{ a (x), \bar{p} (y) \Big\} = \frac{ \delta ( x, y ) }{ 6 a (x) } ,
\end{equation}
where $\delta^{cd}_{ab}:=\delta_{(a}^c\delta_{b)}^d$.
The spatial derivatives of matter fields vanish, $\partial_a \psi_I = 0$.
One may couple a perfect fluid to the metric by including the energy density $\rho$ in the constraint or the action \cite{Brown:1992kc},
\begin{equation}
    C \supset a^3 \rho,
\quad
    L \supset - a^3 \rho,
        \label{eq:methodology_cosmo_action}
\end{equation}
which must satisfy the continuity equation,
\begin{equation}
    \dot{\rho} + 3 \mathcal{H} \rho \left( 1 + w_\rho \right) = 0,
        \label{eq:methodology_cosmo_continuity}
\end{equation}
where $w_\rho$ is the perfect fluid's cosmological equation of state, the ratio of the pressure density to the energy density.


For investigations into whether there are implications for the hypothesised inflationary period in the very early universe, I must define what is considered to be a period of inflation.  The simple definition is when the finite scale factor is both expanding and accelerating, $\dot{a}>0$ and $\ddot{a}>0$.


As said above, loop quantum cosmology with real variables seems to predict a big bounce instead of a big bang or crunch. In this thesis, I take the very literal interpretation of this (as found in ref.~\cite{Cattoen:2005dx}) and define a bounce as a turning point for a finite scale factor, $a>0$, $\dot{a}=0$ and $\ddot{a}>0$.  This definition may be usable, but it is not ideal.  If a bounce does indeed happen when $\beta<0$, as predicted in the literature, then this is when the effective metric signature is Euclidean, when $\dot{a}$ may be a complex number.


Ideally, I would like to extract cosmological observables such as the primordial scalar index to find phenemenological constraints \cite{Peacock:1999ye}.  However, to calculate the power spectra of primordial fluctuations would require adapting the cosmological perturbation theory formalism to ensure it is valid for deformed covariance, something which would probably be highly non-trivial.  Unfortunately, there was not enough time to investigate this.

\chapter{Second order scalar-tensor model and the classical limit}
\label{sec:2ndst}


In this chapter, I derive the general form of a minimally-deformed, non-minimally-coupled scalar-tensor model which includes up to two orders in momenta or time derivatives.  This allows me to demonstrate that the higher order gravity model derived in \secref{sec:methodology_FR} does not deform the constraint algebra or general covariance, and therefore show how the deformed models derived in subsequent chapters are distinct.  For those later chapters, this minimally-deformed model provides a useful reference point.
This chapter is adapted from work I previously published in ref.~\cite{cuttell2018}.

I find the form of the model by deriving restrictions on the constraint using \eqref{eq:dist-eqn_con} and then transform to find the action.  It would be completely equivalent to derive the action first, because the minimally deformed case maintains a linear relationship between velocities and momenta, meaning that the transformation between the action and constraint is trivial.
After finding the constraint and action, I look at some of the cosmological implications in \secref{sec:2ndst_cosmo}, especially the interesting influence of the non-minimal coupling of the scalar field.

I use the structure of the scalar-tensor constraint which is a parameterisation of $F(\four{R})$, \eqref{eq:geo_constraint}, to guide the structure of my general ansatz for a spatial metric coupled to several scalar fields.  I include spatially covariant terms up to second order in momenta or spatial derivatives, and ignore terms linear in momenta,
\begin{equation}
\begin{split}
    C & = C_\0 + C_{(R)} R 
    + C^{(p^2)}_{abcd} p^{ab} p^{cd} 
    + C^{(p\pi_I)} p \pi_I 
\\
    & + C_{(\psi_I'\psi_J')} \partial_a \psi_I \partial^a \psi_J 
    + C_{(\psi_I'')} \Delta \psi_I 
    + C^{(\pi_I\pi_J)} \pi_I \pi_J,
        \label{eq:2ndst_ansatz}
\end{split}
\end{equation}
with summation over $I$ and $J$ implied.
I have included $C_{(\psi_I'\psi_J')}$ because it appears in the constraint for minimally coupled scalar fields \cite[p.~62]{bojowald2010canonical}.
I aimed to define the most general ansatz for a scalar-tensor constraint containing up to two orders in derivatives which is covariant under general spatial diffeomorphisms, as well as under time reversal, and preserves spatial parity.  Each coefficient is potentially a function of $q$ and $\psi_I$, allowing for non-minimal coupling. The spatial indices of $C^{(p^2)}_{abcd}$ only represent different combinations of the metric. The zeroth order term might include terms such as scalar field potentials or perfect fluids, and it behaves as a generalised potential $C_\0=\sqrt{q}\,U(q,\psi_I)$.


\section{Solving the distribution equation}
\label{sec:2ndst_dist-eqn}

I substitute into the distribution equation \eqref{eq:dist-eqn_con} my ansatz for a second order constraint \eqref{eq:2ndst_ansatz}, the diffeomorphism constraint from \eqref{eq:diff_scalar} and \eqref{eq:diff_metric}, and a zeroth order deformation $\beta\left(q,\psi\right)$,
\begin{equation}
\begin{split}
    0 & = \funcdif{C_0(x)}{q_{ab}(y)} \left( 2 p^{cd} C^{(p^2)}_{abcd} + \pi q_{ab} C^{(p\pi)} \right)_y + \funcdif{C_0(x)}{\psi(y)} \left( p \, C^{(p\pi)} + 2 \pi \, C^{(\pi^2)} \right)_y
        \\
    &  + \left\{ 2 \beta \left( \partial_b p^{ab} + \Gamma^a_{bc} p^{bc} \right) - \beta \partial^a \psi \, \pi \right\}_x \partial_{a(x)} \delta(x,y) - \left( x \leftrightarrow y \right),
\end{split}
    \label{eq:2ndst_dist-eqn}
\end{equation}
where $C_0$ is the part of the constraint without momenta.  From here there are two routes to solution, by focusing on either the $p^{ab}$ and $\pi$ components.  I must do both to find all consistency conditions on the coefficients of the Hamiltonian constraint.

\subsection{\texorpdfstring{$p^{ab}$}{Metric momentum} sector}
\label{sec:2ndst_dist-eqn_p}

To proceed to the metric momentum sector, I take \eqref{eq:2ndst_dist-eqn} and find the functional derivative with respect to $p^{ab}(z)$,
\begin{equation}
\begin{split}
    0 & = \left( 2 \funcdif{C_0(x)}{q_{cd}(y)} C^{(p^2)}_{abcd} (y) + \funcdif{C_0(x)}{\psi(y)} C^{(p\pi)}_{ab} (y) \right) \delta(z,y)
        \\
    & 
    + 2 \beta (x) \Big\{ \big( \delta^{c}_{(a} \partial_{b)} \big)_x \delta(z,x) + \Gamma^c_{ab}(x) \delta(z,x) \Big\} \partial_{c(x)} \delta(x,y) - \left( x \leftrightarrow y \right),
\end{split}
    \label{eq:2ndst_dist-eqn_p}
\end{equation}
where I explicitly show the coordinate of the partial derivative as 
$\displaystyle{\partial_{a(y)}:=\partdif{}{y^a}}$ 
because the distinction is important when integrating by parts.
I then proceed by moving derivatives away from $\delta(z,y)$ terms and discarding total derivatives,
\begin{equation}
\begin{split}
    0 & = \left( 2 \funcdif{C_0(x)}{q_{cd}(y)} C^{(p^2)}_{abcd} (y) + \funcdif{C_0(x)}{\psi(y)} C^{(p\pi)}_{ab} (y)
    + 2 \partial_{c(y)} \left[ \big( \beta \delta^{c}_{(a} \partial_{b)} \big)_y \delta (y,x) \right] 
\right. \\  & 
    - 2 \big( \beta \Gamma^c_{ab} \partial_c \big)_y \delta (y,x) \bigg) \delta(z,y) - \left( x \leftrightarrow y \right),
\end{split}
\end{equation}
which can be rewritten as,
\begin{equation}
    0 = A_{ab}(x,y) \delta(z,y) - A_{ab}(y,x) \delta(z,x).
\end{equation}
Integrating over $y$, I find that part of the equation can be combined into a tensor dependent only on $x$,
\begin{equation}
\begin{split}
    0 & = A_{ab}(x,z) - \delta(z,x) \int \mathrm{d}^3 y A_{ab}(y,x),
        \\
    & = A_{ab}(x,z) - \delta(z,x) A_{ab} (x),
    \quad \mathrm{where} \;
    A_{ab}(x) = \int \mathrm{d}^3 y A_{ab} \left(y, x \right).
\end{split}
\end{equation}
Substituting in the definition of $A_{ab}(x,z)$ then relabelling,
\begin{equation}
\begin{split}
    0 & = 2 \funcdif{C_0(x)}{q_{cd}(y)} C^{(p^2)}_{abcd} (y) 
    + \funcdif{C_0(x)}{\psi(y)} C^{(p\pi)}_{ab} (y) 
    + 2 \partial_{c(y)} \left[ \left( \beta \delta^{c}_{(a} \partial_{b)} \right)_y \delta (y,x) \right]
        \\
    &  - 2 \left( \beta \Gamma^c_{ab} \partial_c \right)_y \delta (y,x) - A_{ab}(x) \delta(y,x).
\end{split}
\end{equation}
Multiplying by an arbitrary test tensor $\theta^{ab}\left(y\right)$, then integrating by parts over $y$, I get
\begin{equation}
\begin{split}
    0 & = \theta^{ab} \left( \cdots \right)_{ab}
    + \partial_c \theta^{ab} \left\{ 
        2 C^{(p^2)}_{abde} \partdif{C_0}{q_{de,c}} 
        + 4 \partial_d C^{(p^2)}_{abef} \partdif{C_0}{q_{ef,cd}} 
        + C^{(p\pi)}_{ab} \partdif{C_0}{\psi_{,c}} 
\right. \\ & \left.
        + 2 \partial_d C^{(p\pi)}_{ab} \partdif{C_0}{\psi_{,cd}} 
        + 2 \delta^c_{(a} \partial_{b)} \beta 
        + 2 \beta \Gamma^c_{ab} \right\}
\\ &
    + \partial_{cd} \theta^{ab} \left\{ 
        2 C^{(p^2)}_{abef} \partdif{C_0}{q_{ef,cd}} 
        + C^{(p\pi)}_{ab} \partdif{C_0}{\psi_{,cd}} 
        + 2 \beta \delta^{cd}_{ab} \right\}
    ,
\end{split}
\end{equation}
where I do not need to consider the zeroth derivative terms because they do not produce restrictions on the form of the constraint.  Since $\theta^{ab}$ is arbitrary beyond the symmetry of its indices, each unique contraction of it forms a linearly independent equation.

To calculate the derivatives of $C_0$, I must use the decomposition of the Riemann tensor \eqref{eq:var_riemann_2} and the second covariant derivative of the metric variation expressed in terms of partial derivatives \eqref{eq:var_metric}.  This gives,
\begin{equation}
\begin{gathered}
    \partdif{C_0}{\psi_{,ab}}
    = C_{(\psi'')} q^{ab},
        \quad
    \partdif{C_0}{\psi_{,a}}
    = 2 C_{(\psi^{\prime2})} \partial^a \psi - C_{(\psi'')} \Gamma^a,
        \quad
    \partdif{C_0}{q_{ab,cd}}
    = C_{(R)} \Phi^{abcd},
        \\
    \partdif{C_0}{q_{ab,c}}
    = C_{(\psi'')} \left( \half q^{ab} \partial^c \psi - q^{c(a} \partial^{b)} \psi \right) - C_{(R)} \Phi^{defg} \left( \Gamma^c_{fg} \delta^{ab}_{de} + 4 \delta^{(a}_{(d} \Gamma^{b)}_{e)(f} \delta^c_{g)} \right),
\end{gathered}
    \label{eq:2ndst_C0_derivatives}%
\end{equation}
\begin{sloppypar}
where ${\Phi^{abcd}=Q^{abcd}-q^{ab}q^{cd}}$ as found in \eqref{eq:var_coeff_contract}.  Note that ${Q^{abcd}:=q^{a(c}q^{d)b}}$ and ${\delta^{ab}_{de}:=\delta^{(a}_{d}\delta^{b)}_{e}}$.  I evaluate the coefficient of $\partial_{dc}\theta^{ab}$ and find the linearly independent components,
\end{sloppypar}
\begin{subequations}
\begin{align}
    q_{ab} \partial^2 \theta^{ab} :
    0 & = - 2 C_{(R)} \left( 2 C^{(p^2\p)} + C^{(p^2\x)} \right) + C_{(\psi'')} C^{(p\pi)},
        \label{eq:2ndst_d2epsilon_1} \\
    \partial_{ab} \theta^{ab} :
    0 & = C_{(R)} C^{(p^2\x)} + \beta,
        \label{eq:2ndst_d2epsilon_2}
\end{align}
    \label{eq:2ndst_d2epsilon}%
\end{subequations}
where I have decomposed the constraint coefficient
$C^{(p^2)}_{abcd} = q_{ab} q_{cd} C^{(p^2\p)} + Q_{abcd} C^{(p^2\x)}$.
Then evaluating similarly for $\partial_c\theta^{ab}$,
\begin{subequations}
\begin{align}
\begin{split}
    q_{ab} \partial^c \psi \partial_c \theta^{ab}  :
    0 & = 2 \left( C_{(\psi^{\prime2})} + C_{(\psi'')} \partial_\psi \right) C^{(p\pi)}
    + \left( C_{(\psi'')} - 8 C_{(R)} \partial_\psi \right) C^{(p^2\p)}
\\ &
    + \left( C_{(\psi'')} - 4 C_{(R)} \partial_\psi \right) C^{(p^2\x)},
        \label{eq:2ndst_d1epsilon_1} 
\end{split}
        \\
    \partial_b \psi \partial_a \theta^{ab} :
    0 & = \left( - C_{(\psi'')} + 2 C_{(R)} \partial_\psi \right) C^{(p^2\x)} + \partial_\psi \beta,
        \label{eq:2ndst_d1epsilon_2} \\
    X_b \partial_a \theta^{ab} :
    0 & = C_{(R)} \left( 1 + 2 \partial_q \right) C^{(p^2\x)} + \partial_q \beta,
        \label{eq:2ndst_d1epsilon_5} \\
    X^c q_{ab} \partial_c \theta^{ab} :
    0 & = - 2 C_{(R)} \left( 1 + 4 \partial_q \right) \left( 2 C^{(p^2\p)} + C^{(p^2\x)} \right) 
\\ &
    + C_{(\psi'')} \left( 1 + 4 \partial_q \right) C^{(p\pi)},
        \label{eq:2ndst_d1epsilon_7}
\end{align}
    \label{eq:2ndst_d1epsilon}%
\end{subequations}
where 
$\displaystyle{\partial_\psi:=\partdif{}{\psi}}$, 
$\displaystyle{\partial_q:=\partdif{}{\,\log{q}}}$ and 
${X_a:=q^{bc}\partial_aq_{bc}}$.
Note that the equations for 
${\partial^cq_{ab}\partial_c\theta^{ab}}$, 
${\partial_aq_{bc}\partial^c\theta^{ab}}$ and 
${q_{ab}\partial^dq_{cd}\partial^c\theta^{ab}}$ are not included because they are identical to \eqref{eq:2ndst_d2epsilon}.

Using \eqref{eq:2ndst_d2epsilon_2} to solve for $C^{(p^2\x)}$, then substituting it into \eqref{eq:2ndst_d1epsilon_5}, I find,
\begin{equation}
    \partdif{\,\log{C}_{(R)}}{\,\log{q}} 
    = \half \left( 1 + \partdif{\,\log{\beta}}{\,\log{q}} \right),
\end{equation}
\begin{sloppypar}
which is solved by 
${C_{(R)}\left(q,\psi\right)=f\left(\psi\right)\sqrt{q\,\left|\beta\left(q,\psi\right)\right|}}$,
where $f(\psi)$ is some unknown function.  If I solve \eqref{eq:2ndst_d2epsilon} for $C^{(p^2\p)}$ and $C^{(p^2\x)}$, then substitute them into \eqref{eq:2ndst_d1epsilon_7}, I find a similar equation to the one above for $C_{(R)}$, and therefore 
${C_{(\psi'')}\left(q,\psi\right)=f_{(\psi'')}\left(\psi\right)\sqrt{q\,\left|\beta\left(q,\psi\right)\right|}}$.  
Taking \eqref{eq:2ndst_d1epsilon_2} then substituting in for $C^{(p^2\x)}$, $C_{(R)}$ and $C_{(\psi'')}$, I find that 
${f_{(\psi'')}\left(\psi\right)=-2\partial_\psi f\left(\psi\right)}$,
\end{sloppypar}
\begin{subequations}
\begin{align}
    C_{(R)} & = f \sqrt{ q \, \left| \beta \right| },
        &
    C_{(\psi'')} & = - 2 \partial_\psi f \sqrt{ q \, \left| \beta \right| }, 
        \label{eq:2ndst_C0_sol} \\
    C^{(p^2\x)} & = \frac{-\sigma_\beta}{f} \sqrt{ \frac{ \left| \beta \right| }{ q } },
        &
    C^{(p^2\p)} & = \frac{\sigma_\beta}{2f} \sqrt{ \frac{ \left| \beta \right| }{ q } } - \frac{ \partial_\psi f }{ 2 f } C^{(p\pi)},
        \label{eq:2ndst_Cp2_sol}
\end{align}
    \label{eq:2ndst_p_sector_sol}%
\end{subequations}
where $\sigma_\beta:=\sgn{\beta}$,
which is all the conditions which can be obtained from the metric momentum sector of the distribution equation.  The remaining conditions must be found in the scalar momentum sector.

\subsection{\texorpdfstring{$\pi$}{Scalar momentum} sector}
\label{sec:2ndst_dist-eqn_pi}

Similar to \subsecref{sec:2ndst_dist-eqn_p} above, I take the functional derivative of \eqref{eq:2ndst_dist-eqn} with respect to $\pi(z)$,
\begin{equation}
\begin{split}
    0 & = \left( \funcdif{C_0(x)}{q_{ab}(y)} C^{(p\pi)}_{ab} (y) + 2 \funcdif{C_0(x)}{\psi(y)} C^{(\pi^2)} (y) \right) \delta(z,y)
\\ &
    - \left( \beta \partial^a \psi \partial_a \right)_x \delta(x,y) \delta(z,x)
    - \xty ,
\end{split}
    \label{eq:2ndst_dist-eqn_pi}
\end{equation}
then exchange terms to find the coefficient of $\delta(z,y)$,
\begin{equation}
\begin{split}
    0 & = \left( \funcdif{C_0(x)}{q_{ab}(y)} C^{(p\pi)}_{ab} (y) 
    + 2 \funcdif{C_0(x)}{\psi(y)} C^{(\pi^2)} (y) 
\right. \\ & \left.
    + \left( \beta \partial^a \psi \partial_a \right)_y \delta(y,x) \right) \delta(z,y) 
    - \xty ,
\end{split}
\end{equation}
which can be rewritten as,
\begin{subequations}
\begin{align}
    0 & = A (x,y) \delta(z,y) - A (y,x) \delta(z,x),
        \\
    0 & = A (x,z) - \delta(z,x) \int \mathrm{d}^3 y A (y,x),
        \\
    & = A (x,z) - \delta(z,x) A (x),
        \quad \mathrm{where} \;
    A (x) = \int \mathrm{d}^3 y A \left( y, x \right),
\end{align}%
\end{subequations}
leading to
\begin{equation}
\begin{split}
    0 = \funcdif{C_0(x)}{q_{ab}(y)} C^{(p\pi)}_{ab} (y) 
    + 2 \funcdif{C_0(x)}{\psi(y)} C^{(\pi^2)} (y)
    + \left( \beta \partial^a \psi \partial_a \right)_y \delta(y,x) 
    - A (x) \delta(y,x).
\end{split}%
\end{equation}
Multiplying by an arbitrary test function $\eta(y)$, then integrating by parts over $y$, I get
\begin{equation}
\begin{split}
    0 &= \eta \left( \cdots \right) 
    + \partial_{ab} \eta \left( 
        C^{(p\pi)}_{cd} \partdif{C_0}{q_{cd,ab}} 
        + 2 C^{(\pi^2)} \partdif{C_0}{\psi_{,ab}} 
    \right)
\\ & 
    + \partial_a \eta \left(
        C^{(p\pi)}_{bc} \partdif{C_0}{q_{bc,a}} 
        + 2 \partial_b C^{(p\pi)}_{cd} \partdif{C_0}{q_{cd,ab}} 
        + 2 C^{(\pi^2)} \partdif{C_0}{\psi_{,a}} 
        + 4 \partial_b C^{(\pi^2)} \partdif{C_0}{\psi_{,ab}} 
        - \beta \partial^a \psi 
    \right).
\end{split}
\end{equation}
I then substitute in \eqref{eq:2ndst_C0_derivatives} to find the linearly independent conditions,
\begin{subequations}
\begin{align}
    \partial^2 \eta :
    0 & = C_{(R)} C^{(p\pi)} - C_{(\psi'')} C^{(\pi^2)},
        \label{eq:2ndst_dphi_1} \\
    \partial^a \psi \partial_a \eta :
    0 & = \left( \half C_{(\psi'')} - 4 C_{(R)} \partial_\psi \right) C^{(p\pi)} + 4 \left( C_{(\psi^{\prime2})} + C_{(\psi'')} \partial_\psi \right) C^{(\pi^2)} - \beta,
        \label{eq:2ndst_dphi_2} \\
    X^a \partial_a \eta :
    0 & = C_{(R)} \left( 1 + 4 \partial_q \right) C^{(p\pi)} - C_{(\psi'')} \left( 1 + 4 \partial_q \right) C^{(\pi^2)}.
        \label{eq:2ndst_dphi_3}
\end{align}
    \label{eq:2ndst_dphi}%
\end{subequations}
Note that there is another condition from $\partial^bq_{ab}\partial^a\eta$, but it is identical to \eqref{eq:2ndst_dphi_1}.

I can solve \eqref{eq:2ndst_dphi_1} for $C^{(p\pi)}=C_{(\psi'')}C^{(\pi^2)}/C_{(R)}$, and then substitute into \eqref{eq:2ndst_dphi_2} to find,
\begin{equation}
    0 = C^{(\pi^2)} \left\{ C_{(\psi^{\prime2})} - \partial_\psi C_{(\psi'')} + \frac{C_{(\psi'')}}{C_{(R)}} \left( \partial_\psi C_{(R)} + \frac{C_{(\psi'')}}{8} \right) \right\} - \frac{\beta}{4},
\end{equation}
which I can solve for $C^{(\pi^2)}$, and is the same conclusion I get from \eqref{eq:2ndst_d1epsilon_1} (though I did not explicitly write it above because it is simpler to write it here).  The condition \eqref{eq:2ndst_dphi_3} is solved when I substitute in all my results so far,
\begin{subequations}
\begin{align}
    C^{(\pi^2)} & = \frac{\sigma_\beta}{4} \sqrt{\frac{\left|\beta\right|}{q}} \left\{ 
        \frac{C_{(\psi^{\prime2})}}{\sqrt{q\left|\beta\right|}} 
        + 2 f'' 
        - \frac{3f^{\prime2}}{2f} 
    \right\}^{-1},
        \\
    C^{(p\pi)} & = \frac{ - \sigma_\beta f'}{2f} \sqrt{\frac{\left|\beta\right|}{q}} \left\{     \frac{C_{(\psi^{\prime2})}}{\sqrt{q\left|\beta\right|}} 
        + 2 f'' 
        - \frac{3f^{\prime2}}{2f} 
    \right\}^{-1},
\end{align}%
\end{subequations}
and if I collect all of the coefficients, I find the Hamiltonian constraint,
\begin{equation}
\begin{split}
    C & = \sqrt{q\left|\beta\right|} \Big( f R - 2 f' \Delta \psi  \Big) + C_{(\psi^{\prime2})} \partial_a \psi \partial^a \psi + C_\0
        \\
    &  + \sigma_\beta \sqrt{\frac{\left|\beta\right|}{q}} \left\{ \frac{1}{f} \left( \frac{p^2}{6} - \bp \right) + \frac{1}{4} \left( \pi - \frac{f'}{f} p \right)^2 \left( \frac{C_{(\psi^{\prime2})}}{\sqrt{q\left|\beta\right|}} + 2 f'' - \frac{3f^{\prime2}}{2f} \right)^{-1} \right\},
\end{split}
\end{equation}
so the freedom in any $(3+1)$ dimensional scalar-tensor theory with time symmetry and minimally deformed general covariance comes down to the choice of $f\left(\psi\right)$, $\beta\left(q,\psi\right)$, $C_{(\psi^{\prime2})}\left(q,\psi\right)$ and the zeroth order term $C_\0(q,\psi)$.
It is convenient to make a redefinition, 
${C_{(\psi^{\prime2})}=g\left(q,\psi\right)\sqrt{q\left|\beta\right|}}$,
where I have made the scalar weight and expected dependence on $\beta$ explicit.  It is worth remembering that this is an assumption, and that $g$ could be a function of $\beta$.
It is also convenient to treat the zeroth order term as a general potential, and to extract the scalar density, $C_\0=\sqrt{q}\,U(q,\psi)$.

I find the effective Lagrangian associated with this Hamiltonian constraint by performing a Legendre transformation,
\begin{equation}
\begin{split}
    L & = \sqrt{q\left|\beta\right|} \left\{ 
        f \left( \frac{\mathcal{K}}{\beta} - R \right) 
        + f' \left( \frac{\nu{}v}{\beta} + 2 \Delta \psi \right) 
        + \left( g + 2 f'' \right)\frac{\nu^2}{\beta} 
\right. \\ & \left.
        - g \, \partial_a \psi \partial^a \psi 
        - \frac{U}{\sqrt{\left|\beta\right|}} 
    \right\}.
\end{split}%
\end{equation}
Integrating by parts at the level of the action does not affect the dynamics because it only eliminates boundary terms.  This allows me to find the effective form of the Lagrangian, with a space-time decomposition and without second order time derivatives.  I can also do this in the opposite direction to find the covariant form of the above effective Lagrangian,
\begin{equation}
    L_\mathrm{cov} = \sqrt{q\left|\beta\right|} \left( - f \, {}^{(4,\beta)}\!R - \left( g + 2 f'' \right) \partial_\mu^{(4,\beta)} \psi \, \partial^{\mu}_{(4,\beta)} \psi \right) - \sqrt{q} \, U,
\end{equation}
where the deformed four dimensional Ricci scalar and partial derivative are given by,
\begin{subequations}
\begin{gather}
    {}^{(4,\beta)}\!R = R + \frac{ \sigma_\beta }{\sqrt{\left|\beta\right|}} q^{ab} \mathcal{L}_n \left( \frac{v_{ab}}{\sqrt{\left|\beta\right|}} \right) + \frac{1}{4\beta} v^2 - \frac{3}{4\beta} v^{ab} v_{ab} - \frac{2 \Delta \left( \sqrt{\left|\beta\right|} \,  N\right)}{\sqrt{\left|\beta\right|} \, N},
        \\
    \partial_\mu^{(4,\beta)} \psi \, \partial^{\mu}_{(4,\beta)} \psi = \partial_a \psi \, \partial^a \psi - \frac{1}{\beta} \nu^2.
\end{gather}%
\end{subequations}
If this is compared to \eqref{eq:ricci_decomposition}, I see that the deformation seems to have transformed the effective lapse function 
${N\to\sqrt{\left|\beta\right|}\,N}$, 
and transformed the effective normalisation of the normal vector to 
${g_{\mu\nu}n^\mu{}n^{\nu}=-\sigma_\beta}$.  
Here is where I see the effective signature change which comes from the deformation.

It is useful to take the Lagrangian in covariant form and use it to redefine the coupling functions so that minimal coupling is when the functions are equal to unity, $f=-\half\omega_R$ and $g=-\half\omega_\psi+\omega_R''$,
\begin{equation}
    L_\mathrm{cov} = 
    \half \sqrt{q\left|\beta\right|} \left( 
        \omega_R ( \psi ) \, {}^{(4,\beta)}\!R 
        - \omega_\psi ( q, \psi ) \, \partial_\mu^{(4,\beta)} \psi \, \partial^\mu_{(4,\beta)} \psi \right) 
        - \sqrt{q} \, U \left( q, \psi \right),
        \label{eq:2ndst_lagrangian_covariant}
\end{equation}
so the effective forms of the constraint and Lagrangian are given by,
\begin{subequations}
\begin{align}
\begin{split}
    L & = \half \sqrt{ q \left| \beta \right| } \left\{ \omega_R \left( R - \frac{\mathcal{K}}{\beta} \right) - \omega_R' \left( \frac{\nu{}v}{\beta} + 2 \Delta \psi \right) + \frac{\omega_\psi\nu^2}{\beta} 
\right. \\ & \left.
    - \left( \omega_\psi + 2 \omega_R'' \right) \partial_a \psi \partial^a \psi \right\} 
    - \sqrt{q} \, U,
\end{split}
    \label{eq:2ndst_lagrangian_effective}
        \\
\begin{split}
    C & = \sqrt{q\left|\beta\right|} \left\{ \frac{2\sigma_\beta}{q\omega_R} \left( \bp - \frac{p^2}{6} \right) - \frac{\omega_R}{2} R + \frac{\sigma_\beta}{2q} \left( \pi - \frac{\omega_R'}{\omega_R} p \right)^2 \left( \omega_\psi + \frac{3\omega_R^{\prime2}}{2\omega_R} \right)^{-1} \right.
        \\
    &  \left. + \omega_R' \Delta \psi + \left( \frac{\omega_\psi}{2} + \omega_R'' \right) \partial_a \psi \partial^a \psi \right\} + \sqrt{q} \, U,
\end{split}
    \label{eq:2ndst_constraint_effective}
\end{align}
    \label{eq:2ndst_effective}%
\end{subequations}
which is the main result of this section in its most useful form.

Since I have non-minimal coupling, I am working in the Jordan frame.  I can get to the Einstein frame by making a specific conformal transformation which absorbs the coupling $\omega_R$ by setting $q_{ab}=\omega_R\,\tilde{q}_{ab}$ and $N=\omega_R^{-1/2}\tilde{N}$,
\begin{equation}
    \tilde{L} = \half \sqrt{\tilde{q}\left|\beta\right|}
    \left\{
        \left( \tilde{R} - \frac{\tilde{\mathcal{K}}}{\beta} \right) + \left( \frac{\omega_\psi}{\omega_R} + \frac{3\omega_R^{\prime2}}{2\omega_R^2} \right) \left( \frac{\tilde{\nu}^2}{\beta} - \tilde{q}^{ab} \partial_a \psi \partial_b \psi \right)
    \right\} - \sqrt{\tilde{q}} \left( \frac{U}{\omega_R^2} \right),
        \label{eq:2ndst_lagrangian_einstein}
\end{equation}
where variables with tildes are Einstein-frame quantities.
So the Einstein frame couplings are given by $\tilde{\omega}_R=1$, $\tilde{\omega}_\psi=\left(\omega_\psi\omega_R+3\omega_R^{\prime2}/2\right)/\omega_R^2$, and the potential by $\tilde{U}=U/\omega_R^2$.

When the term `Einstein frame' is used elsewhere in the literature, it often refers to an action which is transformed further so that the effective scalar coupling is also unity.  I can make this transformation to a minimally coupled scalar $\varphi$ by solving the differential equation,
\begin{equation}
    \partdif{ \varphi}{\psi} = \sqrt{\frac{\omega_\psi}{\omega_R} + \frac{3}{2} \left( \frac{\partial_\psi\omega_R}{\omega_R} \right)^2},
\end{equation}
for example, when $\omega_\psi=0$, this is solved by
${
\displaystyle  \varphi \left( \psi \right) = \sqrt{ \frac{3}{2} } \log{ \omega_R \left( \psi \right) } \sgn{ \partial_\psi \log{ \omega_R \left( \psi \right) } }
}$.
For the parameterisation of $F(\four{R})$ given in \secref{sec:methodology_FR}, $\omega_R=\omega\psi$, and the transformation is given by $\psi\left(\varphi\right)\propto{e}^{\varphi\sqrt{2/3}}$ as long as $\psi>0$.


\section{Multiple scalar fields}
\label{sec:2ndst_multiple}

Consider the case of multiple scalar fields.  I start from the distribution equation as before, but label the scalar field variables with an index.  Proceeding like in \secref{sec:2ndst_dist-eqn_p} by taking functional derivatives with respect to $p^{ab}$ and then integrating by parts with test function $\theta^{ab}$, I obtain the conditions,
\begin{subequations}
\begin{align}
    \partial_{ab} \theta^{ab} : 0 & =
    C_{(R)} C^{(p^2\x)} + \beta,
        \label{eq:2ndst_multiple_conditions_p1} 
        \\
    q_{ab} \partial^2 \theta^{ab} : 0 & =
    - 2 C_{(R)} \left( 2 C^{(p^2\p)} + C^{(p^2\x)} \right) + \sum_I C_{(\psi_I'')} C^{(p\pi_I)},
        \label{eq:2ndst_multiple_conditions_p2} 
        \\
    X_b \partial_a \theta^{ab} : 0 & =
    C_{(R)} \left( 1 + 2 \partial_q \right) C^{(p^2\x)} + \partial_q \beta,
        \label{eq:2ndst_multiple_conditions_p3}
        \\
    \partial_b \psi_I \partial_a \theta^{ab} : 0 & = 
    \left( C_{(\psi_I'')} - 2 C_{(R)} \partial_{\psi_I} \right) C^{(p^2\x)} - \partial_{\psi_I} \beta,
        \label{eq:2ndst_multiple_conditions_p4} 
        \\
\begin{split}
    q_{ab} \partial^c \psi_I \partial_c \theta^{ab} : 0 & = 
    \left( C_{(\psi_I'')} - 8 C_{(R)} \partial_{\psi_I} \right) C^{(p^2\p)} + \left( C_{(\psi_I'')} - 4 C_{(R)} \partial_{\psi_I} \right) C^{(p^2\x)}
\\
    + 2 \left( C_{(\psi_I^{\prime2})} 
    \right. & + \left. C_{(\psi_I'')} \partial_{\psi_I} \right) C^{(p\pi_I)}
    + \sum_{J\neq{}I} \left( C_{(\psi_I'\psi_J')} + C_{(\psi_J'')} \partial_{\psi_I} \right) C^{(p\pi_J)}.
\end{split}
        \label{eq:2ndst_multiple_conditions_p9} 
\end{align}
    \label{eq:2ndst_multiple_conditions_p}%
\end{subequations}
I note that there are other independent terms, but they do not produce any extra conditions.
Likewise, if I follow the route taken in \secref{sec:2ndst_dist-eqn_pi}, taking the functional derivative with respect to $\pi_I$ then integrating by parts with test function $\eta_I$, I find the conditions,
\begin{subequations}
\begin{align}
    \partial^2 \eta_I :
    0 & = C_{(R)} C^{(p\pi_I)} - C_{(\psi_I'')} C^{(\pi_I^2)} - \half \sum_{J\neq{}I} C_{(\psi_J'')} C^{(\pi_I\pi_J)},
        \label{eq:2ndst_multiple_conditions_pi1}
\\
\begin{split}
    X^a \partial_a \eta_I :
    0 & = C_{(R)} \left( 1 + 4 \partial_q \right) C^{(p\pi_I)} - C_{(\psi_I'')} \left( 1 + 4 \partial_q \right) C^{(\pi_I^2)}
\\ &
    - \half \sum_{J\neq{}I} C_{(\psi_J'')} \left( 1 + 4 \partial_q \right) C^{(\pi_I\pi_J)},
\end{split}        
        \label{eq:2ndst_multiple_conditions_pi2}
\\
\begin{split}
    \partial^a \psi_I \partial_a \eta_I  :
    0 & = \left( \half C_{(\psi_I'')} - 4 C_{(R)} \partial_{\psi_I} \right) C^{(p\pi_I)} + 4 \left( C_{(\psi_I^{\prime2})} + C_{(\psi_I'')} \partial_{\psi_I} \right) C^{(\pi_I^2)}
\\ &  
    + \sum_{J\neq{}I} \left( C_{(\psi_I'\psi_J')} + 2 C_{(\psi_J'')} \partial_{\psi_I} \right) C^{(\pi_I\pi_J)} - \beta,
\end{split}
        \label{eq:2ndst_multiple_conditions_pi3}
\\
\begin{split}
    \partial^a \psi_{J\neq{}I} \partial_a \eta_I  :
    0 & = \left( \half C_{(\psi_J'')} - 2 C_{(R)} \partial_{\psi_J} \right) C^{(p\pi_I)} + 2 \left( C_{(\psi_I'\psi_J')} + 2 C_{(\psi_I'')} \partial_{\psi_J} \right) C^{(\pi_I^2)}
\\
    + 2 \left( C_{(\psi_J^{\prime2})} \right. & + \left. C_{(\psi_J'')} \partial_{\psi_J} \right) C^{(\pi_I\pi_J)}
    + \sum_{K\neq{}I,J} \left( C_{(\psi_J'\psi_K')} + 2 C_{(\psi_K'')} \partial_J \right) C^{(\pi_I\pi_K)},
\end{split}
    \label{eq:2ndst_multiple_conditions_pi4}
\end{align}
    \label{eq:2ndst_multiple_conditions_pi}%
\end{subequations}
and similar to above, there are other independent terms which do no produce any unique conditions.

To solve this system of equations I must make assumptions, in particular about the relationship between the scalar fields.  One choice might be to assume an $O\left(N\right)$ symmetry, where the coupling and deformation would only depend on the absolute value of the scalar field multiplet 
${\textstyle\left|\psi\right|=\sqrt{\sum_I\psi_I^2}}$, 
and relationships between the $C_{(\psi_I'\psi_J')}$ coefficients could be assumed.

However, I instead choose to take one non-minimally coupled field $\left(\psi,\pi_{\psi}\right)$ and one minimally coupled field $\left(\varphi,\pi_{\varphi}\right)$ with no cross-terms in the spatial derivative sector, ${C_{(\varphi'\psi')}=0}$.  The minimally coupled field only appears in terms other than the potential $U\left(q,\psi,\varphi\right)$ through the deformation function $\beta(q,\psi,\varphi)$.  For example, ${C_{(R)}=C_{(R)}\left(q,\psi,\beta\right)}$.

Solving \eqref{eq:2ndst_multiple_conditions_p1} and \eqref{eq:2ndst_multiple_conditions_p3} gives me,
\begin{equation}
    C_{(R)} = f \left( \psi \right) \sqrt{q \left| \beta \left( q, \psi, \varphi \right) \right|},
        \quad
    C^{(p^2\x)} = \frac{-1}{f\left(\psi\right)} \sqrt{\frac{\left| \beta \left( q, \psi, \varphi \right) \right|}{q}},
\end{equation}
as before.  Substituting these into \eqref{eq:2ndst_multiple_conditions_p2} and \eqref{eq:2ndst_multiple_conditions_p4} gives me,
\begin{equation}
    C_{(\psi'')} = - 2 f' \sqrt{q\left|\beta\right|},
        \quad
    C_{(\varphi'')} = 0,
        \quad
    C^{(p^2\p)} = \frac{\sigma_\beta}{2f} \sqrt{\frac{\left|\beta\right|}{q}} - \frac{f'}{2f} C^{(p\pi_{\psi})},
\end{equation}
and the remaining conditions are,
\begin{subequations}
\begin{gather}
    C^{(p\pi_{\psi})}  = 
    \frac{-\sigma_\beta f'}{2f}  \sqrt{\frac{\left|\beta\right|}{q}} 
    \left\{
        \frac{C_{(\psi^{\prime2})}}{\sqrt{q\left|\beta\right|}} + 2 f'' - \frac{3f^{\prime2}}{2f}
    \right\}^{-1}
        \\
    C^{(\pi_{\varphi}\pi_{\psi})} = 
    \frac{-\partial_\varphi\beta\,\partial_\psi{}f}{4C_{(\varphi^{\prime2})}}
    \left\{
    \frac{
        \frac{2C_{(\psi^{\prime2})}}{\sqrt{q\left|\beta\right|}} \left( 1 - \partdif{\,\log{}C_{(\psi^{\prime2})}}{\,\log{}\beta} \right) + 2 f'' - \frac{3f^{\prime2}}{2f}
    }{
        \left[ \frac{C_{(\psi^{\prime2})}}{\sqrt{q\left|\beta\right|}} + 2 f'' - \frac{3f^{\prime2}}{2f} \right]^2
    }
    \right\},
        \\
    C^{(\pi_{\varphi}^2)} = \frac{\beta}{4C_{(\varphi^{\prime2})}},
        \quad
    C^{(p\pi_{\varphi})} =
        -\frac{f'}{f} C^{(\pi_{\varphi}\pi_{\psi})}.
\end{gather}%
\end{subequations}
I note that the constraint is significantly simpler if I assume 
${C_{(\varphi^{\prime2})}=g_\varphi\left(\psi\right)\sqrt{q\left|\beta\right|}}$ 
and 
${C_{(\psi^{\prime2})}=g_\psi\left(\psi\right)\sqrt{q\left|\beta\right|}}$, 
where $g_\varphi$ and $g_\psi$ are arbitrary functions.
In this case the whole Hamiltonian constraint is
\begin{equation}
\begin{split}
    C & = \sqrt{q\left|\beta\right|} \Big( f R - 2 f' \Delta \psi + g_\varphi \partial_a \varphi \partial^a \varphi + g_\psi \partial_a \psi \partial^a \psi \Big) + \sqrt{q} \, U
        \\
    &  + \sigma_\beta \sqrt{\frac{\left|\beta\right|}{q}} \left\{ \frac{\pi_{\varphi}^2}{4g_\varphi} + \frac{1}{f} \left( \frac{p^2}{6} - \bp \right)
    + \frac{\left( \pi_{\psi} - \frac{f'}{f} p \right) \left( \pi_{\psi} - \frac{f'}{f} p - \frac{f'\partial_\varphi\beta}{\beta{}g_\varphi} \pi_{\varphi} \right)
    }{4 \left( g_\psi + 2 f'' - \frac{3f^{\prime2}}{2f} \right)}
    \right\},
\end{split}
\end{equation}
and the associated Lagrangian density is
\begin{subequations}
\begin{gather}
\begin{split}
    L & = \sqrt{q \left| \beta \right|} \left\{ 
        f \left( \frac{\mathcal{K}}{\beta} - R \right) 
        + f' \left( \frac{\nu_\psi v}{\beta} + 2 \Delta \psi \right) 
        + \left( \frac{\hat{g}_\psi}{h} + \frac{3f^{\prime2}}{2f} \right) \frac{\nu_\psi^2}{\beta} 
\right.  \\ & \left.  
        - g_\psi \partial_a \psi \partial^a \psi 
        + \frac{g_\varphi}{h\beta} \nu_\varphi^2 
        - g_\varphi \partial_a \varphi \partial^a \varphi 
        + \frac{f'\partial_\varphi\beta}{h\beta} \nu_\varphi \nu_\psi 
    \right\} - \sqrt{q} \, U,
\end{split}
        \\
    \hat{g}_{\psi} = g_\psi + 2 f'' - \frac{3f^{\prime2}}{2f},
        \quad \quad
    h = 1 - \frac{f^{\prime2}\partial_\varphi\beta^2}{4g_\varphi\hat{g}_\psi\beta^2}.
\end{gather}%
    \label{eq:2ndst_multiple_lagrangian_rough}%
\end{subequations}
If $\beta$ does not depend on $\varphi$, then this can be simplified greatly, in which case the effective and covariant forms of the Lagrangian are given by,
\begin{subequations}
\begin{align}
\begin{split}
    L & = \half \sqrt{ q \left| \beta \right| } \left\{ \omega_R \left( R - \frac{\mathcal{K}}{\beta} \right) - \omega_R' \left( \frac{\nu_\psi{}v}{\beta} + 2 \Delta \psi \right) + \omega_\varphi \bigg( \frac{\nu_\varphi^2}{\beta} - \partial_a \varphi \partial^a \varphi \bigg) \right.
        \\
    &  \left.
    + \frac{\omega_\psi\nu_\psi^2}{\beta} - \left( \omega_\psi + 2 \omega_R'' \right) \partial_a \psi \partial^a \psi  \right\} - \sqrt{q} \, U,
\end{split}
        \\
    L_\mathrm{cov} & = \half \sqrt{q\left|\beta\right|} \left( \omega_R \, {}^{(4,\beta)}\!R - \omega_\psi \, \partial_\mu^{(4,\beta)} \psi \partial^\mu_{(4,\beta)} \psi - \omega_\varphi \, \partial_\mu^{(4,\beta)} \varphi \partial^\mu_{(4,\beta)} \varphi \right) - \sqrt{q} \, U,
\end{align}%
    \label{eq:2ndst_multiple_lagrangian}%
\end{subequations}
where ${\omega_R=-2f}$, 
${\omega_\psi=2\left(g_\psi+2f''\right)}$, 
${\omega_\varphi=2g_\varphi}$.  Therefore, when I assume that the minimally coupled scalar field can also be considered to be minimally coupled to the deformation function, I find that the action simplifies to the expected form.
It would be interesting to see what effects appear for scalar field multiplets, especially for non-Abelian symmetries, but that is beyond the scope of this study.
Instead, I now turn to studying the cosmological dynamics of my results.


\section{Cosmology}
\label{sec:2ndst_cosmo}

To find the cosmological dynamics, I restrict to a flat, homogeneous, and isotropic metric in proper time ($N=1$).  I also assume that $\beta$ does not depend on the minimally coupled scalar field $\varphi$ for the sake of simplicity.  From \eqref{eq:2ndst_multiple_lagrangian}, I find the Friedmann equation, which can be written in two equivalent forms,
\begin{subequations}
\begin{align}
    \mathcal{H} \left( \omega_R \mathcal{H} + \omega_R' \dot{\psi} \right)
    & =
    \third \left( 
        \frac{\omega_\psi}{2} \dot{\psi}^2 
        + \frac{\omega_\varphi}{2} \dot{\varphi}^2 
        + \sigma_\beta \sqrt{\left|\beta\right|} \, U  
    \right),
        \label{eq:2ndst_cosmo_friedmann_1}
\\
    \left( \omega_R \mathcal{H} + \half \omega_R' \dot{\psi} \right)^2
    & =
    \third \left[ 
        \half \left( \omega_R \omega_\psi 
        + \frac{3}{2} \omega_R^{\prime2} \right) \dot{\psi}^2 
        + \frac{\omega_R \omega_\varphi}{2} \dot{\varphi}^2 
        + \sigma_\beta \omega_R \sqrt{\left|\beta\right|} \, U  
    \right].
        \label{eq:2ndst_cosmo_friedmann_2}
\end{align}
    \label{eq:2ndst_cosmo_friedmann}%
\end{subequations}
From \eqref{eq:2ndst_cosmo_friedmann_2} I see that ${\omega_R\omega_\psi+3\omega_R^{\prime2}/2\geq0}$ and ${\omega_R\omega_\varphi\geq0}$ are necessary when $U\to0$ to ensure real-valued fields.  If I compare this condition to the Einstein frame Lagrangian \eqref{eq:2ndst_lagrangian_einstein}, I can see that it is also the condition which follows from insisting that the scalar field $\psi$ is not ghost-like in that frame.  Similarly, I see that ${\sigma_\beta\omega_R>0}$ is necessary when ${\dot{\psi},\dot{\varphi}\to0}$.

For the reasonable assumption that the minimally coupled field $\varphi$ does not affect the deformation function $\beta$, the only way that field is modified is through a variable maximum phase speed ${c_\varphi^2=\beta}$.  Due to this minimal modification, it does not produce any of the cosmological phenomena I am interested in (bounce, inflation) through any novel mechanism.  Therefore, I will ignore this field for the rest of the chapter.

I find the equations of motion by varying the Lagrangian \eqref{eq:2ndst_multiple_lagrangian} with respect to the fields.  
For the simple undeformed case $\beta=1$ the equations are given by,
\begin{subequations}
\begin{gather}
\begin{gathered}
    \left( \omega_R \omega_\psi + \frac{3}{2} \omega_R^{\prime2} \right) \ddot{\psi} =
    - 3 \dot{\psi} \mathcal{H} \left( \omega_R \omega_\psi + \omega_R^{\prime2} \right) - \omega_R \partial_\psi U + \frac{3}{2} \omega_R \omega_R' \mathcal{H}^2
        \\
    - \half \dot{\psi}^2 \left( \omega_R \omega_\psi' + \frac{3}{2} \omega_R' \omega_\psi + 3 \omega_R' \omega_R'' \right) + \frac{3}{2} \omega_R' \left( 1 + \frac{a}{3} \partdif{}{a} \right) U,
\end{gathered}
    \label{eq:2ndst_cosmo_scalar}
        \\
\begin{gathered}
    \left( \omega_R \omega_\psi + \frac{3}{2} \omega_R^{\prime2} \right) \frac{\ddot{a}}{a} =
    - \frac{1}{2} \mathcal{H}^2 \left( \omega_R \omega_\psi + 3 \omega_R^{\prime2} \right) + \frac{\omega_\psi}{2} \left( 1 + \frac{a}{3} \partdif{}{a} \right) U
        \\
    - \quarter \dot{\psi}^2 \left( \omega_\psi^2 + 2 \omega_\psi \omega_R'' - \omega_\psi' \omega_R' \right) + \half \omega_R' \omega_\psi \dot{\psi} \mathcal{H} - \frac{\omega_R'}{2} \partial_\psi U,
\end{gathered}
    \label{eq:2ndst_cosmo_acceleration}
\end{gather}
    \label{eq:2ndst_cosmo_eom}%
\end{subequations}
\begin{sloppypar}
where I can see from the equations of motion that the model breaks down if 
${\omega_R\omega_\psi+3\omega_R^{\prime2}/2\to0}$ 
because it will tend to cause 
${|\ddot{\psi}|\to\infty}$ and ${|\ddot{a}|\to\infty}$.
\end{sloppypar}

\subsection{Bounce}
\label{sec:2ndst_cosmo_bounce}

I will address the question of whether there are conditions under which there can be a big bounce as defined in \secref{sec:methodology_cosmo}.
I find in \chapref{sec:pert} (and in ref.~\cite{cuttell2014}) that a deformation function which depends on curvature terms can generate a bounce. Elsewhere in the literature on loop quantum cosmology the bounce happens in a regime when $\beta<0$ because the terms depending on curvature or energy density overpower the zeroth order terms \cite{Cailleteau2012a, Mielczarek:2012pf}.  However, I am not including derivatives in the deformation here so the effect would have to come from the non-minimal coupling of the scalar field or the zeroth order deformation.

I take $\dot{a}=0$ for finite $a$, include a deformation and I ignore the minimally coupled field for simplicity.  From the Friedmann equation \eqref{eq:2ndst_cosmo_friedmann} I find,
\begin{equation}
    0 = \frac{\omega_\psi}{2} \dot{\psi}^2 + \sigma_\beta \sqrt{ \left| \beta \right| } \, U,
        \label{eq:2ndst_cosmo_bounce}
\end{equation}
which implies that ${\sigma_\beta\omega_\psi<0}$ for a bounce because otherwise the equation cannot balance for $U>0$ and ${\psi\in\mathbb{R}}$.
Substituting \eqref{eq:2ndst_cosmo_bounce} into the full equation of motion for the scale factor, and demanding that $\ddot{a}>0$ to make it a turning point, I find the following conditions,
\vspace{-\baselineskip}
\begin{subequations}%
\begin{gather}
    \sigma_\beta \omega_\psi < 0,
        \\
    \omega_R \omega_\psi + \frac{3}{2} \omega_R^{\prime2} > 0,
        \\
    \sigma_\beta \sqrt{ \left| \beta \right| } \left( \omega_\psi + 2 \omega_R'' \right) U - \frac{\sigma_\beta\omega_R'}{2\omega_\psi} \partial_\psi \left( \sqrt{ \left| \beta \right| } \omega_\psi U \right) + \frac{a\beta}{6} \partdif{}{a} \left( \frac{ \omega_\psi U }{ \sqrt{ \left| \beta \right| } } \right) > 0,
\end{gather}
    \label{eq:2ndst_cosmo_bounce_conditions}%
\end{subequations}
from which I can determine what the coupling functions, deformation and potential must be for a bounce.
For example, if I look at the minimally coupled case, when ${\omega_R=\omega_\psi=1}$, and assume that $U>0$, I can see that the conditions are given by,
\begin{equation}
    \sigma_\beta < 0,
        \quad
    \partdif{ \, \log \left( \left| \beta \right|^{-1/2} U \right) }{ \, \log{a} } < -6.
\end{equation}
Since I must have ${\beta\to1}$ in the classical limit and ${\sigma_\beta<0}$ at the moment of the bounce, then $\beta$ must change sign at some point.  Therefore, a universe which bounces purely due to a zeroth order deformation must have effective signature change.
Another example is obtained by assuming scale independence and choosing ${\beta=1}$ and ${U>0}$.  In this case the bounce conditions become,
\begin{equation}
    \omega_\psi < 0,
        \quad
    \omega_\psi \omega_R + \frac{3}{2} \omega_R^{\prime2} > 0,
        \quad
    \omega_\psi + 2 \omega_R'' - \half \omega_R' \partial_\psi \log \left( \omega_\psi U \right) > 0,
\end{equation}
which I can use to find a model which bounces purely due to a scale-independent non-minimally coupled scalar.  I present this model in \subsecref{sec:2ndst_cosmo_BS}.

\subsection{Inflation}
\label{sec:2ndst_cosmo_inflation}

Now consider the inflationary dynamics.
For simplicity I assume that inflation will come from a scenario similar to slow-roll inflation with possible enhancements coming from the non-minimal coupling or the deformation.
The conditions for slow-roll inflation are,
\begin{equation}
    \dot{\psi}^2 \ll U, 
        \quad
    \left| \ddot{\psi} \right| \ll \left| \dot{\psi} \mathcal{H} \right|,
        \quad
    \left| \dot{\mathcal{H}} \right| \ll \mathcal{H}^2,
        \label{eq:2ndst_slow-roll_conditions}
\end{equation}
assuming the couplings, potential and deformation are scale independent and the deformation is positive, I get the following slow roll equations,
\begin{subequations}
\begin{align}
    \mathcal{H} & \simeq
    \sqrt{ \frac{ \beta^{1/2} U }{ 3 \omega_R } },
        \\
    \dot{\psi} & \simeq
    - \sqrt{ \frac{ \beta^{1/2} U }{ 3 \omega_R } }
    \left( \frac{ \partial_\psi \log \left( \frac{U}{ \beta^{1/2} \omega_R^2 } \right) }{ \frac{\omega_\psi}{\omega_R} + \frac{\omega_R^{\prime2}}{\omega_R^2} + \frac{\beta'\omega_R'}{2\beta\omega_R} } \right),
\end{align}%
    \label{eq:2ndst_slow-roll_equations}%
\end{subequations}
and define the slow-roll parameters,
\begin{equation}
    \epsilon : = \frac{-\dot{\mathcal{H}}}{\mathcal{H}^2},
        \quad
    \eta : = \frac{-\ddot{\mathcal{H}}}{\mathcal{H}\dot{\mathcal{H}}},
        \quad
    \zeta : = \frac{-\ddot{\psi}}{\dot{\psi}\mathcal{H}},
        \label{eq:2ndst_slow-roll_parameters_definition}
\end{equation}
which, under slow-roll conditions are given by,
\begin{subequations}
\begin{align}
    \epsilon & \simeq \frac{ \partial_\psi \log \left( \frac{\beta^{1/2}U}{\omega_R} \right) \partial_\psi \log \left( \frac{U}{\beta^{1/2}\omega_R^2} \right) }{ 2 \left( \frac{\omega_\psi}{\omega_R} + \frac{\omega_R^{\prime2}}{\omega_R^2} + \frac{\beta'\omega_R'}{2\beta\omega_R}  \right) }
        \label{eq:2ndst_slow-roll_epsilon} \\
    \eta & \simeq \left( \frac{ \partial_\psi \log \left( \frac{U}{\beta^{1/2}\omega_R^2} \right) }{ \frac{\omega_\psi}{\omega_R} + \frac{\omega_R^{\prime2}}{\omega_R^2} + \frac{\beta'\omega_R'}{2\beta\omega_R} } \right) \partial_\psi \log{\epsilon} + 2 \epsilon,
        \label{eq:2ndst_slow-roll_eta} \\
    \zeta & \simeq
    \partial_\psi \left(
    \frac{\partial_\psi \log \left( \frac{U}{\beta^{1/2}\omega_R^2} \right) }{ \frac{\omega_\psi}{\omega_R} + \frac{\omega_R^{\prime2}}{\omega_R^2} + \frac{\beta'\omega_R'}{2\beta\omega_R} }
    \right)
    + \epsilon,
        \label{eq:2ndst_slow-roll_zeta}%
\end{align}%
        \label{eq:2ndst_slow-roll_parameters}%
\end{subequations}
where a prime indicates a partial derivative with respect to $\psi$, i.e. $\beta'=\partial_\psi\beta$.
The slow-roll regime ends when the absolute value of any of these three parameters approaches unity.

Defining $\mathcal{N}$ to mean the number of e-folds from the end of inflation, 
${a\left(t\right)=a_\mathrm{end}e^{-\mathcal{N}\left(t\right)}}$, 
I find that,
\begin{equation}
    \mathcal{N} = - \int_{t_\mathrm{end}}^t \mathrm{d} t \mathcal{H} = - \int_{\psi_\mathrm{end}}^\psi \mathrm{d} \psi \frac{\mathcal{H}}{\dot{\psi}},
        \label{eq:2ndst_efolds_general}
\end{equation}
and using the slow-roll approximation,
\begin{equation}
    \mathcal{N} \simeq \int_{\psi_\mathrm{end}}^\psi \mathrm{d} \psi \frac{ \frac{\omega_\psi}{\omega_R} + \frac{\omega_R^{\prime2}}{\omega_R^2} + \frac{\beta'\omega_R'}{2\beta\omega_R}  }{\partial_\psi \log \left( \frac{ U }{ \beta^{1/2} \omega_R^2} \right) },
        \label{eq:2ndst_efolds}
\end{equation}
which can be solved once I specify the form of the couplings, deformation and potential.
I cannot find equations for observables such as the spectral index $n_s$ because it would require investigating how the cosmological perturbation theory is modified in the presence of non-minimal coupling and deformed general covariance.
Beyond this, it is difficult to make general statements about the dynamics unless I restrict to a given model, so I will now consider some models and discuss their specific dynamics.

\subsection{Geometric scalar model}
\label{sec:2ndst_cosmo_geo}

As demonstrated in the previous chapter, \secref{sec:methodology_FR}, the geometric scalar model comes from parameterising $\textstyle{}F\left({}^{(4)}\!R\right)$ gravity so that the additional degree of freedom of the scalar curvature is instead embodied in a non-minimally coupled scalar field $\psi$ \cite{Deruelle:2009pu, Deruelle2010}.
Its couplings are given by $\omega_R=\psi$ and $\omega_\psi=0$.  This model is a special case of the Brans-Dicke model, which has $\omega_\psi=\omega_0/\psi$, when the Dicke coupling constant $\omega_0$ vanishes.
I can add in a minimally coupled scalar field with $\omega_\varphi=1$ and thereby see the effect of this scalar-tensor gravity on the matter sector.  However, I set $\omega_\varphi=0$ because it does not significantly affect my results.

The effective action for this model is given by,
\begin{subequations}
\begin{gather}
\begin{split}
    L_\mathrm{geo} & = \half \sqrt{q\left|\beta\right|} \left\{ 
        \psi \left( R - \frac{\mathcal{K}}{\beta} \right) 
        - \frac{\nu_\psi v}{\beta} 
        - 2 \Delta \psi 
    \right\} 
    - \sqrt{q} \, U \left( \psi \right),
\end{split}
        \\
    U \left( \psi \right) = 
    \frac{\psi}{2} \left( F^{\,\prime} \right)^{-1} \left( \psi \right) 
    - \half F \left( \left( F^{\,\prime} \right)^{-1} \left( \psi \right) \right),
\end{gather}%
\end{subequations}
where $F$ refers to the 
${\textstyle{}F\left(\Rfour\right)}$ 
function which has been parameterised.  The equations of motion when ${\beta\to1}$ are given by,
\begin{subequations}
\begin{gather}
    \mathcal{H} \left( \psi \mathcal{H} + \dot{\psi} \right) 
    = \third U ,
        \\
    \frac{\ddot{a}}{a} 
    = - \mathcal{H}^2  + \frac{1}{3} \partdif{ U }{ \psi } ,
        \\
    \ddot{\psi} 
    = -2 \dot{\psi} \mathcal{H} + \psi \mathcal{H}^2 + \left( 1 + \frac{a}{3} \partdif{}{a} - \frac{2\psi}{3} \partdif{}{\psi} \right) U,
\end{gather}%
\end{subequations}
from which I can see that the scalar field has very different dynamics compared to minimally coupled scalars.  This reflects its origin as a geometric degree of freedom rather than a purely matter field.

\begin{figure}[t]
	\begin{center}
	{\subfigure[Scale factor (Logarithmic)]{
	    \label{fig:2ndst_geo_scale-factor-log}
		\includegraphics[height=0.24\textwidth]{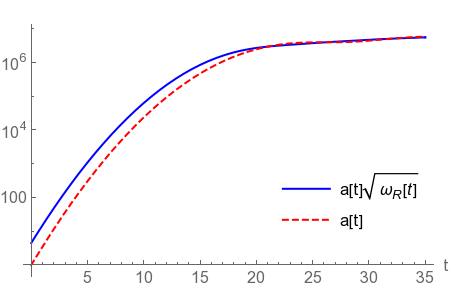}}}
	{\subfigure[Scale factor]{
		\label{fig:2ndst_geo_scale-factor}
		\includegraphics[height=0.24\textwidth]{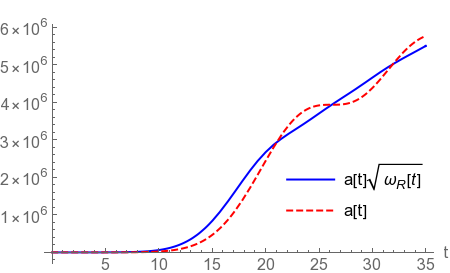}}}
	{\subfigure[Scalar field]{
	    \label{fig:2ndst_geo_scalar}
		\includegraphics[height=0.24\textwidth]{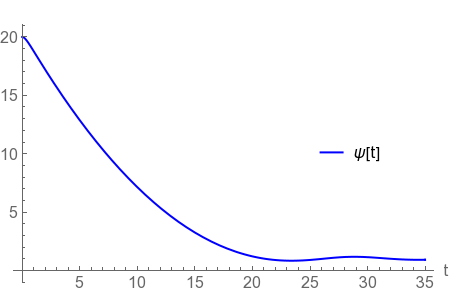}}}
	\end{center}
	\caption[Geometric scalar inflation]{Inflation from the geometric scalar model version of the Starobinsky model through slow-roll of the non-minimally coupled scalar field.  For the scale factor, I compare the Jordan and Einstein frames because the coupling causes the former to oscillate unusually.  Initial conditions, $a=1$, ${\psi=20}$, ${\dot{\psi}=0}$, $M=1$.}
    	\label{fig:2ndst_geo}
\end{figure}

Looking at inflation, the geometric scalar model with a potential corresponding to the Starobinsky model,
\begin{equation}
    F \left( \Rfour \right) = \Rfour + \frac{1}{2M^2} \Rfour^2
    \quad \to \quad
    U = \frac{M^2}{4} \left( \psi - 1 \right)^2,
\end{equation}
can indeed cause inflation through a slow-roll of the scalar field down its potential.  The non-minimal coupling of the scalar to the metric also causes the scale factor to oscillate unusually, however.  It is interesting to compare in \figref{fig:2ndst_geo} the scale factor in the Jordan frame, $a$, and the conformally transformed scale factor in the Einstein frame, ${\tilde{a}=a\sqrt{\omega_R}}$.
Assuming ${\psi>1}$ during inflation, the slow-roll parameters \eqref{eq:2ndst_slow-roll_parameters} are given by,
\begin{equation}
    \epsilon \simeq
    \frac{\psi+1}{\left(\psi-1\right)^2},
    \quad
    \eta \simeq
    \frac{-2}{\psi^2-1},
    \quad
    \zeta \simeq
    \frac{1}{\psi-1},
\end{equation}
\begin{sloppypar}
so the slow-roll regime of inflation ends at ${\psi\approx3}$ when ${\epsilon\to1}$.
The equation for the number of e-folds of inflation in the slow-roll regime \eqref{eq:2ndst_efolds} is given by
${
\displaystyle
    \mathcal{N} \simeq \half \left( \psi - \psi_\mathrm{end} - \log{\frac{\psi}{\psi_\mathrm{end}}} \right)
}$.
\end{sloppypar}

\subsection{Non-minimally enhanced scalar model}
\label{sec:2ndst_cosmo_nes}

\begin{figure}[t]
	\begin{center}
	{\subfigure[]{
	    \label{fig:2ndst_NES-exclusion}
		\includegraphics[height=0.39\textwidth]{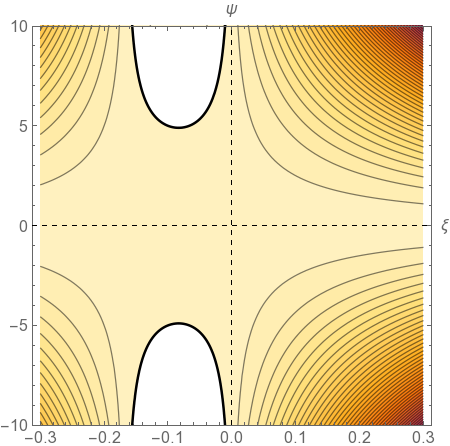}}}
	{\subfigure[]{
	    \label{fig:2ndst_NES-ghostzones}
	    \includegraphics[height=0.39\textwidth]{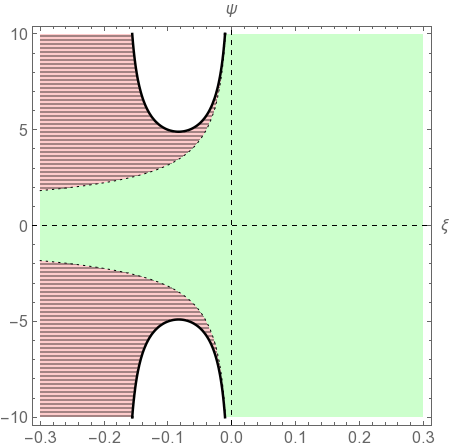}}}
	\end{center}
	\caption[Non-minimally enhanced scalar model parameter-space]{A contour plot of ${\omega_R\omega_\psi+3\omega_R^{\prime2}/2}$ for the non-minimally enhanced scalar model is shown in \subref{fig:2ndst_NES-exclusion}. In \subref{fig:2ndst_NES-ghostzones}, the red region is when the metric becomes ghost-like (when ${\omega_R<0}$). In both, the white regions are forbidden because it is where ${\omega_R\omega_\psi+3\omega_R^{\prime2}/2<0}$, implying imaginary fields. The green region is the region of well-behaved evolution.}
	    \label{fig:2ndst_NES}
\end{figure}

Unlike the geometric scalar model considered above, the non-minimally enhanced scalar model (NES) from \cite{Nozari:2010uu}, takes a scalar field from the matter sector and introduces a non-minimal coupling rather than extracting a degree of freedom from the gravity sector.  The coupling functions are given by
${\omega_R=1+\xi\psi^2}$, ${\omega_\psi=1}$ and ${\omega_\varphi=0}$. The strength of the quadratic non-minimal coupling is determined by the constant $\xi$.
The deformed effective Lagrangian for this model is given by,
\begin{equation}
\begin{split}
    L_\mathrm{NES} & = \sqrt{q\left|\beta\right|} \left\{ \half \left( 1 + \xi \psi^2 \right) \left( R - \frac{\mathcal{K}}{\beta} \right) + \half \left( \frac{\nu_\psi^2}{\beta} - \partial_a \psi \partial^a \psi \right)
    \right.
        \\
    &  \left. 
    - 2 \xi \left( \frac{\psi \nu_\psi v}{2\beta} + \psi \Delta \psi + \partial_a \psi \partial^a \psi \right) \right\} - \sqrt{q} \, U \left( \psi \right).
\end{split}%
\end{equation}
For some negative values of $\xi$, there are values of $\psi$ which are forbidden if I am to keep my variables real, shown in \figref{fig:2ndst_NES}.

\begin{figure}[t]
    \begin{center}
    {\subfigure[]{
        \label{fig:2ndst_nes_inflation_numerical}
            \includegraphics[height=0.32\textwidth]{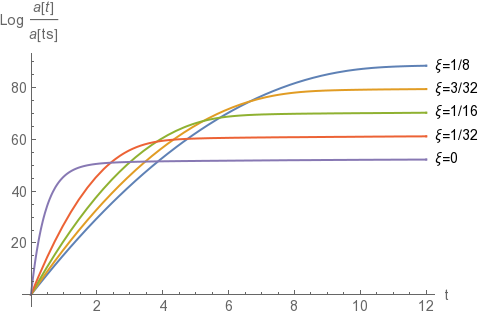}}}
    {\subfigure[]{
        \label{fig:2ndst_nes_inflation_efolds}
            \includegraphics[height=0.32\textwidth]{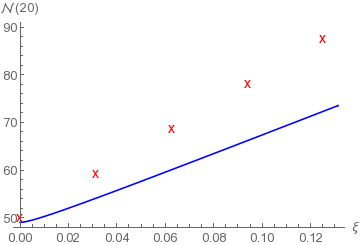}}}
    \end{center}
    \caption[Non-minimally enhanced scalar inflation]{For the non-minimally enhanced scalar model with ${U=\psi^4/4}$, \subref{fig:2ndst_nes_inflation_numerical} shows numerical solutions of inflation for different coupling strengths. Initial conditions, ${\psi=20}$, ${\dot{\psi}=0}$, ${\mathcal{H}>0}$.  In \subref{fig:2ndst_nes_inflation_efolds}, $\mathcal{N}$ for ${\psi=20}$ is compared for the numerical solutions (red crosses) and the analytical solution in the slow-roll approximation \eqref{eq:2ndst_efolds_NES} (blue line).}
    \label{fig:2ndst_nes_inflation}
\end{figure}

The equations of motion for this model when it is undeformed are given by,
\begin{subequations}
\begin{gather}
    \left( 1 + \xi \psi^2 \right) \mathcal{H}^2 + 2 \xi \psi \dot{\psi} \mathcal{H}
    = \third \left( \half \dot{\psi^2} + U \right),
        \label{eq:2ndst_cosmo_nes_friedmann} \\
\begin{split}
   \left( 1 + \left( 1 + 6 \xi \right) \xi \psi^2 \right) \frac{\ddot{a}}{a} 
    & = \frac{-1}{2} \mathcal{H}^2 \left( 1 + \left( 1 + 12 \xi \right) \xi \psi^2 \right) - \frac{1+4\xi}{4} \dot{\psi}^2
        \\
    & + \xi \psi \dot{\psi} \mathcal{H} + \half \left( 1 + \frac{a}{3} \partdif{}{a} \right) U + \xi \psi \partial_\psi U,
\end{split}
        \label{eq:2ndst_cosmo_nes_acceleration} \\
\begin{split}
    \left( 1 + \left( 1 + 6 \xi \right) \xi \psi^2 \right) \ddot{\psi}
    & = - 3 \dot{\psi} \mathcal{H} \left( 1 + \left( 1 + 4 \xi \right) \xi \psi^2 \right) - \left( 1 + \xi \psi^2 \right) \partial_\psi U
        \\
    & + 3 \xi \psi \left( \left( 1 + \xi \psi^2 \right) \mathcal{H}^2 - \frac{1+4\xi}{2} \dot{\psi}^2 + U + \frac{a}{3} \partdif{U}{a} \right).
\end{split}
        \label{eq:2ndst_cosmo_nes_scalar}
\end{gather}
    \label{eq:2ndst_cosmo_ns}%
\end{subequations}
and I proceed to use them to consider this model's inflationary dynamics.  For a power-law potential $U=\frac{\lambda}{n}|\psi|^n$ and ${\xi>0}$, the slow-roll parameter which reaches unity first is $\epsilon$ at 
${\displaystyle \psi_\mathrm{end} \simeq \frac{\pm n}{\sqrt{ 2 + n \left( 6 - n \right) \xi }}}$.
The number of e-folds from the end of inflation is given by,
\begin{equation}
    \mathcal{N}_\mathrm{NES} \left( \psi \right) \simeq \int^\psi_{\psi_\mathrm{end}} \mathrm{d} \varphi
    \frac{ \varphi \left( 1 + \left( 1 + 4 \xi \right) \xi \varphi^2 \right) }{ \left( 1 + \xi \varphi^2 \right) \left( n + \left( n - 4 \right) \xi \varphi^2 \right)},
\end{equation}
and if I specify that $n=4$, I find
\begin{equation}
    \mathcal{N}_\mathrm{NES} \simeq \frac{1+4\xi}{8} \psi^2 - 1 + \half \log{\frac{1+12\xi}{\left(1+4\xi\right)\left(1+\xi\psi^2\right)}},
        \label{eq:2ndst_efolds_NES}
\end{equation}
and the presence of $\xi$ in the dominant first term shows how the non-minimal coupling enhances the amount of inflation.  If I compare this result to numerical solutions in \figref{fig:2ndst_nes_inflation}, I see this effect.

The slow-roll approximation works less well as $\xi$ increases.  I can see this when I look at \figref{fig:2ndst_nes_inflation_efolds} where I compare the slow-roll approximation to when I numerically determine the end of inflation, i.e. when ${\epsilon=-\dot{\mathcal{H}}/\mathcal{H}^2=1}$.

I must be wary when dealing with this model, because the coupling can produce an effective potential which is not bounded from below.  If I substitute the Friedmann equation \eqref{eq:2ndst_cosmo_nes_friedmann} into \eqref{eq:2ndst_cosmo_nes_acceleration} and \eqref{eq:2ndst_cosmo_nes_scalar} I can find effective potential terms. These terms are those which do not vanish when all time derivatives are set to zero, and I can infer what bare potential they effectively behave like.  If the bare potential is ${U=\lambda\psi^2/2}$, then the effective potential term in the scalar equation behaves like
\begin{equation}
    U_\psi = \frac{-\lambda\psi^2}{2\left(1+6\xi\right)} + \frac{\lambda \left( 1 + 3 \xi \right)}{\xi \left( 1 + 6 \xi \right)^2} \log{\left( 1 + \left( 1 + 6 \xi \right) \xi \psi^2 \right)},
\end{equation}
which is not bounded from below when ${\xi>0}$ and ${\lambda>0}$ and is therefore unstable.  More generally, there are local maxima in the effective potential at ${\displaystyle\psi=\pm\sqrt{\frac{n}{\xi\left(4-n\right)}}}$, so for ${\xi>0}$ the model is stable for bare potentials which are of quartic order or higher.

\subsection{Bouncing scalar model}
\label{sec:2ndst_cosmo_BS}

\begin{figure}[t]
	\begin{center}
	{\subfigure[Scale factor]{
		\label{fig:2ndst_scalar-bounce_scale-factor}
		\includegraphics[width=0.46\textwidth]{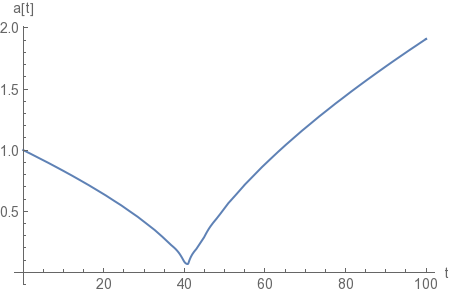}}}
	{\subfigure[Scalar]{
	    \label{fig:2ndst_scalar-bounce_scalar-field}
		\includegraphics[width=0.46\textwidth]{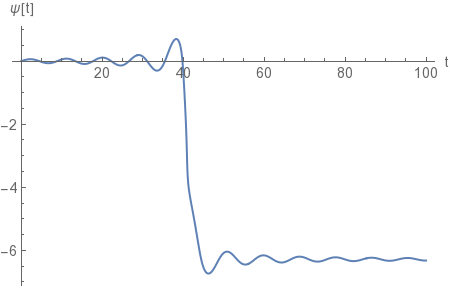}}}
	{\subfigure[Scalar coupling (zoomed)]{
	    \label{fig:2ndst_scalar-bounce_coupling}
		\includegraphics[width=0.46\textwidth]{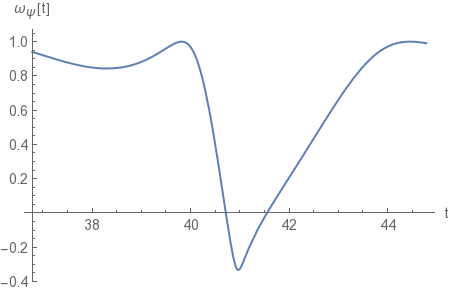}}}
	\end{center}
	\caption[Bouncing scalar]{Cosmological bounce generated by non-minimally coupled scalar field with ${b}=2$ and ${\textstyle{}U=\sin^2\left(\psi/2\right)}$. Initial conditions, $\psi=0$, $\dot{\psi}=1/25$, $\mathcal{H}<0$}
    	\label{fig:2ndst_scalar-bounce}
\end{figure}

As I said in \subsecref{sec:2ndst_cosmo_bounce}, I have taken the bounce conditions and constructed a model which bounces purely from the non-minimal coupling.
This model consists of a non-minimally coupled scalar with periodic symmetry.  My couplings are given by ${\omega_R=\cos{\psi}}$ and ${\displaystyle{}\omega_\psi=\frac{1+{b}\cos{\psi}}{1+{b}}}$, where ${b}$ is some real constant, and for simplicity I ignore deformations and the minimally coupled scalar field.  The bouncing scalar model Lagrangian in covariant and effective forms are given by,
\begin{subequations}
\begin{gather}
    L_\mathrm{BS,cov} = \sqrt{q} \left( \frac{\cos{\psi}}{2} \Rfour - \frac{1+{b}\cos{\psi}}{2\left(1+{b}\right)} \partial_\mu \psi \partial^\mu \psi - U \right),
        \\
\begin{aligned}
    L_\mathrm{BS} & = \frac{\sqrt{q}}{2} \bigg( \cos{\psi} \left( R - \mathcal{K} \right) + \sin{\psi} \left( \nu v + 2 \Delta \psi \right) + \left( \frac{1 + {b} \cos{\psi}}{1+{b}} \right) \nu^2 
        \\
    & 
    + \left( \frac{ \left(2+{b}\right) \cos{\psi} - 1}{1+{b}} \right) \partial_a \psi \partial^a \psi - 2 U \bigg).
\end{aligned}%
\end{gather}%
\end{subequations}
As confirmed by numerically evolving the equations of motion, I know from the bouncing conditions \eqref{eq:2ndst_cosmo_bounce_conditions} that this model will bounce when ${b}>1$ because then there is a value of $\psi$ for which ${\omega_\psi<0}$.
As I show in \figref{fig:2ndst_scalar-bounce}, the collapsing universe excites the scalar field so much that it `tunnels' through to another minima of the potential.  The bounce happens when the field becomes momentarily ghost-like, when ${\omega_\psi<0}$.

I can construct other models which produce a bounce purely through non-minimal coupling by having any ${U\left(\psi\right)}$ with multiple minima and couplings of the approximate form ${\omega\sim1-U}$.  However, to ensure the scalar does not attempt to tunnel through the potential to infinity and thereby not prevent collapse, the coupling functions must become negative only for values of $\psi$ between stable minima.  For example, for the $\mathbb{Z}_2$ potential
${
    U \left( \psi \right) = \lambda \left( \psi^2 - 1 \right)^2
}$,
couplings which are guaranteed to produce a bounce are
${
\omega_R \left( \psi \right) = \omega_\psi \left( \psi \right) = 1 - e^{-\psi^2} \, U \left( \psi \right)
}$
when $\lambda > 1$.


\section{Summary}
\label{sec:2ndst_summary}

In this chapter I have presented my calculation of the most general action for a second-order non-minimally coupled scalar-tensor model which satisfies a minimally deformed general covariance.  
I presented a similar calculation which involves multiple scalar fields.  I showed how the magnitude of the deformation can be removed by a transformation of the lapse function, but the sign of the deformation and the associated effective signature change cannot be removed.

I explored the background dynamics of the action, in particular showing the conditions required for either a big bounce or a period of slow-roll inflation.  By specifying the free functions I showed how to regain well-known models from my general action.  In particular I discussed the geometric scalar model, which is a parameterisation of ${\textstyle{}F\left({}^{(4)}\!R\right)}$ gravity and related to the Brans-Dicke model; and I discussed the non-minimally enhanced scalar model of a conventional scalar field with quadratic non-minimal coupling to the curvature.

I presented a model which produces a cosmological bounce purely through non-minimal coupling of a periodic scalar field to gravity.  I also provided the general method of producing similar models without a periodic symmetry.
I did not consider in detail the effect that the deformation has on the cosmological dynamics.  However, I did show that a big bounce which is purely due to a zeroth order deformation necessarily involves effective signature change.

Perhaps most importantly, I have established the minimally-deformed low-curvature limit that the subsequent chapters refer to.

\chapter{Fourth order perturbative gravitational action}
\label{sec:pert}

As I showed in \secref{sec:methodology_order_action}, the deformed action doesn't seem to naturally have a cut-off for higher powers of derivatives, and it must either be considered completely in general or treated perturbatively as a polynomial expansion.  In this chapter I will treat it perturbatively in order to find the lowest order corrections which are non-trivial.  This chapter is mostly adapted from a previously published paper \cite{cuttell2014}.

Firstly, I solve the distribution equation for the deformed gravitational action in \secref{sec:pert_dist-sol}.  Then I specify the variables used to construct the action and thereby find the conditions restricting its form in \secref{sec:pert_conditions}.  Afterwards, I progressively restrict the action when it is perturbatively expanded to fourth order in derivatives \secref{sec:pert_evaluate}.  Finally, I investigate the cosmological consequences of the results in \secref{sec:pert_cosmo}.

\section{Solving the action's distribution equation}
\label{sec:pert_dist-sol}

The general deformed action must satisfy the distribution equation \eqref{eq:dist-eqn_act},
\begin{equation}
    0 = 
    \funcdif{ L (x) }{ q_{ab} (y) } v_{ab} (y)
    + \sum_I \funcdif{ L (x) }{ \psi_I (y) } \nu_I (y)
    + \left( \beta D^a \partial_a \right)_x \delta \left( x, y \right)
    - \xty.
    \label{eq:pert_dist-eqn}
\end{equation}
I restrict to the case when there is only a metric field, for which the diffeomorphism constraint is given by \eqref{eq:diff_metric},
\begin{equation}
    D^a = - 2 \nabla_b p^{ab}
    = - 2 \left( \delta^a_{(b} \partial_{c)} + \Gamma^a_{bc} \right) \partdif{ L }{ v_{bc} }.
    \label{eq:pert_diffeomorphism}
\end{equation}
Firstly, I integrate \eqref{eq:pert_dist-eqn} by parts to move spatial derivatives from $L$ and onto the delta functions.  I discard the surface term and find,
\begin{equation}
\begin{split}
    0 & = \funcdif{ L (x) }{ q_{ab} (y) } v_{ab} (y)
    - 2 \left( \beta \partdif{ L }{ v_{bc} } \Gamma^a_{bc} \partial_a \right)_x \delta ( x, y )
\\ &
    + 2 \left( \partdif{ L }{ v_{ab} } \partial_b \right)_x \left[ \left( \beta \partial_a \right)_x \delta ( x, y ) \right]
    - \xty,
\end{split}
\end{equation}
from this I take the functional derivative with respect to $v_{ab}(z)$ (after relabelling the other indices),
\begin{equation}
\begin{split}
    0 & =
    \funcdif{ L (x) }{ q_{ab} (y) } \delta ( y, z )
    + \left\{
        \funcdif{ \partial L (x) }{ q_{cd} (y) \partial v_{ab} (x) } v_{cd} (y)
\right. \\ & \left.
        + 2 \left[ 
            \partdif{}{v_{ab}} \left( \partial_d \beta \partdif{ L }{ v_{cd} } - \beta \partdif{ L }{ v_{de} } \Gamma^c_{de} \right) \partial_c 
            + \partdif{}{ v_{ab} } \left( \beta \partdif{ L }{ v_{cd} } \right) \partial_{cd}
        \right]_x \delta ( x, y )
    \right\} \delta ( x, z )
\\ & 
   + 2 \left( \partdif{ \beta_{,d} }{ v_{ab,e} } \partdif{ L }{ v_{cd} } \right)_x \partial_{c(x)} \delta ( x, y ) \partial_{d(x)} \delta ( x, z )
    - \xty.
\end{split}
\end{equation}
I move the derivative from $\delta(x,z)$ and exchange some terms using the $\xty$ symmetry to find it in the form,
\begin{equation}
    0 = A^{ab} ( x , y ) \delta( y , z ) - A^{ab} ( y , x ) \delta( x , z ),
    \label{eq:pert}
\end{equation}
where,
\begin{equation}
\begin{split}
    A^{ab} ( x , y ) & = \funcdif{ L (x) }{ q_{ab} (y) }
    - v_{cd} (x) \funcdif{ \partial L (y) }{ q_{cd} (x) \partial v_{ab} (y) }
    + 2 \left\{ \partdif{}{ v_{ab} } 
        \left( \beta \partdif{ L }{ v_{de} } \Gamma^c_{de} - \partial_d \beta \partdif{ L }{ v_{cd} } \right) \partial_c
\right. \\ & \left.
    - \partdif{}{ v_{ab} } \left( \beta \partdif{ L }{ v_{cd} } \right) \partial_{cd}
        + \partial_e \left( \partdif{ \beta_{,d} }{ v_{ab,e} } \partdif{ L }{ v_{cd} } \right) \partial_c
    \right\}_y \delta ( y, x ).
\end{split}
\end{equation}
Integrating over $y$, I find that part of the equation can be combined into a tensor dependent only on $x$,
\begin{equation}
\begin{split}
    0 & = A^{ab}(x,z) - \delta(z,x) \int \mathrm{d}^3 y A^{ab}(y,x),
        \\
    & = A^{ab}(x,z) - \delta(z,x) A^{ab} (x),
    \quad \mathrm{where} \;
    A^{ab}(x) = \int \mathrm{d}^3 y A^{ab} \left(y, x \right).
\end{split}%
\end{equation}
Substituting in the definition of $A^{ab}(x,z)$ then relabelling,
\begin{equation}
\begin{split}    
    0 &= \funcdif{ L (x) }{ q_{ab} (y) }
    - v_{cd} (x) \funcdif{ \partial L (y) }{ q_{cd} (x) \partial v_{ab} (y) }
    + 2 \left\{ \partdif{}{ v_{ab} } 
        \left( \beta \partdif{ L }{ v_{de} } \Gamma^c_{de} - \partial_d \beta \partdif{ L }{ v_{cd} } \right) \partial_c
\right. \\ & \left.
    - \partdif{}{ v_{ab} } \left( \beta \partdif{ L }{ v_{cd} } \right) \partial_{cd}
        + \partial_e \left( \partdif{ \beta_{,d} }{ v_{ab,e} } \partdif{ L }{ v_{cd} } \right) \partial_c
    \right\}_y \delta ( y, x )
    - A^{ab} (x) \delta ( x, y ).
\end{split}
\end{equation}
To find this in terms of one independent variable, I multiply by the test tensor $\theta_{ab}(y)$ and integrate by parts over $y$,
\begin{equation}
\begin{split}
    0 & = \partdif{ L }{ q_{ab} } \theta_{ab} 
    + \partdif{ L }{ q_{ab,c} } \partial_c \theta_{ab}
    + \partdif{ L }{ q_{ab,cd} } \partial_{cd} \theta_{ab}
    - v_{cd} \partdif{^2 L }{ q_{cd} \partial v_{ab} } \theta_{ab} 
\\ & 
    + v_{cd} \partial_e \left( \partdif{^2 L }{ q_{cd,e} \partial v_{ab} } \theta_{ab} \right)
    - v_{cd} \partial_{ef} \left( \partdif{^2 L }{ q_{cd,ef} \partial v_{ab} } \theta_{ab} \right)
\\ & 
    + 2 \partial_c \left\{
        \theta_{ab} \partdif{}{ v_{ab} } \left( \partial_d \beta \partdif{ L }{ v_{cd} } - \beta \partdif{ L }{ v_{de} } \Gamma^c_{de} \right)
        - \theta_{ab} \partial_e \left( \partdif{ \beta_{,d} }{ v_{ab,e} } \partdif{ L }{ v_{cd} } \right)
    \right\}
\\ & 
    + 2 \partial_{cd} \left\{ \theta_{ab} \partdif{ \beta_{,e} }{ v_{ab,(c} } \partdif{ L }{ v_{d)e} }
    - \theta_{ab} \partdif{}{ v_{ab} } \left( \beta \partdif{ L }{ v_{cd} } \right)
    \right\}
    - A^{ab} \theta_{ab}.
\end{split}
\end{equation}
Then collecting derivatives of $\theta_{ab}$,
\begin{equation}
\begin{split}
    0 & = \theta_{ab} \left( \cdots \right)^{ab}
    + \partial_c \theta_{ab} \left\{
        \partdif{ L }{ q_{ab,c} } 
        + v_{de} \partdif{^2 L }{ q_{de,c} \partial v_{ab} } - 2 v_{ef} \partial_d \left( \partdif{^2 L }{ q_{ef,cd} \partial v_{ab} } \right)
\right. \\ & \left.
        + 2 \partdif{}{ v_{ab} } \left( \partial_d \beta \partdif{ L }{ v_{cd} } - \beta \partdif{ L }{ v_{de} } \Gamma^c_{de} \right)
        - 4 \partial_d \left[ \partdif{}{ v_{ab} } \left( \beta \partdif{ L }{ v_{cd} } \right) \right]
        + 2 \partial_e \left( \partdif{ \beta_{,d} }{ v_{ab,c} } \partdif{ L }{ v_{de} } \right)
    \right\}
\\ & 
    + \partial_{cd} \theta_{ab} \left\{ 
        \partdif{ L }{ q_{ab,cd} }
        - v_{ef} \partdif{^2 L }{ q_{ef,cd} \partial v_{ab} }
        - 2 \partdif{}{ v_{ab} } \left( \beta \partdif{ L }{ v_{cd} } \right)
        + 2 \partdif{ \beta_{,e} }{ v_{ab,(c} } \partdif{ L }{ v_{d)e} }
    \right\},
\end{split}
\end{equation}
where I have discarded the terms containing $\theta_{ab}$ without derivatives, because they do not provide any restrictions on the form of the action.
This is simplified by noting that $\partial_c$ and $\displaystyle{\partdif{}{v_{ab}}}$ commute, and that 
$\displaystyle{\partdif{\beta_{,e}}{v_{ab,c}}=\delta^c_e\partdif{\beta}{v_{ab}}}$.  
Therefore, the solution is given by,
\begin{equation}
\begin{split}
    0 & = \theta_{ab} \left( \cdots \right)^{ab}
    + \partial_c \theta_{ab} \left\{
        \partdif{ L }{ q_{ab,c} } 
        + v_{de} \partdif{^2 L }{ q_{de,c} \partial v_{ab} } - 2 v_{ef} \partial_d \left( \partdif{^2 L }{ q_{ef,cd} \partial v_{ab} } \right)
\right. \\ & \left.
        - 2 \Gamma^c_{de} \partdif{}{ v_{ab} } \left( \beta \partdif{ L }{ v_{de} } \right)
        - 2 \partial_d \beta \partdif{^2 L }{ v_{ab} \partial v_{cd} }
        - 4 \beta \partial_d \left( \partdif{^2 L }{ v_{ab} \partial v_{cd} } \right)
\right. \\ & \left.
        - 2 \partdif{ \beta }{ v_{ab} } \partial_d \left( \partdif{ L }{ v_{cd} } \right)
    \right\}
    + \partial_{cd} \theta_{ab} \left\{ 
        \partdif{ L }{ q_{ab,cd} }
        - v_{ef} \partdif{^2 L }{ q_{ef,cd} \partial v_{ab} }
        - 2 \beta \partdif{^2 L }{ v_{ab} \partial v_{cd} }
    \right\}.
\end{split}
    \label{eq:pert_dist-eqn-sol}
\end{equation}
At this point I need to make some assumptions about the form of the action before I can use this equation to restrict its form.


\section{Finding the conditions on the action}
\label{sec:pert_conditions}

Firstly, the variables used for the action and deformation must be determined.  I am considering only the spatial metric field $q_{ab}$ and its normal derivative $v_{ab}$, and for simplicity I am only considering tensor contractions which contain up to second order in derivatives, as previously stated in \secref{sec:methodology_FR}.  The only covariant quantities I can form up to second order in derivatives from the spatial metric are the determinant $q=\det{q_{ab}}$ and the Ricci curvature scalar $R$.  The normal derivative can be split into its trace and traceless components, ${v_{ab}=v^\T_{ab}+\third{}vq_{ab}}$, so it can form scalars from the trace $v$ and a variety of contractions of the traceless tensor $v^\T_{ab}$.  However, to second order I only need to consider ${w:=Q^{abcd}v^\T_{ab}v^\T_{cd}=v^\T_{ab}v_\T^{ab}}$.

Substituting these variables into \eqref{eq:pert_dist-eqn-sol}, the resulting equation contains a series of unique tensor combinations.  The test tensor $\theta_{ab}$ is completely arbitrary so the coefficient of each unique tensor contraction with it must independently vanish if the whole equation is to be satisfied.

Firstly, I focus on the terms depending on the second order derivative $\partial_{cd}\theta_{ab}$. I evaluate each individual term in \appref{sec:pert_extras}.
Substituting \eqref{eq:pert_d2theta_components} into \eqref{eq:pert_dist-eqn-sol}, I find the following independent conditions,
\begin{subequations}
\begin{align}
    q^{ab} \partial^2 \theta_{ab} : 0 & =
    \partdif{ L }{ R } 
    - \frac{2v}{3} \partdif{^2 L }{ R \partial v }
    + 2 \beta \left( \partdif{^2 L }{ v^2 } - \frac{2}{3} \partdif{ L }{ w } \right),
\label{eq:pert_d2theta_1} \\
    Q^{abcd} \partial_{cd} \theta_{ab} : 0 & =
    \partdif{ L }{ R } - 4 \beta \partdif{ L }{ w },
\label{eq:pert_d2theta_2} \\
    q^{ab} v_\T^{cd} \partial_{cd} \theta_{ab} : 0 & =
    \partdif{^2 L }{ R \partial v }
    + 4 \beta \partdif{^2 L }{ w \partial v },
\label{eq:pert_d2theta_3} \\
    v_\T^{ab} \partial^2 \theta_{ab} : 0 & =
    \frac{v}{3} \partdif{^2 L }{ R \partial w } 
    - \beta \partdif{^2 L }{ v \partial w },
\label{eq:pert_d2theta_4} \\
    v_\T^{ab} v_\T^{cd} \partial_{cd} \theta_{ab} : 0 & =
    \partdif{^2 L }{ R \partial w }
    + 4 \beta \partdif{^2 L }{ w^2 }.
\label{eq:pert_d2theta_5}
\end{align}
    \label{eq:pert_d2theta}%
\end{subequations}
Before I analyse these equations, I will find the conditions from the first order derivative part of \eqref{eq:pert_dist-eqn-sol}.  There are many complicated tensor combinations that need to be considered, so for convenience I define 
$X_a:=q^{bc}\partial_{a}q_{bc}$ and $Y_a:=q^{bc}\partial_{c}q_{ab}$.
I evaluate the individual terms in \appref{sec:pert_extras}.
When I substitute the results \eqref{eq:pert_dtheta_components} into \eqref{eq:pert_dist-eqn-sol}, I once again find a series of unique tensor combinations with their own coefficient which vanishes independently.  Most of these conditions are the same as those found in \eqref{eq:pert_d2theta} so I won't bother duplicating them again here.  However, I do find the following new conditions,
\begin{subequations}
\begin{align}
    X^a \partial^b \theta_{ab} : 0 & =
    \partdif{ L }{ R } 
    - 4 \left( \partial_q \beta + 2 \beta \partial_q \right) \partdif{ L }{ w },
\label{eq:pert_dtheta_1} \\
\begin{split}
    q^{ab} X^c \partial_c \theta_{ab} : 0 & =
    \frac{-1}{2} \partdif{ L }{ R }
    + \frac{ v }{ 3 } \left( 4 \partial_q - 1 \right)
    \partdif{^2 L }{ v \partial R }
    + \partdif{ \beta }{ v } \left( 1 - 2 \partial_q \right) \partdif{ L }{ v }
\\ &
    + \left( \beta - 2 \partial_q \beta - 4 \beta \partial_q \right) \left( \partdif{^2 L }{ v^2 } - \frac{2}{3} \partdif{ L }{ w } \right),
\end{split}
\label{eq:pert_dtheta_2} \\
\begin{split}
    v_\T^{ab} X^c \partial_c \theta_{ab} : 0 & =
    \frac{v}{3} \left( 4 \partial_q - 1 \right) \partdif{^2 L }{ w \partial R }
    + \partdif{ \beta }{ w } \left( 1 - 2 \partial_q \right) \partdif{ L }{ v }
\\ &
    + \left( \beta - 2 \partial_q \beta - 4 \beta \partial_q \right) \partdif{^2 L }{ v \partial w },
\end{split}
\label{eq:pert_dtheta_3} \\
    q^{ab} v_\T^{cd} X_d \partial_c \theta_{ab} : 0 & =
    \left( 1 - 2 \partial_q \right) \partdif{^2 L }{ v \partial R }
    - 4 \left( \partial_q \beta + 2 \beta \partial_q \right) \partdif{^2 L }{ v \partial w }
    - 4 \partdif{ \beta }{ v } \partial_q \partdif{ L }{ w },
\label{eq:pert_dtheta_4} \\
    v_\T^{ab} v_\T^{cd} X_d \partial_c \theta_{ab} : 0 & =
    \left( 1 - 2 \partial_q \right) \partdif{^2 L }{ w \partial R } 
    - 4 \left( \partial_q \beta + 2 \beta \partial_q \right) \partdif{^2 L }{ w^2 }
    - 4 \partdif{ \beta }{ w } \partial_q \partdif{ L }{ w },
\label{eq:pert_dtheta_5} \\
    q^{ab} v_\T^{cd} Y_d \partial_c \theta_{ab} : 0 & =
    2 \beta \partdif{^2 L }{ v \partial w }
    + \partdif{ \beta }{ v } \partdif{ L }{ w },
\label{eq:pert_dtheta_6} \\
    v_\T^{ab} v_\T^{cd} Y_d \partial^c \theta_{ab} : 0 & =
    2 \beta \partdif{^2 L }{ w^2 }
    + \partdif{ \beta }{ w } \partdif{ L }{ w },
\label{eq:pert_dtheta_7}
\end{align}%
\begin{align}
    \partial^a F \partial^b \theta_{ab} : 0 & = \left( \partdif{ \beta }{ F } + 2 \beta \partdif{}{ F } \right) \partdif{ L }{ w },
\label{eq:pert_dtheta_8} \\
\begin{split}
    q^{ab} \partial^c F \partial_c \theta_{ab} : 0 & =
    \frac{2v}{3} \partdif{^3 L }{ F \partial v \partial R } 
    - \partdif{ \beta }{ v } \partdif{^2 L }{ F \partial v }
\\ &
    - \left( \partdif{ \beta }{ F } + 2 \beta \partdif{}{F} \right) \left( \partdif{^2 L }{ v^2 } - \frac{2}{3} \partdif{ L }{ w } \right),
\end{split}
\label{eq:pert_dtheta_9} \\
    v_\T^{ab} \partial^c F \partial_c \theta_{ab} : 0 & =
    \frac{2v}{3} \partdif{^3 L }{ F \partial w \partial R } 
    - \partdif{ \beta }{ w } \partdif{^2 L }{ F \partial v }
    - \left( \partdif{ \beta }{ F } + 2 \beta \partdif{}{F} \right) \partdif{^2 L }{ v \partial w },
\label{eq:pert_dtheta_10} \\
    q^{ab} v_\T^{cd} \partial_d F \partial_c \theta_{ab} : 0 & =
    \half \partdif{^3 L }{ F \partial v \partial R } 
    + \partdif{ \beta }{ v } \partdif{^2 L }{ F \partial w }
    + \left( \partdif{ \beta }{ F } + 2 \beta \partdif{}{F} \right) \partdif{^2 L }{ v \partial w },
\label{eq:pert_dtheta_11} \\
    v_\T^{ab} v_\T^{cd} \partial_d F \partial_c \theta_{ab} : 0 & =
    \half \partdif{^3 L }{ F \partial w \partial R } 
    + \partdif{ \beta }{ w } \partdif{^2 L }{ F \partial w }
    + \left( \partdif{ \beta }{ F } + 2 \beta \partdif{}{F} \right) \partdif{^2 L }{ w^2 },
\label{eq:pert_dtheta_12}
\end{align}
    \label{eq:pert_dtheta}%
\end{subequations}
where $F\in\{v,w,R\}$.

By this point, I have accumulated all conditions on the form of the Lagrangian for my choice of variables.  The next step is to try and consolidate them.


\section{Evaluating the fourth order perturbative action}
\label{sec:pert_evaluate}

For this section, I construct an ansatz for the action and deformation that is explicit in being a perturbative expansion. For each time derivative above the classical solution, I include the small parameter $\varepsilon$, and consider up to $\mathcal{O}\left(\varepsilon^2\right)$.  I consider two orders because in models of loop quantum cosmology which have deformed covariance, the holonomy corrections to the action expand into even powers of time derivatives \cite{Mielczarek:2011ph, Cailleteau2013}.  Therefore, considering a fourth order action and a second order deformation should include the nearest higher-order terms in an expansion of those holonomy functions.
Therefore I write,
\begin{subequations}
\begin{align}
\begin{split}
    L & = L_0 + L_{(v)} v + L_{(w)} w + L_{(v^2)} v^2 
    + \varepsilon \left( L_{(vw)} v w + L_{(v^3)} v^3 \right)
\\ &
    + \varepsilon^2 \left( L_{(w^2)} w^2 + L_{(v^2w)} v^2 w + L_{(v^4)} v^4 \right)
    + \mathcal{O}(\varepsilon^3),
\label{eq:pert_ansatz_1_lag}
\end{split}
\\
    \beta & = \beta_0 + \varepsilon \beta_{(v)} v 
    + \varepsilon^2 \left( \beta_{(v^2)} v^2 + \beta_{(w)} w \right)
    + \mathcal{O}(\varepsilon^3),
\label{eq:pert_ansatz_1_def}
\end{align}%
    \label{eq:pert_ansatz_1}%
\end{subequations}
where each coefficient is potentially a function of $q$ and $R$.

I take the condition from $Q^{abcd}\partial_{cd}\theta_{ab}$, \eqref{eq:pert_d2theta_2} and truncate to $\mathcal{O}(\varepsilon^2)$.  Separating different powers of $v$ and $w$, it gives the following conditions for the Lagrangian coefficients,
\begin{subequations}
\begin{gather}
\begin{aligned}
    \varepsilon^2 w^2 & : \, \partial_R L_{(w^2)} = 0,
&
    \varepsilon^2 v^2 w & : \partial_R L_{(v^2w)} = 0,
&
    \varepsilon^2 v^4 & : \partial_R L_{(v^4)} = 0,
\\
    \varepsilon v w & : \partial_R L_{(vw)} = 0,
&
    \varepsilon v^3 & : \partial_R L_{(v^3)} = 0,
        \label{eq:pert_prelim_conditions_A}
\end{aligned}
\\
\begin{aligned}
    w & : \partial_R L_{(w)} = 4 \varepsilon^2 \left( \beta_{(w)} L_{(w)} + 2 \beta_0 L_{(w^2)} \right),
\\
    v^2 & : \partial_R L_{(v^2)} = 4 \varepsilon^2 \left( \beta_{(v^2)} L_{(w)} + \beta_{(v)} L_{(vw)} + \beta_0 L_{(v^2w)} \right),
\\
    v & : \partial_R L_{(v)} = 4 \varepsilon \left( \beta_{(v)} L_{(w)} + \beta_0 L_{(vw)} \right).
        \label{eq:pert_prelim_conditions_B}
\end{aligned}
\end{gather}%
    \label{eq:pert_prelim_conditions}%
\end{subequations}
So from the five conditions in \eqref{eq:pert_prelim_conditions_A}, one can see that terms with three or four time derivatives must not contain any spatial derivatives.  From the three conditions in \eqref{eq:pert_prelim_conditions_B}, one can see that including $R$ in these coefficients requires including a factor of $\varepsilon$ for every combined derivative order above two.  Therefore, the spatial derivatives must be treated equally with time derivatives when one is performing a perturbative expansion, as expected.
So I can now further expand the ansatz to include explicit factors of $R$,
\begin{subequations}
\begin{align}
\begin{split}
    L & = L_\0 + L_{(v)} v 
    + L_{(w)} w + L_{(v^2)} v^2 + L_{(R)} R
    + \varepsilon \left( L_{(vw)} v w + L_{(v^3)} v^3
\right. \\ & \left.
    + L_{(vR)} v R \right)
    + \varepsilon^2 \left( L_{(w^2)} w^2 + L_{(v^2w)} v^2 w + L_{(v^4)} v^4 + L_{(wR)} w R 
\right. \\ & \left.
    + L_{(v^2R)} v^2 R + L_{(R^2)} R^2 \right)
    + \mathcal{O}(\varepsilon^3),
\label{eq:pert_ansatz_2_lag}
\end{split}
\\
    \beta & = \beta_\0 + \varepsilon \beta_{(v)} v 
    + \varepsilon^2 \left( \beta_{(v^2)} v^2 + \beta_{(w)} w + \beta_{(R)} R \right)
    + \mathcal{O}(\varepsilon^3),
\label{eq:pert_ansatz_2_def}
\end{align}%
    \label{eq:pert_ansatz_2}%
\end{subequations}
where each coefficient is potentially a function of $q$.
I now substitute this ansatz into the conditions found for the action so that its form can be progressively restricted.
Looking once again at the condition from $Q^{abcd}\partial_{cd}\theta_{ab}$ \eqref{eq:pert_d2theta_2}, one finds it is satisfied by the following solutions,
\begin{subequations}
\begin{align}
    \0 & : L_{(w)} = \frac{ L_{(R)} }{ 4 \beta_\0 },
\label{eq:pert_conditions_A1} \\
    \varepsilon v & : L_{(vw)} = \frac{ 1 }{ 4 \beta_\0^2 } \Big( \beta_\0 L_{(vR)} - 4 \beta_{(v)} L_{(R)} \Big),
\label{eq:pert_conditions_A2} \\
    \varepsilon^2 R & : L_{(wR)} = \frac{ 1 }{ 4 \beta_\0^2 } \Big( 2 \beta_\0 L_{(R^2)} - 2 \beta_{(R)} L_{(R)} \Big),
\label{eq:pert_conditions_A3} \\
    \varepsilon^2 v^2 & : L_{(v^2w)} = \frac{ 1 }{ 4 \beta_\0^3 } \Big\{ \beta_\0^2 L_{(v^2R)} - \beta_\0 \beta_{(v)} L_{(vR)} + \left( \beta_{(v)}^2 - \beta_\0 \beta_{(v^2)} \right) L_{(R)} \Big\},
\label{eq:pert_conditions_A4} \\
    \varepsilon^2 w & : L_{(w^2)} = \frac{ 1 }{ 32 \beta_\0^3 } \Big\{ 2 \beta_\0 L_{(R^2)} - \left( \beta_{(R)} + 4 \beta_\0 \beta_{(w)} \right) L_{(R)} \Big\},
\label{eq:pert_conditions_A5}%
\end{align}%
    \label{eq:pert_conditions_A}%
\end{subequations}
and then looking at the condition from $v_\T^{ab}\partial^2\theta_{ab}$, \eqref{eq:pert_d2theta_4},
\begin{subequations}
\begin{align}
    \varepsilon & : L_{(vR)} = \frac{ \beta_{(v)} L_{(R)} }{ \beta_\0 },
\label{eq:pert_conditions_B1} \\
    \varepsilon^2 v & : L_{(v^2R)} = \frac{ 1 }{ 6 \beta_\0^2 } \Big\{ 2 \beta_\0 L_{(R^2)} + \left( 6 \beta_\0 \beta_{(v^2)} - \beta_{(R)} \right) L_{(R)} \Big\},
\label{eq:pert_conditions_B2}
\end{align}%
    \label{eq:pert_conditions_B}%
\end{subequations}
where \eqref{eq:pert_conditions_B1} and \eqref{eq:pert_conditions_A2} combine to give $L_{(vw)}=0$.
Then looking at the condition from $q^{ab}v_\T^{cd}\partial_{cd}\theta_{ab}$, \eqref{eq:pert_d2theta_3}
\begin{subequations}
\begin{align}
    \varepsilon & : \beta_{(v)} = 0,
\label{eq:pert_conditions_C1} \\
    \varepsilon^2 v & : L_{(R^2)} = \frac{ L_{(R)} }{ 2 \beta_\0 } \left( \beta_{(R)} - 3 \beta_\0 \beta_{(v^2)} \right),
\label{eq:pert_conditions_C2}
\end{align}
    \label{eq:pert_conditions_C}%
\end{subequations}
one can see that $L_{(vR)}=0$ and therefore all the third order terms all vanish.
Looking at the condition from $q^{ab}\partial^2\theta_{ab}$, \eqref{eq:pert_d2theta_1},
\begin{subequations}
\begin{align}
    \0 & : L_{(v^2)} = \frac{ - L_{(R)} }{ 6 \beta_\0 },
\label{eq:pert_conditions_D1} \\
    \varepsilon v & : L_{(v^3)} = 0,
\label{eq:pert_conditions_D2} \\
    \varepsilon^2 w & : \beta_{(v^2)} = \frac{-2}{3} \beta_{(w)},
\label{eq:pert_conditions_D3} \\
    \varepsilon^2 v^2 & : L_{(v^4)} = \frac{ - \beta_{(w)} L_{(R)} }{ 36 \beta_\0^2 },
\label{eq:pert_conditions_D4}
\end{align}
    \label{eq:pert_conditions_D}%
\end{subequations}
and then from $X^a\partial^b\theta_{ab}$, \eqref{eq:pert_dtheta_1},
\begin{subequations}
\begin{align}
    \0 & : L_{(R)} = f \sqrt{q\left|\beta_\0\right|},
\label{eq:pert_conditions_E1} \\
    \varepsilon^2 R & : \beta_{(w)} = b - \frac{ \beta_{(R)} }{ 4 \beta_\0 },
\label{eq:pert_conditions_E2}
\end{align}
    \label{eq:pert_conditions_E}%
\end{subequations}
where $f$ and $b$ arise as integration constants.  From $q^{ab}X^c\partial_c\theta_{ab}$, \eqref{eq:pert_dtheta_2},
\begin{equation}
    \varepsilon v : L_{(v)} = \xi \sqrt{q},
    \label{eq:pert_conditions_F}
\end{equation}
where $\xi$ is also an integration constant.
Finally, the condition from $\partial^aR\partial^b\theta_{ab}$, \eqref{eq:pert_dtheta_8}, means that
\begin{equation}
    \varepsilon : b = 0.
    \label{eq:pert_conditions_G}
\end{equation}
From this point on the remaining equations don't provide any new conditions on the Lagrangian coefficients.

To make sure the classical limit of the result matches the action found in \chapref{sec:2ndst}, I set $f=\omega/2$, and replace the normal derivatives with the standard extrinsic curvature contraction 
$\displaystyle{\mathcal{K}=\frac{v^2}{6}-\frac{w}{4}}$.
Therefore, the fourth order perturbative gravitational action is given by,
\begin{equation}
    L = L_\0 + \xi v \sqrt{q}
    + \frac{\omega}{2} \sqrt{ q \left| \beta_\0 \right| } \left\{ 
        R - \frac{ \mathcal{K} }{ \beta_\0 }
        - \frac{ \varepsilon^2 \beta_{(R)} }{ 4 \beta_\0 } \left( R + \frac{ \mathcal{K} }{ \beta_\0 } \right)^2
    \right\}
    + \mathcal{O} \left( \varepsilon^3 \right),
        \label{eq:pert_action_sol}
\end{equation}
with the associated deformation
\begin{equation}
    \beta = \beta_\0 
    + \varepsilon^2 \beta_{(R)} \left( 
    R + \frac{ \mathcal{K} }{ \beta_\0 } \right) 
    + \mathcal{O} \left( \varepsilon^3 \right).
        \label{eq:pert_def_sol}
\end{equation}
So the remaining freedom in the action comes down to the constants $\xi$ and $\omega$, the functions $\beta_\0$ and $\beta_{(R)}$.  There is also a term which doesn't affect the kinematic structure and acts like a generalised notion of a potential, so can be rewritten as $L_\0(q)=-\sqrt{q}\,U(q)$.


\section{Cosmology}
\label{sec:pert_cosmo}

In this section I find the cosmological implications of the nearest order corrections coming from the deformation to general covariance.  Since it is a perturbative expansion, the results when the corrections become large should be taken to be indicative rather than predictive.

I restrict to a flat FLRW metric as in \secref{sec:methodology_cosmo},
\begin{equation}
    L = -a^3 U(a) - \frac{3 \sigma_\0 \omega a^3}{ N^2 \sqrt{ \left| \beta_\0 \right| }} \mathcal{H}^2 
    \left( 1 + \frac{ 3 \varepsilon^2 \beta_2 }{ 2 N^2 \beta_\0 } \mathcal{H}^2 \right)
    + \mathcal{O} \left( \varepsilon^3 \right),
	\label{eq:pert_cosmo_lagrangian}
\end{equation}
where $a$ is the scale factor, $\mathcal{H} = \dot{a}/a$ is the Hubble expansion rate, $\sigma_\0:=\sgn{\beta_\0}$, and $\beta_2=\beta_{(R)}/\beta_\0$ is the coefficient of $\mathcal{K}$ in the deformation.

I couple this to matter with energy density $\rho$ and pressure density $P = w_\rho \rho$.  I Legendre transform the effective Lagrangian to find the Hamiltonian.  Imposing the Hamiltonian constraint $C \approx 0$ gives us 
\begin{equation}
    \frac{1}{N^2} \mathcal{H}^2 \left( 1 + \frac{ 9 \varepsilon^2 \beta_2 }{ 2 \beta_\0 N^2 } \mathcal{H}^2 \right) = \frac{ \sigma_\0 }{ 3 \omega } \sqrt{ \left| \beta_\0 \right| } \, U,
\end{equation}
which can be solved to find the modified Friedmann equation,
\begin{equation}
    \frac{1}{N^2} \mathcal{H}^2 = \frac{ 2 \sigma_\0 \sqrt{ \left| \beta_\0 \right| } }{ 3 \omega ( 1 + \alpha ) } U,
	\label{eq:pert_cosmo_mod_friedmann}
\end{equation}
where the correction factor is
\begin{equation} 
    \alpha := \sqrt{ 1 + \frac{ 6 \varepsilon^2 \beta_2 }{ \omega \sqrt{ \left| \beta_\0 \right| } } \, U}.
	\label{eq:pert_cosmo_correction_factor}
\end{equation}
Going back to the effective Lagrangian, and varying it with respect to the scale factor, I find the Euler-Lagrange equation of motion.  When I substitute in Eq.~\eqref{eq:pert_cosmo_mod_friedmann}, I get the acceleration
equation
\begin{equation}
\begin{split}
    \frac{\ddot{a}}{aN^2} = \frac{ \sigma_\0 \sqrt{ \left| \beta_\0 \right| } }{ 6 \alpha } \, U & \left\{ 2 + \partdif{\, \log U }{\, \log a } + 2 \partdif{\,\log N }{\,\log a} + \half \partdif{ \, \log \beta_\0 }{ \, \log a }
\right. \\ & \left.
    + 2 \left( \frac{ \alpha - 1 } { \alpha + 1 } \right) \left[ 1 + \partdif{ \, \log N }{ \, \log a} - \half \partdif{}{\,\log a}\log \left( \frac{ \beta_2 }{ \beta_\0 } \right) \right] \right\}.
	\label{eq:pert_cosmo_mod_acceleration}
\end{split}
\end{equation}
If I take a perfect fluid, then $U=\rho$, where $\rho$ is the fluid's energy density, which satisfies the continuity equation
\begin{equation}
	\dot{\rho} + 3 \mathcal{H} \rho (1+w_\rho) = 0.
	\label{eq:pert_cosmo_continuity}
\end{equation}
where $w_\rho$ is the perfect fluid's equation of state.
Note that there are corrections to the matter sector due to the modified constraint algebra \cite{bojowald_radiation_2007, bojowald_quantum_2013}, as shown for scalar fields in other chapters.
However, these have not been included here, as it is not known how the deformation would affect a perfect fluid.

Since $\varepsilon$ is a small parameter, it can be used to expand Eq.~\eqref{eq:pert_cosmo_mod_friedmann},
\begin{equation}
    \frac{1}{N^2}\mathcal{H}^2 = \frac{ \sigma_\0 \sqrt{ \left| \beta_\0 \right| } }{ 3 \omega } \rho \left( 1 + \frac{ 3 \varepsilon^2 \beta_2 }{ \omega \sqrt{ \left| \beta_\0 \right| } } \rho \right) + \mathcal{O} \left( \varepsilon^3 \right),
	\label{eq:pert_cosmo_mod_friedmann_bounce}
\end{equation}
and expanding the bracket in Eq.~\eqref{eq:pert_cosmo_mod_acceleration} to first order, it can be seen that $\ddot{a}/a>0$ when $w_\rho<w_a$, where
\begin{equation}
    w_a = \frac{-1}{3} \left\{ 1 - \half \partdif{\,\log \beta_\0 }{\,\log a} 
    + \frac{ 6 \varepsilon^2 \beta_2 }{ \omega \sqrt{ \left| \beta_\0 \right| } } \rho
    \left[ 1 - \half \partdif{}{\,\log a}\log \left( \frac{ \beta_2 }{ \beta_\0 } \right) \right] \right\},
	    \label{eq:pert_cosmo_eqn_state_acceleration}
\end{equation}
having set $N=1$, so this is applicable for cosmic time.

When $\beta_2<0$, the modified Friedmann equation \eqref{eq:pert_cosmo_mod_friedmann_bounce} suggests a big bounce rather than a big bang at high energy density, since $\dot{a} \to 0$ when $a>0$ and $\ddot{a}>0$ is possible when $\rho \to \rho_c$  where
\begin{equation}
    \rho_c = \frac{ \omega \sqrt{ \left| \beta_\0 \right| } }{ 6 \varepsilon^2 \left| \beta_2 \right| }.
\end{equation}
This requires either $\rho_c$ to be constant, or for it to diverge at a slower rate than $\rho$ as $a \to 0$.

Let me emphasise that the bounce is found considering only holonomy corrections manifesting as higher-order powers of of second-order derivatives and not considering ignoring higher-order derivatives. The equations \eqref{eq:pert_cosmo_mod_friedmann_bounce} and \eqref{eq:pert_cosmo_eqn_state_acceleration} have been expanded to leading order in $\beta_2$, so I should be cautious about the regime of their validity.  Note that the Lagrangian is also an expansion; $\beta_2$ is a coefficient of the fourth order term and appears only linearly, I conclude that there is no good reason why I should have more trust in equations such as \eqref{eq:pert_cosmo_mod_friedmann} or \eqref{eq:pert_cosmo_mod_acceleration} simply because they contain higher orders.  In Ref.~\cite{Ashtekar2006}, Ashtekar,
Pawlowski and Singh write their effective Friedmann equation with leading order corrections (which is the same as
\eqref{eq:pert_cosmo_mod_friedmann_bounce}) and say that it holds surprisingly well even for $\rho \approx \rho_c$, the regime when the perturbative expansion should break down (I should note that their work refers only to the case where $w_\rho = 1$, and does not say whether this is true generally).


\subsection{Linking the \texorpdfstring{$\beta$}{deformation} function to LQC}
\label{subsec:cosmo_beta}

I need to know $\beta_\0(a)$ and $\beta_2(a)$ in order to make progress beyond this point, so I compare my results to those found in previous investigations.  In Ref.~\cite{Cailleteau2013}, Cailleteau, Linsefors and Barrau have found information about the correction function when inverse-volume and holonomy effects are both included in a perturbed FLRW system.  Their equation (Eq.~$(5.18)$ in Ref.~\cite{Cailleteau2013}) gives (rewritten slightly)
\begin{equation}
    \beta(a,\dot{a}) = f(a) \Sigma(a,\dot{a}) \partdif{^2}{\dot{a}^2} \left\{ \gamma_\0(a,\dot{a}) \left( \frac{\sin[\BI \mu(a) \dot{a}]}{\BI \mu(a)} \right)^2 \right\},
        \label{eq:pert_cosmo_beta_barrau_general}
\end{equation}
where $\BI$ is the Barbero-Immirzi parameter, $\gamma_\0$ is the function which contains information about inverse-volume corrections, $\Sigma(a,\dot{a})$ depends on the form of $\gamma_\0$, and $f(a)$ is left unspecified.  I just consider the case where $\gamma_\0=\gamma_\0(a)$, in which case $\Sigma = 1/\left(2\sqrt{\gamma_\0}\right)$ and $\mu = a^{\delta-1}\sqrt{\gamma_\0 \area}$. The constant $\area$ is usually interpreted as being the ``area gap'' derived in loop quantum gravity.  I leave $\delta$ unspecified for now, because different quantisations of loop quantum cosmology give it
equal to different values in the range $[0,1]$.
Equation~\eqref{eq:pert_cosmo_beta_barrau_general} now becomes
\begin{equation}
	\beta = f \sqrt{\gamma_\0}
		\cos \left(
			2 \BI \sqrt{\gamma_\0 \area} a^{\delta} \mathcal{H}
		\right),
	\label{eq:pert_cosmo_beta_barrau_specific}
\end{equation}
The ``old dynamics'' or ``$\mu_0$ scheme'' corresponds to $\delta = 1$, and the favoured ``improved dynamics'' or ``$\bar{\mu}$ scheme'' corresponds to $\delta = 0$ \cite{sakellariadou_lattice_2007-1,   sakellariadou_lattice_2007}.  In the semi-classical regime, $\mathcal{H} \sqrt{\area} \ll 1$, so I can Taylor expand this equation for the correction function to get
\begin{equation}
	\beta = f \sqrt{\gamma_\0}
	 - 2 \BI^2 \area a^{2\delta} f (\gamma_\0)^{3/2} \mathcal{H}^2
	 + \mathcal{O} \left( \area^2 \right).
	 \label{eq:pert_cosmo_beta_barrau_expand}
\end{equation}
The way that $\gamma_\0$ is defined is that it multiplies the background gravitational term in the Hamiltonian constraint relative to the classical form.  Since I am assuming $\gamma_\0 = \gamma_\0(a)$, I can isolate it by taking the Lagrangian \eqref{eq:pert_cosmo_lagrangian} and setting
$\beta_2=0$.  If I then Legendre transform to find a Hamiltonian expressed in terms of the momentum of the scale factor, I find that it is proportional to $\sqrt{\left|\beta_\0\right|}$. Thus, I conclude that $\gamma_\0=\sqrt{\left|\beta_\0\right|}$ when $\gamma_\0$ is just a function of the scale factor.  Using this to compare \eqref{eq:pert_cosmo_beta_barrau_expand} to \eqref{eq:pert_def_sol},
\begin{equation}
	\beta = \beta_\0
	    + 6 \varepsilon^2 \beta_2 \mathcal{H}^2
		+ \mathcal{O} \left( \varepsilon^3 \right),
	    \label{eq:pert_cosmo_beta_background}
\end{equation}
I find that $f=\sigma_\0\left|\beta_\0\right|^{3/4}$, and therefore $f = \sigma_\0 \gamma_\0^{3/2}$.
From this, I can now deduce the form of the coefficient for the higher-order corrections,
\begin{equation}
	\varepsilon^2 \beta_2 = \frac{-\sigma_\0}{3} \BI^2 \area a^{2\delta} \gamma_\0^3.
\end{equation}
The exact form of $\gamma_\0(a)$ is uncertain, and the possible forms that have been found also contain quantisation ambiguities.  The form given by Bojowald in Ref.~\cite{bojowald_loop_2004} is
\begin{equation}
	\gamma_\0 = \frac{3r^{1-l}}{2l} \left\{
		\frac{(r+1)^{l+2}-|r-1|^{l+2}}{l+2} -r
		\frac{(r+1)^{l+1}-\sgn{r-1}|r-1|^{l+1}}{l+1} \right\},
\end{equation} 
where $l\in (0,1)$, $r = a^2 / a_{\star}^2$ and $a_{\star}$ is the characteristic scale of the inverse-volume corrections, related to the discreteness scale.  I will only use the asymptotic expansions of this function, namely 
\begin{equation}
	\gamma_\0 \approx \left\{
\begin{aligned}
	& 1 + \frac{(2-l)(1-l)}{10} \left( \frac{a}{a_{\star}} \right)^{-4},
    & {\rm if } \; a \gg a_{\star}
        \\
    & \frac{3}{1+l} \left( \frac{a}{a_{\star}} \right)^{2(2-l)},
    & {\rm if } \; a
        \ll a_{\star}
\end{aligned}
    \right.
	    \label{eq:pert_cosmo_inv_vol_cases}
\end{equation}
and even then I will only take $\gamma_\0 \approx 1$ for $a \gg a_{\star}$, since the correction quickly becomes vanishingly small.
I replace the area gap with a dimensionless parameter $\tilde{\area}=\area\omega$ which is of order unity.  The modified Friedmann equation \eqref{eq:pert_cosmo_mod_friedmann_bounce} is now given by
\begin{equation}
	\mathcal{H}^2 =
	\frac{ \sigma_\0 \gamma_\0 }{ 3 \omega } \rho \left(
		1 - \frac{ \sigma_\0 \BI^2 \tilde{\area} }{3\omega^2} a^{2\delta}
		\gamma_\0^2 \rho \right),
	\label{eq:pert_cosmo_mod_friedmann_planck}
\end{equation}
which I need to compare for different types of matter.  First of all I will consider a perfect fluid, and then I will consider a scalar field with a power-law potential.


\subsection{Perfect fluid}
\label{subsec:cosmo_fluid}

I consider the simple case of a perfect fluid.  Solving the continuity equation \eqref{eq:methodology_cosmo_continuity} gives us the energy density as a function of the scale factor,
\begin{equation}
	\rho(a) = \rho_0 a^{-3(1+w_\rho)}.
\end{equation}
To investigate whether there can be a big bounce, I insert this into Eq.~\eqref{eq:pert_cosmo_mod_friedmann_planck}, which becomes of the form
\begin{equation}
	\mathcal{H}^2 \propto a^{-3(1+w_\rho)} \left(
		1 - \frac{ \BI^2 \tilde{\area} }{ 3 \omega^2 } \rho_0 
		a^{\Theta} \right),
	\label{eq:pert_cosmo_mod_friedmann_fluid}
\end{equation}
where $\Theta$ depends on which regime of \eqref{eq:pert_cosmo_inv_vol_cases} we are in, namely
\begin{equation}
	\Theta = \left\{
\begin{aligned}
	& 2 \delta - 3(1+w_\rho),
		& {\rm if } \; a \gg a_{\star},
		    \\
	& 2 \delta + 4(2-l) - 3(1+w_\rho),
		& {\rm if } \; a \ll a_{\star},
\end{aligned}
    \right.
	    \label{eq:pert_cosmo_theta_bounce_cases}
\end{equation}
and I simply ignored the constant coefficients for $a \ll a_{\star}$.
Whether a bounce happens depends on whether $\mathcal{H} \to 0$ when $a>0$, which would happen if the higher-order correction in the modified Friedmann equation became dominant for small values of $a$, i.e. if $\Theta <0$, which is also required to match the classical limit.  The reason this is required is because $\rho$ needs to diverge faster than $\rho_c$ as $a \to 0$ in order for there to be a bounce.  This will happen when $w_\rho > w_b$, where
\begin{equation}
    w_b = \left\{
\begin{aligned}
	& -1 + \frac{2}{3} \delta,
		& {\rm if } \; a \gg a_{\star} \\
	& -1 + \frac{2}{3} \delta + \frac{4}{3}(2-l),
		& {\rm if } \; a \ll a_{\star}
\end{aligned}
    \right.
	    \label{eq:pert_cosmo_wb_bounce_cases}
\end{equation}
which means that, if the bounce does not happen in the $a \gg a_{\star}$ regime, the inverse-volume corrections make the bounce \emph{less} likely to happen.  If I use the favoured value of $\delta = 0$, and assume $l = 1$, then $w_b = 1/3$ and so $w_\rho$ still needs to be greater than that found for radiation in order for there to be a
bounce.  A possible candidate for this would be a massless (or kinetic-dominated) scalar field, where $w_\rho =1$.

Another aspect to investigate is whether the conditions for inflation are modified.  Taking \eqref{eq:pert_cosmo_eqn_state_acceleration}, I see that acceleration happens when $w_\rho<w_a$, where
\begin{equation}
	w_a = \left\{
\begin{aligned}
	& - \frac{1}{3} + \frac{ 2 \BI^2 \tilde{\area} }{ 9 \omega^2 } ( 1 - \delta ) \rho_0 a^{\Theta},
    & {\rm if} \; a \gg a_{\star}
        \\
    & 1 - \frac{2l}{3} - \frac{ 2 \BI^2 \tilde{\area} }{ \omega^2 a_{\star}^{4(2-l)} } \frac{ 1 + \delta - l } { ( 1 + l )^2 } \rho_0 a^{\Theta},
    & {\rm if } \; a \ll a_{\star}
\end{aligned}
    \right.
	    \label{eq:pert_cosmo_wa_bounce_cases}
\end{equation}
so the range of values of $w_\rho$ which can cause accelerated expansion is indeed modified.  Holonomy-type corrections increase the range since $\Theta\leq0$, and so may inverse-volume corrections. However, the latter also seems to include a cut-off when the last term of Eq.~\eqref{eq:pert_cosmo_wa_bounce_cases} in the $a \ll a_{\star}$ regime dominates. Since a bounce requires $\dot{a}=0$ and $\ddot{a}>0$, the condition $w_b<w_\rho<w_a$ must be satisfied and so it must happen before the cut-off dominates if it is to happen at all.


\subsection{Scalar field}
\label{subsec:cosmo_scalar}

I now investigate the effects that the inverse-volume and holonomy corrections can have when I couple gravity to an undeformed scalar field.  In this case, the energy and pressure densities are given by
\begin{equation}
	\rho = \half \dot{\varphi}^2 + U(\varphi),
	\qquad
	P = \half \dot{\varphi}^2 - U(\varphi),
	\label{eq:pert_cosmo_scalar_eos}
\end{equation}
and the continuity equation gives us the equation of motion for the scalar field,
\begin{equation}
	\ddot{\varphi} + 3 \mathcal{H} \dot{\varphi} + U' = 0,
	\label{eq:pert_cosmo_scalar_eom}
\end{equation}
where 
$\displaystyle{ U' = \partdif{U}{\varphi} }$.

Let us investigate the era of slow-roll inflation.  Using the assumptions $|\ddot{\varphi}/U'| \ll 1$ and $\half \dot{\varphi}^2 \ll U$, I have the slow-roll equations,
\begin{subequations}
\begin{align}
	& \displaystyle{\dot{\varphi} = \frac{-U'}{3 \mathcal{H}}},
	\label{eq:pert_cosmo_slowroll_phidot}\\
	& \displaystyle{\mathcal{H}^2 = \frac{ \sigma_\0
            \gamma_\0 }{ 3 \omega } U \left( 1 - \frac{ \sigma_\0 \BI^2 \tilde{\area} }{ 3 \omega^2 }
            a^{ 2 \delta }
          \gamma_\0^2 U \right)}.
	\label{eq:pert_cosmo_slowroll_hubble}
\end{align}
    \label{eq:pert_cosmo_slowroll}%
\end{subequations}
If I substitute \eqref{eq:pert_cosmo_slowroll_hubble} into \eqref{eq:pert_cosmo_slowroll_phidot}, take the derivative with respect to time and substitute in \eqref{eq:pert_cosmo_slowroll_hubble} and \eqref{eq:pert_cosmo_slowroll_phidot} again, I find
\begin{equation}
	\frac{\ddot{\varphi}}{U'} = \frac{1}{3} \eta,
	\qquad
	\frac{\dot{\varphi}^2}{2U} = \frac{1}{3} \epsilon,
\end{equation}
where the slow-roll parameters are
\begin{subequations}
\begin{align}
    \eta & :=
    \displaystyle{ \frac{1}{1-\varsigma} \left( \frac{ \omega }{ \gamma_\0 } \frac{U''}{U} -(1-2\varsigma) \epsilon + \chi - \delta \varsigma \right),}
        \label{eq:pert_cosmo_slowroll_eta} \\
    \epsilon & :=
    \displaystyle{ \frac{1}{1-\varsigma} \frac{ \omega }{ 2 \gamma_\0} \left(\frac{U'}{U}\right)^2, } 
        \label{eq:pert_cosmo_slowroll_epsilon} \\
    \chi & :=
    \displaystyle{ \frac{1-3\varsigma}{2} \partdif{\,\log \gamma_\0}{\,\log a} }
        \label{eq:pert_cosmo_slowroll_chi} \\
    \varsigma & := 
    \displaystyle{ \frac{ \BI^2 \tilde{\area} }{ 3 \omega^2 } a^{2\delta} \gamma_\0^2 U,}
        \label{eq:pert_cosmo_slowroll_sigma}%
\end{align}%
\end{subequations}
and the conditions for slow-roll inflation are
\begin{equation}
	|\eta | \ll 1,
        \quad
	\epsilon \ll 1,
        \quad
	|\chi | \ll 1,
        \quad
	|\varsigma | \ll 1.
	\label{eq:pert_cosmo_slow_roll_parameters}
\end{equation}
I would like to investigate how these semi-classical effects affect the number of e-folds of the scale factor during inflation.  The number of e-folds before the end of inflation $\mathcal{N}(\varphi)$ is defined by $a(\varphi) = a_{\rm end} e^{-\mathcal{N}(\varphi)}$, where
\begin{equation}
	\mathcal{N}(\varphi)
	= -\int_{\varphi_{\rm end}}^{\varphi} {\rm d}\varphi
		\frac{\mathcal{H}}{\dot{\varphi}}
	=
		\int_{\varphi_{\rm end}}^{\varphi} {\rm d}\varphi
		\frac{ \gamma_\0 U }{ \omega U'}
		\left( 1 - \frac{ \BI^2 \tilde{\area} }{ 3 \omega^2 }
		a^{2\delta} \gamma_\0^2 U \right).
\end{equation} 
If I remove the explicit dependence on $a$ from the integral by setting $\delta = 0$ and $\gamma_\0 = 1$ (i.e. taking only a certain form of holonomy corrections and ignoring inverse-volume corrections), and choose a power-law potential
\begin{equation}
	U(\varphi) = \frac{\lambda}{n} \varphi^n =
	\frac{\tilde{\lambda}}{n} \varphi^n \omega^{2-\frac{n}{2}},
	\label{eq:pert_cosmo_scalar_potential}
\end{equation}
where $\tilde{\lambda} >0$ and $n/2 \in \mathbb{N}$, then the number of e-folds before the end of inflation is
\begin{equation}
	\mathcal{N}(\varphi) = \frac{ 1 }{ 2 n \omega }
	\left( \varphi^2 - \varphi_{\rm end}^2 \right)
-\frac{ \BI^2 \tilde{\area} \tilde{\lambda} }{ 3 n^2 (n+2) \omega^{1+\frac{n}{2}}}
	\left( \varphi^{2+n} - \varphi_{\rm end}^{2+n} \right).
\end{equation}
If I take the approximation that slow-roll inflation is valid beyond the regime specified by \eqref{eq:pert_cosmo_slow_roll_parameters}, then I can calculate a value for the maximum number of e-folds by starting inflation at the big bounce,
\begin{equation}
\begin{split}
	\mathcal{N}_{\rm max} & = 
	\frac{ 1 }{ 2 n } \left\{
	\left( \frac{ 3 n }{ \BI^2 \tilde{\area} \tilde{\lambda} } \right)^{\frac{2}{n}}
	- \frac{ \varphi_\mathrm{end}^2 }{ \omega } \right\}
\\ &
    - \frac{ \BI^2 \tilde{\area} \tilde{\lambda} }{ 3 n^2 ( n + 2 ) } \left\{
	\left( \frac{ 3 n }{ \BI^2 \tilde{\area} \tilde{\lambda} } \right)^{1+\frac{2}{n}}
	- \left( \frac{ \varphi_\mathrm{end}^2 }{ \omega }\right)^{1+\frac{n}{2}} \right\},
\end{split}
\end{equation}
and if I can assume $\varphi_{\rm end}^2/\omega \ll 1$, then
\begin{equation}
	\mathcal{N}_{\rm max} = \frac{1}{2(n+2)} \left(
	\frac{ 3 n }{ \BI^2 \tilde{\area} \tilde{\lambda}} \right)^{2/n}.
\end{equation}

\begin{figure}[t]
	\begin{center}
	{\subfigure[Full phase space for $U(\varphi) = \lambda \varphi^2 /2$]{
		\label{fig:scalar_phase_2_full}
		\includegraphics[width = 0.40\textwidth]{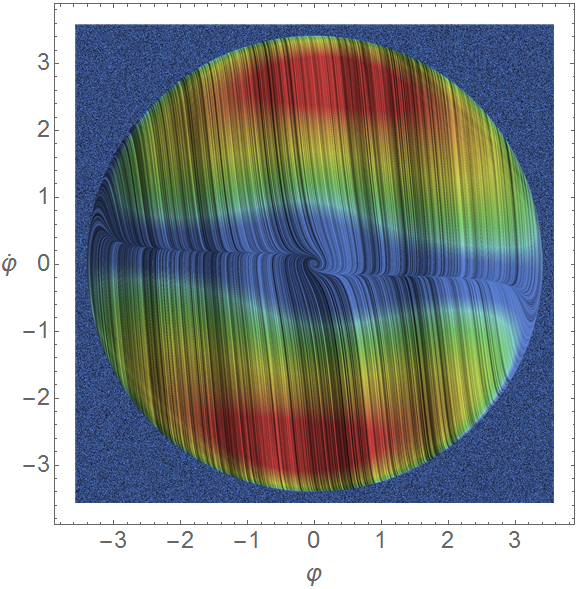}}}
	{\subfigure[Attractor solution for $U(\varphi) = \lambda \varphi^2 /2$]{
		\label{fig:scalar_phase_2_zoom}
		\includegraphics[width = 0.40\textwidth]{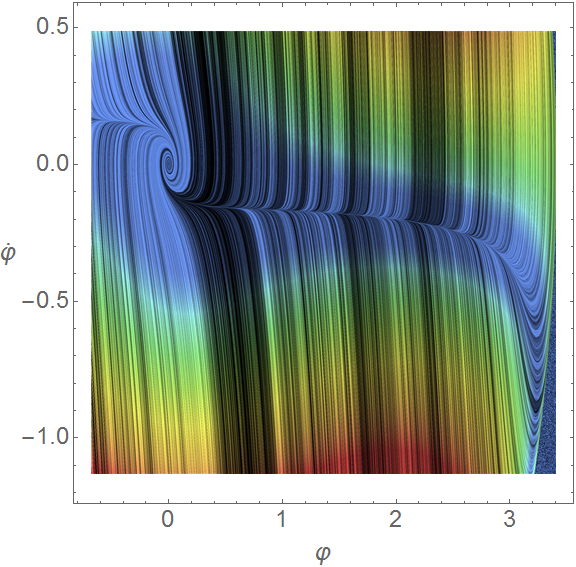}}}
\\
	{\subfigure[Full phase space for $U(\varphi) = \lambda \varphi^4 /4$]{
		\label{fig:scalar_phase_4_full}
		\includegraphics[width = 0.40\textwidth]{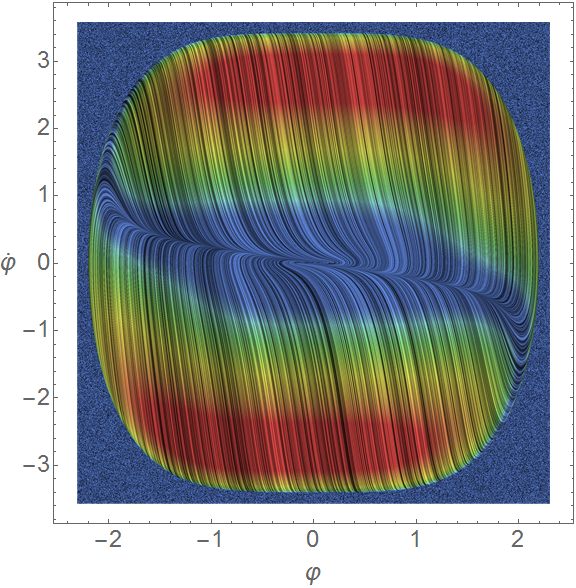}}}
	{\subfigure[Attractor solution for $U(\varphi) = \lambda \varphi^4 /4$]{
		\label{fig:scalar_phase_4_zoom}
		\includegraphics[width = 0.40\textwidth]{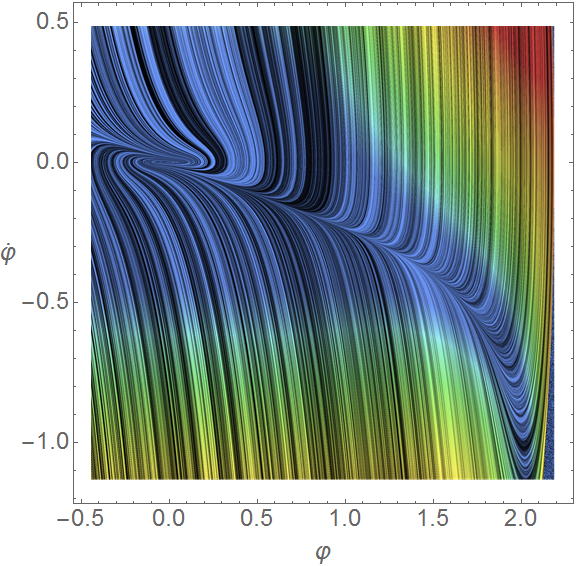}}}
	\end{center}
\caption[Phase space trajectories for undeformed scalar coupled to 4th order deformed gravity]{Line integral convolution plots showing trajectories in phase space for a scalar field with potential $\lambda \varphi^n / n$ with holonomy corrections.  The hue at each point indicates the magnitude of the vector $(\dot{\varphi},\ddot{\varphi})$, with blue indicating low values.  The trajectories do not extend outside of the region \eqref{eq:pert_cosmo_scalar_region}.  The attractor solution (the trajectory approached by a wide range of inital conditions) is well approximated by \eqref{eq:pert_cosmo_scalar_attractor}, corresponding to slow-roll inflation. I use $\tilde{\lambda}=(8\pi_{\circ})^{(4-n)/2}$, $\tilde{\area}=\sqrt{3}\BI/4$, $\delta = 0$, $\gamma_\0 = 1$. Plots are in Planck units, $\omega=1/8\pi_{\circ}$}
	\label{fig:scalar_phase}
\end{figure}

Let us now find the attractor solutions for slow-roll inflation. Substituting the Hubble parameter \eqref{eq:pert_cosmo_mod_friedmann_planck} into the equation of motion for the scalar field \eqref{eq:pert_cosmo_scalar_eom}, I obtain
\begin{equation}
	\ddot{\varphi} 
	+ \dot{\varphi} \sqrt{
	    \frac{ 3 \gamma_\0}{ \omega } \left(
	        \half \dot{\varphi}^2 + U \right) 
	    \left\{ 1 - \frac{ \BI^2 \tilde{\area} }{ 3 \omega^2 } a^{2 \delta} \gamma_\0^2 \left( \half \dot{\varphi}^2 + U \right) \right\}
	} 
    + U' = 0.
\end{equation}
I can remove the explicit scale-factor dependence of the equation by setting $\delta = 0$ and $\gamma_\0 = 1$ (the same assumptions as I used to find $\mathcal{N}$).  Then substituting in the
power-law potential \eqref{eq:pert_cosmo_scalar_potential} I get
\begin{equation}
    \ddot{\varphi} + \dot{\varphi} \sqrt{ \frac{ 3 }{ \omega }
  \left( \half \dot{\varphi}^2 + \frac{\lambda}{n} \varphi^n \right)
  \left\{ 1 - \frac{ \BI^2 \tilde{\area} }{ 3 \omega^2 }
    \left( \half \dot{\varphi}^2 + \frac{\lambda}{n} \varphi^n \right)
    \right\} } + \lambda \varphi^{n-1} = 0,
\end{equation}
which is applicable only for the region where $\rho$ is below a critical value,
\begin{equation}
	1 - \frac{ \BI^2 \tilde{\area} }{ 3 \omega^2 } \left( \half
\dot{\varphi}^2 + \frac{\lambda}{n} \varphi^n \right) >0,
	\label{eq:pert_cosmo_scalar_region}
\end{equation}
otherwise $\mathcal{H}$ and $\dot{\varphi}$ are complex.  I use this equation to plot phase space trajectories in \figref{fig:scalar_phase}.

I can find the slow-roll attractor solution for $|\ddot{\varphi} \varphi^{1-n} / \lambda| \ll 1$ and $\half \dot{\varphi}^2 \ll \frac{\lambda}{n} \varphi^n$,
\begin{equation}
	\dot{\varphi} \approx - \sqrt{ \frac{ n \lambda \omega }{ 3 } } \varphi^{\frac{n}{2}-1} \left( 1 - \frac{ \BI^2
  \tilde{\area} \lambda }{ 3 n \omega^2 } \varphi^n \right)^{-1/2},
	\label{eq:pert_cosmo_scalar_attractor}
\end{equation}
where the term in the bracket is the correction to the classical solution.  Looking at \figref{fig:scalar_phase_2_zoom} and \ref{fig:scalar_phase_4_zoom}, I conclude that the attractor
solutions diverge from a linear relationship as they approach the boundary.

\begin{figure}[t]
	\begin{center}
	{\subfigure[Accelerating values of $w_\rho$ for $U(\varphi) =
            \lambda \varphi^2 /2$]{
		\label{fig:scalar_eos_2}
		\includegraphics[width = 0.45\textwidth]{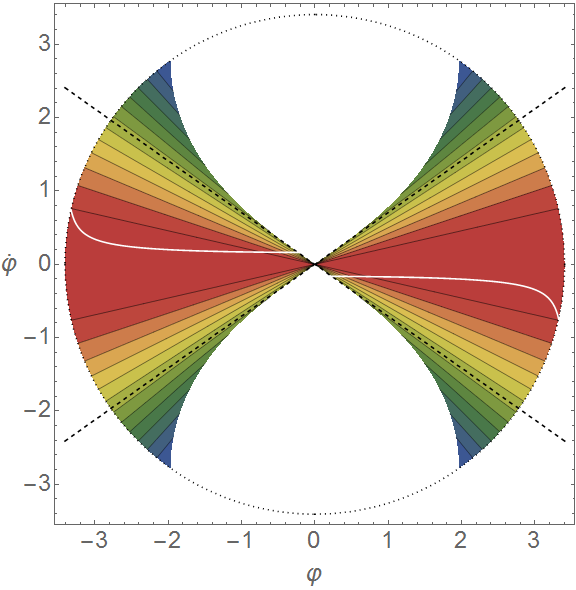}}}
        {\subfigure[Accelerating values of $w_\rho$ for $U(\varphi) =
            \lambda \varphi^4 /4$]{
		\label{fig:scalar_eos4}
		\includegraphics[width = 0.45\textwidth]{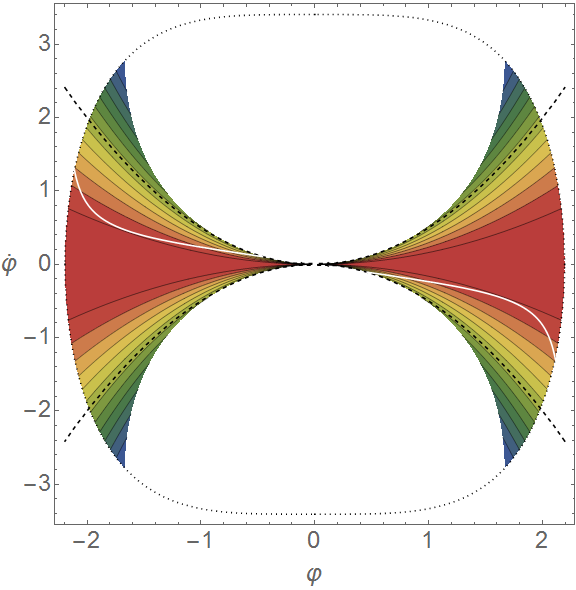}}}
	\end{center}
\caption[Accelerating regions of phase space for undeformed scalar coupled to 4th order deformed gravity]{Contour plots showing the region in scalar phase space satisfying the condition for accelerated expansion when holonomy corrections are included \eqref{eq:pert_cosmo_scalar_acceleration}.  The dashed line indicates the classical acceleration condition $w_a = -1/3$ and the dotted line indicates the bounce boundary.  The white line indicates the slow-roll solution \eqref{eq:pert_cosmo_scalar_attractor}. The contours indicate the value of $w_\rho$ by their colour, and the most blue contour is for $w_\rho \approx 0.2$. I use $\tilde{\lambda}=(8\pi_{\circ})^{(4-n)/2}$, $\tilde{\area}=\sqrt{3}\BI/4$, $\delta = 0$, $\gamma_\0 = 1$. Plots are in Planck units, $\omega=1/8\pi_{\circ}$.}
	\label{fig:scalar_eos}
\end{figure}

The condition for acceleration for the case I am considering here is
\begin{equation}
	w_\rho < w_{a} = \frac{-1}{3} \left\{
		1 - \frac{ 2 \BI^2 \tilde{\area} }{ 3 \omega^2 } \left(
			\half \dot{\varphi}^2 + \frac{\lambda}{n} \varphi^n
		\right)
	\right\} 
	\label{eq:pert_cosmo_scalar_acceleration}
\end{equation}
where we can define the effective equation of state as $w_\rho=P(\varphi)/\rho(\varphi)$ using \eqref{eq:pert_cosmo_scalar_eos}.
I plot in \figref{fig:scalar_eos} this region on the phase space of the scalar field to see how accelerated expansion can happen in a wider range than in the classical case.  In order to be able to solve the equations and make plots, I have neglected non-zero values of $\delta$ and non-unity values of $\gamma_\0$. It may be that in these cases the big bounce and inflation are no longer inevitable, as was found for the perfect fluid.

\section{Discussion}
\label{sec:pert_discussion}

In this chapter, I calculated the general conditions on a deformed action which has been formed from the variables $\left(q,v,w,R\right)$.  I then found the nearest-order curvature corrections coming from the deformation by solving these conditions for a fourth order action.
I found that these corrections can act as a repulsive gravitational effect which may produce a big bounce.  

When coupling gravity to a perfect fluid, the effects that the quantum corrections have depend on the equation of state, but inflation and a big bounce are possible.
I coupled deformed gravity to an undeformed scalar in this preliminary investigation into higher order curvature corrections.  I investigated slow-roll inflation and a big bounce in the presence of this scalar field.  In \chapref{sec:allst}, I find that scalar fields must be deformed in much the same way as the metric.  Therefore, these results might be interesting on some level, but cannot be taken too literally.  Unfortunately, there was simply not enough time to research the fully deformed cases, hence why this material remains.

\chapter{Deformed scalar-tensor constraint to all orders}
\label{sec:allst}

In this chapter I find the most general gravitational constraint which satisfies the deformed constraint algebra.  To find the constraint is easier than finding the action, so I also include a non-minimally coupled scalar field in order to find the most general deformed scalar-tensor constraint.  This material has not been previously published.

As stated in \chapref{sec:methodology}, I am not looking for models with degrees of freedom beyond a simple scalar-tensor model.  Since actions which contain Riemann tensor squared contractions introduce additional tensor degrees of freedom \cite{Deruelle2010}, I automatically do not consider such terms here.  This means I only need to expand the constraint using variables which are tensor contractions containing up to two orders of spatial derivatives or up to two in momenta.  It also means I do not need to consider spatial derivatives of momenta in the constraint.
Therefore, for a metric tensor field ${\left(q_{ab},p^{cd}\right)}$ and a scalar field ${\left(\psi,\pi\right)}$, I expand the constraint into the following variables,
\begin{equation}
\begin{gathered}
    q = \det{q_{ab}},
        \quad
    p = q_{ab} p^{ab},
        \quad
    \bp = Q_{abcd} p_\T^{ab} p_\T^{cd},
        \quad
    R,
        \\
    \psi,
        \quad
    \pi,
        \quad
    \Delta := q^{ab} \nabla_a \nabla_b \psi 
    = \partial^2 \psi - q^{ab} \Gamma^c_{ab} \partial_c \psi,
        \quad
    \gamma := q^{ab} \nabla_a \psi \nabla_b \psi = \partial^a \psi \partial_a \psi,
\end{gathered}
        \label{eq:allst_variables}
\end{equation}
where 
${p_\T^{ab}:=p^{ab}-\third{}pq^{ab}}$
is the traceless part of the metric momentum.
Therefore, I start with the constraint given by ${C=C(q,p,\bp,R,\psi,\pi,\Delta,\gamma)}$. 
I must solve the distribution equation again to find the equations which restrict the form of the constraint.  The calculations in this chapter generalise those presented in \chapref{sec:2ndst} where the minimally deformed scalar-tensor constraint was regained from the constraint algebra.

\section{Solving the distribution equation}
\label{sec:dist-eqn-sol}

Starting from \eqref{eq:dist-eqn_con}, I have the general distribution equation for a Hamiltonian constraint, without derivatives of the momenta, which depends on a metric tensor and a scalar field,
\begin{equation}
    0 = 
    \funcdif{ C (x) }{ q_{ab} (y) } 
    \left. \partdif{ C }{ p^{ab} } \right|_y 
    + \funcdif{ C (x) }{ \psi (y) }
    \left. \partdif{ C }{ \pi } \right|_y
    - \left( \beta D^a \partial_a \right)_x \delta \left( x, y \right) - \left( x \leftrightarrow y \right).
        \label{eq:allst_dist-eqn}
\end{equation}
To solve this I will take the functional derivative with respect to a momentum variable, manipulate a few steps and then integrate with a test tensor to find several equations which the constraint must satisfy.  Since I have two fields, I must do this procedure twice.  The first route I consider will be where I take the derivative with respect to the metric momentum.

\subsection{\texorpdfstring{$p^{ab}$}{Metric momentum} route}
\label{sec:dist-eqn-sol-metric}

Starting from the distribution equation \eqref{eq:allst_dist-eqn}, relabel indices, then take the functional derivative with respect to $p^{ab}(z)$,
\begin{equation}
\begin{split}
    0 & =
    \funcdif{ C (x) }{ q_{cd} (y) } \left. \partdif{^2 C }{ p^{ab} \partial p^{cd} } \right|_y \delta ( z, y )
    + \funcdif{ \partial C (x) }{ q_{cd} (y) \partial p^{ab} (x) } \left. \partdif{ C }{ p^{cd} } \right|_y \delta ( z, x )
\\ & 
    + \funcdif{ C (x) }{ \psi (y) } \left. \partdif{^2 C }{ p^{ab} \partial \pi } \right|_y \delta ( z, y )
    + \funcdif{ \partial C (x) }{ \psi (y) \partial p^{ab} (x) } \left. \partdif{ C }{ \pi } \right|_y \delta ( z, x )
\\ & 
    - \partial_{c(x)} \delta ( x, y ) 
    \left(
        \partdif{ ( \beta D^c ) }{ p^{ab} } + \beta \partdif{ D^c }{ p^{ab}_{,d} } \partial_d
    \right)_x \delta ( z, x)
    - \xty.
\end{split}
\end{equation}
Move derivatives and discard surface terms so that it is
reorganised into the form,
\begin{equation}
    0 = 
    A_{ab} ( x, y ) \delta ( z, y )
    - A_{ab} ( y, x ) \delta ( z, x ),
    \label{eq:allst_dist-eqn_Aab}
\end{equation}
where,
\begin{equation}
\begin{split}
    A_{ab} ( x, y ) & =
    \funcdif{ C (x) }{ q_{cd} (y) } \left. \partdif{^2 C }{ p^{ab} \partial p^{cd} } \right|_y
    - \funcdif{ \partial C (y) }{ q_{cd} (x) \partial p^{ab} (y) } \left. \partdif{ C }{ p^{cd} } \right|_x
\\ &
    + \funcdif{ C (x) }{ \psi (y) } \left. \partdif{^2 C }{ p^{ab} \partial \pi } \right|_y
    - \funcdif{ \partial C (y) }{ \psi (x) \partial p^{ab} (y) } \left. \partdif{ C }{ \pi } \right|_x
\\ & 
    + \left( \partdif{ ( \beta D^c ) }{ p^{ab} } \partial_c \right)_y \delta ( y, x )
    - \partial_{d(y)} \left\{ \left( \beta \partdif{ D^c }{ p^{ab}_{,d} } \partial_c \right)_y \delta ( y, x ) \right\}.
\end{split}
    \label{eq:allst_Aab}
\end{equation}
If I take \eqref{eq:allst_dist-eqn_Aab} and integrate over $y$, I can find $A_{ab}(x,y)$ in terms of a function dependent on only a single independent variable,
\begin{equation}
    0 = A_{ab} ( x, z ) - A_{ab} (x) \delta ( z, x ),
    \quad
    \mathrm{where,} \;
    A_{ab} (x) = \int \mathrm{d}^3 y A_{ab} ( y, x ).
\end{equation}
I then multiply this by an arbitrary, symmetric test tensor $\theta^{ab}(z)$, integrate over $z$, and separate out different orders of derivatives of $\theta^{ab}$,
\begin{equation}
\begin{split}
    0 & = 
    \theta^{ab} \left( \cdots \right)_{ab}
    + \partial_c \theta^{ab} \left\{
        \partdif{ C }{ q_{ef,c} } \partdif{^2 C }{ p^{ab} \partial p^{ef} }
        + 2 \partdif{^2 C }{ q_{ef,cd} } \partial_d \left( \partdif{^2 C }{ p^{ab} p^{ef} } \right)
\right. \\ & \left.
        + \partdif{ C }{ p^{ef} } \partdif{^2 C }{ q_{ef,c} \partial p^{ab} }
        - 2 \partdif{ C }{ p^{ef} } \partial_d \left( \partdif{^2 C }{ q_{ef,cd} \partial p^{ab} } \right)
        + \partdif{ C }{ \psi_{,c} } \partdif{^2 C }{ p^{ab} \partial \pi }
\right. \\ & \left.
        + 2 \partdif{ C }{ \psi_{,cd} } \partial_d \left( \partdif{^2 C }{ p^{ab} \partial \pi } \right)
        + \partdif{ C }{ \pi } \partdif{^2 C }{ \psi_{,c} \partial p^{ab} }
        - 2 \partdif{ C }{ \pi } \partial_d \left( \partdif{^2 C }{ \psi_{,cd} \partial p^{ab} } \right)
\right. \\ & \left.
        - \partdif{ ( \beta D^c ) }{ p^{ab} }
        - \partial_d \left( \beta \partdif{ D^d }{ p^{ab}_{,c} } \right)
    \right\}
    + \partial_{cd} \theta^{ab} \left\{ 
        \partdif{ C }{ q_{ef,cd} } \partdif{^2 C }{ p^{ab} \partial p^{ef} }
\right. \\ & \left.
        - \partdif{ C }{ p^{ef} } \partdif{^2 C }{ q_{ef,cd} \partial p^{ab} }
        + \partdif{ C }{ \psi_{,cd} } \partdif{^2 C }{ p^{ab} \partial \pi }
        - \partdif{ C }{ \pi } \partdif{^2 C }{ \psi_{,cd} \partial p^{ab} }
        - \beta \partdif{ D^c }{ p^{ab}_{,d} }
    \right\}.
\end{split}
    \label{eq:allst_dist-eqn-sol-metric}
\end{equation}
As done in previous chapters, I disregard the term zeroth order derivative of $\theta^{ab}$ because it does not provide useful information.

\begin{sloppypar}
Before I can attempt to interpret this equation, I must first separate out all the different tensor combinations that there are.  Because $\theta^{ab}$ is arbitrary, the coefficients of each unique tensor combination must vanish independently.
When I substitute in ${C=C(q,p,\bp,R,\psi,\pi,\Delta,\gamma)}$, there are many complicated tensor combinations that need to be considered, so for convenience I define
${X_a:=q^{bc}\partial_{a}q_{bc}}$.
\end{sloppypar}
I evaluate each term in the $\partial_{cd}\theta^{ab}$ bracket, and write them  in \eqref{eq:allst_extras_d2theta}, in \appref{sec:allst_extras}.
So the linearly independent terms depending on $\partial_{cd}\theta^{ab}$ produce the following conditions,
\begin{subequations}
\begin{align}
    \partial_{ab} \theta^{ab} : 0 & =
        \partdif{ C }{ R } \partdif{ C }{ \bp } + \beta,
    \label{eq:allst_d2theta_1}
\\
\begin{split}
    q_{ab} \partial^2 \theta^{ab} : 0 & =
        \partdif{ C }{ p } \partdif{^2 C }{ p \partial R }
        - \partdif{ C }{ R } \left(
            \partdif{^2 C }{ p^2 } 
            + \third \partdif{ C }{ \bp }
        \right)
    \\ &
        + \half \partdif{ C }{ \Delta } \partdif{^2 C }{ \pi \partial p }
        - \half \partdif{ C }{ \pi } \partdif{^2 C }{ p \partial \Delta },
\end{split}
    \label{eq:allst_d2theta_2} 
\\
    q_{ab} p_\T^{cd} \partial_{cd} \theta^{ab} : 0 & =
        \partdif{ C }{ R } \partdif{^2 C }{ p \partial \bp }
        - \partdif{ C }{ \bp } \partdif{^2 C }{ p \partial R },
    \label{eq:allst_d2theta_3}
\\
    p^\T_{ab} \partial^2 \theta^{ab} : 0 & =
        \partdif{ C }{ p } \partdif{^2 C }{ \bp \partial R }
        - \partdif{ C }{ R } \partdif{^2 C }{ p \partial \bp }
        + \half \partdif{ C }{ \Delta } \partdif{^2 C }{ \pi \partial \bp }
        - \half \partdif{ C }{ \pi } \partdif{^2 C }{ \bp \partial \Delta },
    \label{eq:allst_d2theta_4}
\\
    p^\T_{ab} p_\T^{cd} \partial_{cd} \theta^{ab} : 0 & =
        \partdif{ C }{ R } \partdif{^2 C }{ \bp^2 }
        - \partdif{ C }{ \bp } \partdif{^2 C }{ \bp \partial R }.
    \label{eq:allst_d2theta_5}
\end{align}
    \label{eq:allst_d2theta}%
\end{subequations}
I then evaluate each term in the ${\partial_c\theta^{ab}}$ bracket of \eqref{eq:allst_dist-eqn-sol-metric} and write them in \eqref{eq:allst_extras_dtheta}.
There are many unique terms which should be considered here, but in this case most of these are already solved by a constraint which satisfies \eqref{eq:allst_d2theta}.  
So the equations containing new information are,
\begin{subequations}
\begin{align}
    \partial_a \psi \partial_b \theta^{ab} : 0 & =
        \left( 
            2 \partdif{ C }{ R } \partial_\psi
            - \partdif{ C }{ \Delta }
        \right) \partdif{ C }{ \bp }
        + \partial_\psi \beta,
    \label{eq:allst_dtheta_1}
\\  
\begin{split}
    q_{ab} \partial^c \psi \partial_c \theta^{ab} : 0 & =
            \left( \half \partdif{ C }{ \Delta } - 4 \partdif{ C }{ R } \partial_\psi \right) \left( \partdif{^2 C }{ p^2 } + \third \partdif{ C }{ \bp } \right)
            + \half \partdif{ C }{ \Delta } \partdif{ C }{ \bp }
        \\ &
            + \partdif{ C }{ p } \left( \half \partdif{^2 C }{ p \partial \Delta } + 4 \partial_\psi \partdif{^2 C }{ p \partial R } \right)
            + 2 \left( \partdif{ C }{ \gamma } + \partdif{ C }{ \Delta } \partial_\psi \right) \partdif{^2 C }{ \pi \partial p }
        \\ &
            + 2 \partdif{ C }{ \pi } \left( \partdif{^2 C }{ p \partial \gamma } - \partial_\psi \partdif{^2 C }{ p \partial \Delta } \right)
            - \pi \partdif{ \beta }{ p },
\end{split}
    \label{eq:allst_dtheta_2}
\\
\begin{split}
    p^\T_{ab} \partial^c \psi \partial_c \theta^{ab} : 0 & =
            \left( \half \partdif{ C }{ \Delta } - 4 \partdif{ C }{ R } \right) \partdif{^2 C }{ p \partial \bp }
            + \partdif{ C }{ p } \left( \half \partdif{^2 C }{ \bp \partial \Delta } + 4 \partial_\psi \partdif{^2 C }{ \bp \partial R } \right)
    \\
            + 2 \left( \partdif{ C }{ \gamma } 
            \!\! \right. & \left.
            + \, \partdif{ C }{ \Delta } \partial_\psi \right) \partdif{^2 C }{ \pi \partial \bp }
            + 2 \partdif{ C }{ \pi } \left( \partdif{^2 C }{ \bp \partial \gamma } - \partial_\psi \partdif{^2 C }{ \bp \partial \Delta } \right)
            - \pi \partdif{ \beta }{ \bp },
\end{split}
    \label{eq:allst_dtheta_3}
\\
    q_{ab} p_\T^{cd} \partial_d \psi \partial_c \theta^{ab} : 0 & =
        \left( 2 \partdif{ C }{ R } \partial_\psi - \partdif{ C }{ \Delta } \right) \partdif{^2 C }{ p \partial \bp }
        - \partdif{ C }{ \bp } \left( 2 \partial_\psi \partdif{^2 C }{ p \partial R } + \partdif{^2 C }{ p \partial \Delta } \right)
    \label{eq:allst_dtheta_4}
\\  
    p^\T_{ab} p_\T^{cd} \partial_d \psi \partial_c \theta^{ab} : 0 & =
        \left( 2 \partdif{ C }{ R } \partial_\psi - \partdif{ C }{ \Delta } \right) \partdif{^2 C }{ \bp^2 }
        - \partdif{ C }{ \bp } \left( 2 \partial_\psi \partdif{^2 C }{ \bp \partial R } + \partdif{^2 C }{ \bp \partial \Delta } \right)
    \label{eq:allst_dtheta_5}
\end{align}
\begin{align}
    X_a \partial_b \theta^{ab} : 0 & =
        \partdif{ C }{ R } \left( 1 + 2 \partial_q \right) \partdif{ C }{ \bp }
        + \partial_q \beta,
    \label{eq:allst_dtheta_6}
\\  
\begin{split}
    q_{ab} X^c\partial_c \theta^{ab} : 0 & =
        \partdif{ C }{ p } \left( 4 \partial_q - 1 \right) \partdif{^2 C }{ p \partial R }
        - \partdif{ C }{ R } \left( 4 \partial_q + 1 \right) \left( \partdif{^2 C }{ p^2 } + \third \partdif{ C }{ \bp } \right)
    \\ &
        + \half \partdif{ C }{ \pi } \left( 1 - 4 \partial_q \right) \partdif{^2 C }{ \Delta \partial p }
        + \half \partdif{ C }{ \Delta } \left( 1 + 4 \partial_q \right) \partdif{^2 C }{ \pi \partial p }
        - \third p \partdif{ \beta }{ p },
\end{split}
    \label{eq:allst_dtheta_7}
\\
\begin{split}
    p^\T_{ab} X^c \partial_c \theta^{ab} : 0 & =
        \partdif{ C }{ p } \left( 4 \partial_q - 1 \right) \partdif{^2 C }{ \bp \partial R }
        - \partdif{ C }{ R } \left( 4 \partial_q + 1 \right) \partdif{^2 C }{ \bp \partial p }
            \\ &
        + \half \partdif{ C }{ \pi } \left( 1 - 4 \partial_q \right) \partdif{^2 C }{ \bp \partial \Delta }
        + \half \partdif{ C }{ \Delta } \left( 1 + 4 \partial_q \right) \partdif{^2 C }{ \bp \partial \pi }
        - \third p \partdif{ \beta }{ \bp },
\end{split}
    \label{eq:allst_dtheta_8}
\\
    q_{ab} p_\T^{cd} X_d \partial_c \theta^{ab} : 0 & =
        \partdif{ C }{ R } \left( 1 + 2 \partial_q \right) \partdif{^2 C }{ p \partial \bp }
        + \partdif{ C }{ \bp } \left( 1 - 2 \partial_q \right) \partdif{^2 C }{ p \partial R },
    \label{eq:allst_dtheta_9}
\\  
    p^\T_{ab} p_\T^{cd} X_d \partial_c \theta^{ab} : 0 & =
        \partdif{ C }{ R } \left( 1 + 2 \partial_q \right) \partdif{^2 C }{ \bp^2 }
        + \partdif{ C }{ \bp } \left( 1 - 2 \partial_q \right) \partdif{^2 C }{ \bp \partial R },
    \label{eq:allst_dtheta_10}
\end{align}
\begin{align}
    \partial_a F \partial_b \theta^{ab} : 0 & =
        2 \partdif{ C }{ R } \partdif{^2 C }{ F \partial \bp } + \partdif{ \beta }{ F },
    \label{eq:allst_dtheta_11}
\\  
\begin{split}
    q_{ab} \partial^c F \partial_c \theta^{ab} : 0 & =
        2 \partdif{ C }{ p } \partdif{^3 C }{ F \partial p \partial R }
        - 2 \partdif{ C }{ R } \partdif{}{F} \left( \partdif{^2 C }{ p^2 } + \third \partdif{ C }{ \bp } \right)
    \\ &
        + \partdif{ C }{ \Delta } \partdif{^3 C }{ F \partial p \partial \pi }
        - \partdif{ C }{ \pi } \partdif{^3 C }{ F \partial p \partial \Delta }
        + \third \delta_F^p \partdif{ \beta }{ p },
\end{split}
    \label{eq:allst_dtheta_12}
\\
\begin{split}
    p^\T_{ab} \partial^c F \partial_c \theta^{ab} : 0 & =
        2 \partdif{ C }{ p } \partdif{^3 C }{ F \partial \bp \partial R }
        - 2 \partdif{ C }{ R } \partdif{^3 C }{ F \partial p \partial \bp }
    \\ &
        + \partdif{ C }{ \Delta } \partdif{^3 C }{ F \partial \bp \partial \pi }
        - \partdif{ C }{ \pi } \partdif{^3 C }{ F \partial \bp \partial \Delta }
        + \third \delta_F^p \partdif{ \beta }{ \bp },
\end{split}
    \label{eq:allst_dtheta_13}
\\
    q_{ab} p_\T^{cd} \partial_d F \partial_c \theta^{ab} : 0 & = 
        \partdif{ C }{ R } \partdif{^3 C }{ F \partial p \partial \bp }
        - \partdif{ C }{ \bp } \partdif{^3 C }{ F \partial p \partial R },
    \label{eq:allst_dtheta_14}
\\
    p^\T_{ab} p_\T^{cd} \partial_c F \theta^{ab} : 0 & =
        \partdif{ C }{ R } \partdif{^3 C }{ F \partial \bp^2 }
        - \partdif{ C }{ \bp } \partdif{^3 C }{ F \partial \bp \partial R },
    \label{eq:allst_dtheta_15}
\end{align}%
    \label{eq:allst_dtheta}%
\end{subequations}
where $F\in\{p,\bp,R,\Delta,\gamma\}$.
These conditions strongly restrict the form of the constraint, but before I attempt to consolidate them I must find the conditions coming from the scalar field.

\subsection{\texorpdfstring{$\pi$}{Scalar momentum} route}
\label{sec:dist-eqn-sol-scalar}

Similar to the calculation using the metric momentum, I return to the distribution equation \eqref{eq:allst_dist-eqn} and take the functional derivative with respect to $\pi(z)$,
\begin{equation}
\begin{split}
    0 & =
    \funcdif{ C (x) }{ q_{ab} (y) } \left. \partdif{^2 C }{ \pi \partial p^{ab} } \right|_y \delta ( z, y )
    + \funcdif{ \partial C (x) }{ q_{ab} (y) \partial \pi (x) } \left. \partdif{ C }{ p^{ab} } \right|_y \delta ( z, x )
        \\ &
    + \funcdif{ C (x) }{ \psi (y) } \left. \partdif{^2 C }{ \pi^2 } \right|_y \delta ( z, y )
    + \funcdif{ \partial C (x) }{ \psi (y) \partial \pi (x) } \left. \partdif{ C }{ \pi } \right|_y \delta ( z, x )
        \\ &
    - \delta ( z, x ) \left( \partdif{ ( \beta D^a ) }{ \pi } \partial_a \right)_x \delta ( x, y )
    - \xty,
\end{split}
\end{equation}
which can be rewritten as,
\begin{equation}
    0 = A ( x, y ) \delta ( z, y ) - A ( y, x ) \delta ( z, x ),
    \label{eq:allst_Adist}
\end{equation}
where,
\begin{equation}
\begin{split}
    A ( x, y ) & =
    \funcdif{ C (x) }{ q_{ab} (y) } \left. \partdif{^2 C }{ \pi \partial p^{ab} } \right|_y
    - \funcdif{ \partial C (y) }{ q_{ab} (x) \partial \pi (y) } \left. \partdif{ C }{ p^{ab} } \right|_x
    + \funcdif{ C (x) }{ \psi (y) } \left. \partdif{^2 C }{ \pi^2 } \right|_y
\\ &
    - \funcdif{ \partial C (y) }{ \psi (x) \partial \pi (y) } \left. \partdif{ C }{ \pi } \right|_x
    + \left( \partdif{ ( \beta D^a ) }{ \pi } \partial_a \right)_y \delta ( y, x ),
\end{split}
\end{equation}
and similar to above, \eqref{eq:allst_Adist} can be solved to find $0=A(x,z)-A(x)\delta(x,z)$.  Multiply this by a test scalar field $\eta(z)$ and integrate over $z$,
\begin{equation}
\begin{split}
    0 & = 
    \eta \left( \cdots \right)
    + \partial_a \eta \left\{ 
        \partdif{ C }{ q_{cd,a} } \partdif{^2 C }{ \pi \partial p^{cd} }
        + 2 \partdif{ C }{ q_{cd,ab} } \partial_b \left( \partdif{^2 C }{ \pi \partial p^{cd} } \right)
        + \partdif{ C }{ p^{cd} } \partdif{^2 C }{ q_{cd,a} \partial \pi }
\right. \\ & \left.
        - 2 \partdif{ C }{ p^{cd} } \partdif{^2 C }{ q_{cd,ab} \partial \pi }
        + \partdif{ C }{ \psi_{,a} } \partdif{^2 C }{ \pi^2 }
        + 2 \partdif{ C }{ \psi_{,ab} } \partial_b \left( \partdif{^2 C }{ \pi^2 } \right)
        + \partdif{ C }{ \pi } \partdif{^2 C }{ \psi_{,a} \partial \pi}
\right. \\ & \left.
        - 2 \partdif{ C }{ \pi } \partial_b \left( \partdif{^2 C }{ \psi_{,ab} \partial \pi } \right)
        - \partdif{ ( \beta D^a ) }{ \pi }
    \right\}
\\ & 
    + \partial_{ab} \eta \left\{
        \partdif{ C }{ q_{cd,ab} } \partdif{^2 C }{ \pi \partial p^{cd} }
        - \partdif{ C }{ p^{cd} } \partdif{^2 C }{ q_{cd,ab} \partial \pi }
        + \partdif{ C }{ \psi_{,ab} } \partdif{^2 C }{ \pi^2 }
        - \partdif{ C }{ \pi } \partdif{^2 C }{ \psi_{,ab} \partial \pi }
    \right\}.
\end{split}
    \label{eq:allst_dist-eqn-sol-scalar}
\end{equation}
I evaluate each of the terms for $\partial_{ab}\eta$, and write them in \eqref{eq:allst_extras_d2eta}.
From these, I find the independent equations,
\begin{subequations}
\begin{align}
    \partial^2 \eta : 0 & =
        \partdif{ C }{ p } \partdif{^2 C }{ \pi \partial R }
        - \partdif{ C }{ R } \partdif{^2 C }{ \pi \partial p }
        + \half \partdif{ C }{ \Delta } \partdif{^2 C }{ \pi^2 }
        - \half \partdif{ C }{ \pi } \partdif{^2 C }{ \Delta \partial \pi },
    \label{eq:allst_d2eta_1}
\\
    p_\T^{ab} \partial_{ab} \eta : 0 & =
        \partdif{ C }{ R } \partdif{^2 C }{ \pi \partial \bp }
        - \partdif{ C }{ \bp } \partdif{^2 C }{ \pi \partial R }.
    \label{eq:allst_d2eta_2}
\end{align}
    \label{eq:allst_d2eta}%
\end{subequations}
Then, I evaluate all the terms for $\partial_a\eta$, and write them in \eqref{eq:allst_extras_deta}.
Therefore, ignoring terms solved by \eqref{eq:allst_d2eta}, the equations I get from $\partial_a\eta$ are,
\begin{subequations}
\begin{align}
\begin{split}
    \partial^a \psi \partial_a \eta : 0 & =
        \left( 
            \half \partdif{ C }{ \Delta}
            - 4 \partdif{ C }{ R } \partial_\psi
        \right) \partdif{^2 C }{ \pi \partial p }
        + \partdif{ C }{ p } \left(
            \half \partdif{^2 C }{ \Delta \partial \pi } + 4 \partial_\psi \partdif{^2 C }{ R \partial \pi }
        \right)
    \\ 
        + 2 \left(
            \partdif{ C }{ \gamma }
            \!\! \right. & \left.
            + \partdif{ C }{ \Delta } \partial_\psi
        \right) \partdif{^2 C }{ \pi^2 }
        + 2 \partdif{ C }{ \pi } \left( 
            \partdif{^2 C }{ \gamma \partial \pi }
            - \partial_\psi \partdif{^2 C }{ \Delta \partial \pi }
        \right)
        - \left( \beta + \pi \partdif{ \beta }{ \pi }\right),
\end{split}
    \label{eq:allst_deta_1}
\\
    p_\T^{ab} \partial_b \psi \partial_a \eta : 0 & =
        \left(
            \partdif{ C }{ R } \partial_\psi
            - \half \partdif{ C }{ \Delta }
        \right) \partdif{^2 C }{ \pi \partial \bp }
        - \partdif{ C }{ \bp }
        \left(
            \partial_\psi \partdif{^2 C }{ \pi \partial R }
            + \half \partdif{^2 C }{  \pi \partial \Delta }
        \right),
    \label{eq:allst_deta_2}
\\
\begin{split}
    X^a \partial_a \eta : 0 & =
        \partdif{ C }{ p } \left( 4 \partial_q - 1 \right) \partdif{^2 C }{ \pi \partial R }
        - \partdif{ C }{ R } \left( 4 \partial_q + 1 \right) \partdif{^2 C }{ \pi \partial p }
    \\ &
        + \half \partdif{ C }{ \Delta } \left( 1 + 4 \partial_q \right) \partdif{^2 C }{ \pi^2 }
        + \half \partdif{ C }{ \pi } \left( 1 - 4 \partial_q \right) \partdif{^2 C }{ \pi \partial \Delta }
        - \third p \partdif{ \beta }{ \pi },
\end{split}
    \label{eq:allst_deta_3}
\\
    p_\T^{ab} X_b \partial_a \eta : 0 & =
        \partdif{ C }{ R } \left( 1 + 2 \partial_q \right) \partdif{^2 C }{ \pi \partial \bp }
        + \partdif{ C }{ \bp } \left( 1 - 2 \partial_q \right) \partdif{^2 C }{ \pi \partial R },
    \label{eq:allst_deta_4}
\\
\begin{split}
    \partial^a F \partial_a \eta : 0 & =
        \partdif{ C }{ p } \partdif{^3 C }{ F \partial \pi \partial R }
        - \partdif{ C }{ R } \partdif{^3 C }{ F \partial \pi \partial R }
        + \half \partdif{ C }{ \Delta } \partdif{^3 C }{ F \partial \pi^2 }
    \\ &
        - \half \partdif{ C }{ \pi } \partdif{^3 C }{ F \partial \pi \partial \Delta }
        + \frac{1}{6} \delta_F^p \partdif{ \beta }{ \pi },
\end{split}
    \label{eq:allst_deta_5}
\\
    p_\T^{ab} \partial_b F \partial_a \eta : 0 & =
        \partdif{ C }{ R } \partdif{^3 C }{ F \partial \pi \partial \bp }
        - \partdif{ C }{ \bp } \partdif{^3 C }{ F \partial \pi \partial R },
    \label{eq:allst_deta_6}
\end{align}
    \label{eq:allst_deta}%
\end{subequations}
where $F \in \{ p, \bp, R, \Delta, \gamma \}$.
Now that I have all of the conditions restricting the form of the constraint, I can move on to consolidating and interpreting them.

\section{Solving for the constraint}
\label{sec:allst_solving}

Now I have the full list of equations, I seek to find the restrictions on the form of $C$ they impose.  Firstly, I use the condition from $\partial_{ab}\theta^{ab}$, \eqref{eq:allst_d2theta_1} to find
\begin{equation}
    \partdif{ C }{ R } = - \beta \left( \partdif{ C }{ \bp } \right)^{-1},
    \label{eq:allst_solving_1}
\end{equation}
which I substitute into the equation from $p^\T_{ab}p_\T^{cd}\partial_{cd}\theta^{ab},$ \eqref{eq:allst_d2theta_5},
\begin{equation}
\begin{split}
    0 & = 
    \partdif{ C }{ R } \partdif{^2 C }{ \bp^2 }
    - \partdif{ C }{ \bp } \partdif{^2 C }{ \bp \partial R }
\\ &
    = - 2 \beta \left( \partdif{ C }{ \bp } \right)^{-1} \partdif{^2 C }{ \bp^2 }
    + \partdif{ \beta }{ \bp }
\\ &
    = \beta \partdif{}{ \bp } \log \left\{ \beta \left( \partdif{ C }{ \bp } \right)^{-2} \right\},
\end{split}
    \label{eq:allst_solving_2}
\end{equation}
and because $\beta\to1$ in the classical limit and so cannot vanish generally, I find that,
\begin{equation}
    \beta = b_1 \, \left( \partdif{ C }{ \bp } \right)^2,
        \quad \mathrm{where} \;
    \partdif{ b_1 }{ \bp } = 0.
    \label{eq:allst_solving_3}
\end{equation}
Substituting this back into \eqref{eq:allst_solving_1} gives me
$\displaystyle{\partdif{C}{R}=-b_1\partdif{C}{\bp}}$, and from this I can find the first restriction on the form of the constraint,
\begin{equation}
\begin{gathered}
    C ( q, p, \bp, R, \psi, \pi, \Delta, \gamma ) = 
    C_1 (q, p, \psi, \pi, \Delta, \gamma, \chi_1 ),
\\
    \mathrm{where} \;
    \chi_1 := \bp - \int_0^R b_1 ( q, p, x, \psi, \pi, \Delta, \psi ) \mathrm{d} x.
\end{gathered}
    \label{eq:allst_solving_4}
\end{equation}
Substituting this into the condition from $\partial_aF\partial_b\theta^{ab}$, \eqref{eq:allst_dtheta_11}, gives
\begin{equation}
    0 = \partdif{ b_1 }{ F } \left( \partdif{ C_1 }{ \chi_1 } \right)^2,
    \quad 
    \mathrm{for} \;
    F \in \{ p, \bp, R, \pi, \Delta, \gamma \},
    \label{eq:allst_solving_5}
\end{equation}
and therefore $b_1$ must only be a function of $q$ and $\psi$. Substituting this into \eqref{eq:allst_solving_4} leads to ${\chi_1=\bp-b_1R}$.
Turning to the condition from $X_a\partial_b\theta^{ab}$, \eqref{eq:allst_dtheta_6}, I find
\begin{equation}
    0 = \left( \partdif{ C_1 }{ \chi_1 } \right)^2 \left( \partial_q - 1 \right) b_1 ,
    \label{eq:allst_solving_6}
\end{equation}
which is solved by ${b_1(q,\psi)=q\,b_2(\psi)}$.  This is as expected because it means both terms in $\chi_1$ have a density weight of two.
From this I see that the condition coming from ${\partial_a\psi\partial_b\theta^{ab}}$, \eqref{eq:allst_dtheta_1}, gives
\begin{equation}
    0 = \partdif{ C_1 }{ \chi_1 } \left( 
    q b_2' \partdif{ C_1 }{ \chi_1 }
    - \partdif{ C_1 }{ \Delta }
    \right)
    \label{eq:allst_solving_7}
\end{equation}
which provides further restrictions on the form of the constraint,
\begin{equation}
    C = C_2 ( q, p, \psi, \pi, \gamma, \chi_2 ),
    \quad
    \chi_2 := \bp - q \left( b_2 R - b_2' \Delta \right).
    \label{eq:allst_solving_8}
\end{equation}
Look at the condition from $p^\T_{ab}\partial^2\theta^{ab}$, \eqref{eq:allst_d2theta_4},
\begin{equation}
\begin{split}
    0 & = \partdif{ C }{ p } \partdif{^2 C }{ \bp \partial R }
    - \partdif{ C }{ R } \partdif{^2 C }{ p \partial \bp }
    + \half \partdif{ C }{ \Delta } \partdif{^2 C }{ \pi \partial \bp }
    - \half \partdif{ C }{ \pi } \partdif{^2 C }{ \bp \partial \Delta },
\\ &
    = q b_2 \left\{
    - \partdif{ C_2 }{ p } \partdif{^2 C_2 }{ \chi_2^2 }
    + \partdif{ C_2 }{ \chi_2 } \partdif{^2 C }{ p \partial \chi_2 }
    + \frac{b_2'}{2b_2} \left(
        \partdif{ C_2 }{ \chi_2 } \partdif{^2 C_2 }{ \pi \partial \chi_2}
        - \partdif{ C_2 }{ \pi } \partdif{^2 C_2 }{ \chi_2^2 }
    \right) \right\}
\\ &
    = q b_2 \partdif{ C_2 }{ \chi_2 } \left( \partdif{ C_2 }{ p } + \frac{ b_2' }{ 2 b_2 } \partdif{ C_2 }{ \pi } \right) \partdif{}{ \chi_2 } \log \left\{
     \left( \partdif{ C_2 }{ p } + \frac{ b_2' }{ 2 b_2 } \partdif{ C_2 }{ \pi } \right) \left( \partdif{ C_2 }{ \chi_2 } \right)^{-1}
    \right\},
\end{split}
    \label{eq:allst_solving_9}
\end{equation}
and because $b_2$ is a non-zero constant in the classical limit, this can be integrated to find
\begin{equation}
    \partdif{ C_2 }{ p } 
    + \frac{ b_2' }{ 2 b_2 } \partdif{ C_2 }{ \pi }
    = g_1 ( q, p, \psi, \pi, \gamma ) \partdif{ C_2 }{ \chi_2 },
    \label{eq:allst_solving_10}
\end{equation}
where $g_1$ is a unknown function arising as an integration constant, and needs to be determined.
This provides a further restriction on the form of the constraint,
\begin{equation}
\begin{gathered}
    C = C_3 \left( q, \psi, \gamma, \Pi, \chi_3 \right),
        \quad
    \Pi := \pi - \frac{ b_2' }{ 2 b_2 } p,
        \\
    \chi_3 := P - q \left( b_2 R - b_2' \Delta \right)
    + \int_0^p g_1 \left( q, x, \psi, \Pi + \frac{ b_2' }{ 2 b_2 } x, \gamma \right) \mathrm{d} x.
\end{gathered}
    \label{eq:allst_solving_11}
\end{equation}
Substituting this into the condition from $\partial^2\eta$, \eqref{eq:allst_d2eta_1}, gives
\begin{equation}
    0 = q \, b_2 \left( \partdif{ C_3 }{ \chi_3 } \right)^2 \partdif{}{\pi} g_1 \left( q, p, \psi, \pi, \gamma \right),
    \label{eq:allst_solving_12}
\end{equation}
and therefore,
\begin{equation}
    \chi_3 = P - q \left( b_2 R - b_2' \Delta \right)
    + \int_0^p g_1 \left( q, x, \psi, \gamma \right) \mathrm{d} x.
    \label{eq:allst_solving_13}
\end{equation}
Evaluating the condition from $q_{ab}\partial^2\theta^{ab}$, \eqref{eq:allst_d2theta_2}, gives
\begin{equation}
    0 = \third q b_2 \left( \partdif{ C_3 }{ \chi_3 } \right)^2 \left( 1 + 3 \partdif{}{ p } \right) g_1 \left( q, p, \psi, \gamma \right),
    \label{eq:allst_solving_14}
\end{equation}
which can be integrated to find 
$g_1=g_2\left(q,\psi,\gamma\right)-p/3$
and therefore \eqref{eq:allst_solving_13} becomes,
\begin{equation}
    \chi_3 = \bp - \frac{1}{6} p^2 + g_2 \, p - q \left( b_2 R - b_2' \Delta \right).
    \label{eq:allst_solving_15}
\end{equation}
Then look at the condition from $p^\T_{ab}X^c\partial_c\theta^{ab}$, \eqref{eq:allst_dtheta_8}, from which can be found
\begin{equation}
    0 = 2 q b_2 \partdif{ C_3 }{ \chi_3 } \partdif{^2 C_3 }{ \chi_3^2 } \left( 2 \partial_q - 1 \right) g_2,
    \label{eq:allst_solving_16}
\end{equation}
which can be solved by,
${g_2 \left( q, \psi, \gamma \right) = \sqrt{q} \, g_3 \left( \psi, \gamma \right)}$
if we assume that 
$\displaystyle{\partdif{^2C_3}{\chi_3^2}\neq0}$ 
generally, which is true for any deformation dependent on curvature
$\displaystyle{\partdif{\beta}{\chi_3}\neq0}$.
This is what is expected for the density weight of each term in $\chi_3$ to match.

I now look at the condition for $p^\T_{ab} \partial^c \gamma \partial_c \theta^{ab}$, which is \eqref{eq:allst_dtheta_13} with $F=\gamma$, 
\begin{equation}
    0 = q^{3/2} b_2 \partdif{ g_3 }{ \gamma } \partdif{ C_3 }{ \chi_3 } \partdif{^2 C_3 }{ \chi_3^2 },
    \label{eq:allst_solving_17}
\end{equation}
which is true when $g_3 = g_3 \left( \psi \right)$.

At this point it gets harder to progress further as I have done so far.  To review, I have restricted the constraint and deformation to the forms,
\begin{equation}
\begin{gathered}
    C \left( q, p, \bp, R, \psi, \pi, \Delta, \gamma \right) = C_3 \left( q, \psi, \Pi, \gamma, \chi_3 \right),
        \quad
    \beta = q \, b_2 \left( \psi \right) \left( \partdif{ C_3 }{ \chi_3 } \right)^2,
        \\
    \Pi = \pi - \frac{b_2'}{2b_2} p,
        \quad
    \chi_3 = \bp - \frac{1}{6} p^2 + p \sqrt{q} \, g_3 \left( \psi \right) - q \left( b_2 R - b_2' \Delta \right),
\end{gathered}
    \label{eq:allst_solving_review}
\end{equation}
which satisfies all the conditions in \eqref{eq:allst_d2theta}, \eqref{eq:allst_d2eta}, \eqref{eq:allst_deta} and \eqref{eq:allst_dtheta} apart from the conditions for $q_{ab}\partial^c\psi\partial_c\theta^{ab}$, \eqref{eq:allst_dtheta_2}, and  $\partial^a\psi\partial_a\eta$, \eqref{eq:allst_deta_1}.  
As it stands, these conditions are not easy to solve.


\subsection{Solving the fourth order constraint to inform the general case}
\label{sec:allst_fourth}

To break this impasse, I use a test ansatz for the constraint which contains up to four orders in momenta,
\begin{equation}
\begin{split}
    C_3 & \to C_0 + C_{(\Pi)} \Pi + C_{(\Pi^2)} \Pi^2 + C_{(\Pi^3)} \Pi^3 + C_{(\Pi^4)} \Pi^4
\\ & 
    + C_{(\chi)} \chi_3 + C_{(\chi^2)} \chi_3^2 + C_{(\Pi\chi)} \Pi \chi_3 + C_{(\Pi^2\chi)} \Pi^2 \chi_3,
\end{split}
\end{equation}
where each coefficient is an unknown function to be determined dependent on $q$, $\psi$ and $\gamma$.  There is an asymmetric term included in $\chi_3$ determined by the function $g_3\left(\psi\right)$, so I do not restrict myself to only even orders of momenta, unlike \secref{sec:2ndst}.

Substituting this into \eqref{eq:allst_deta_1}, I can separate out the multiplier of each unique combination of variables as an independent equation.  For each of the terms which are the multipliers of 5 or 6 orders of momenta, I find a condition specifying that the constraint coefficients for terms 3 or 4 orders of momenta must not depend on $\gamma$, e.g. 
${\displaystyle{\partdif{}{\gamma}C_{(\chi^2)}=0}}$, 
${\displaystyle{\partdif{}{\gamma}C_{(\Pi^3)}=0}}$.  
Since $\gamma$ depends on two spatial derivatives, I see that each term in the constraint must not depend on a higher order of spatial derivatives than it does momenta.  If I include higher orders of spatial derivatives in the ansatz, I quickly find them ruled out in a similar fashion.  Therefore, I use this information to further expand my ansatz,
\begin{equation}
\begin{split}
    C_3 & \to C_\0 + C_{(\gamma)} \gamma + C_{(\gamma^2)} \gamma^2 + C_{(\Pi)} \Pi + C_{(\Pi\gamma)} \Pi \gamma + C_{(\Pi^2)} \Pi^2 
\\ &
    + C_{(\Pi^2\gamma)} \Pi^2 \gamma + C_{(\Pi^3)} \Pi^3 + C_{(\Pi^4)} \Pi^4 + C_{(\chi)} \chi_3 + C_{(\chi\gamma)} \chi_3 \gamma + C_{(\chi^2)} \chi_3^2
\\ &
    + C_{(\Pi\chi)} \Pi \chi_3 + C_{(\Pi^2\chi)} \Pi^2 \chi_3,
\end{split}
\end{equation}
where each coefficient is now an unknown function of $q$ and $\psi$.

One can find all the necessary conditions from \eqref{eq:allst_deta_1}, for which the solution also satisfies \eqref{eq:allst_dtheta_2}.  I will show a route which can taken to progressively restrict $C$.  The condition coming from $\bp^2$ is solved if
\begin{equation}
    C_{(\Pi^2\chi)} = \half C_{(\chi^2)} \left( \frac{ C_{(\chi\gamma)} }{ 2 q b_2 C_{(\chi^2)} } + \frac{ 7 b_2^{\prime2} }{ 8 b_2^2 } - \frac{ b_2'' }{ b_2 } \right)^{-1},
\end{equation}
the condition from $\gamma^2$ is solved by,
\begin{equation}
    C_{(\Pi^2\gamma)} = \quarter C_{(\chi\gamma)} \left( \frac{ 2 C_{(\gamma^2)} }{ b_2 C_{(\chi\gamma)} } + \frac{ 7 b_2^{\prime2} }{ 8 b_2^2 } - \frac{ b_2'' }{ b_2 } \right)^{-1},
\end{equation}
the condition from $\gamma\bp$ is solved by,
\begin{equation}
    C_{(\gamma^2)} = \frac{ C_{(\chi\gamma)}^2 }{ 4 C_{(\chi^2)} },
\end{equation}
the condition from $\pi^4$ is solved by,
\begin{equation}
    C_{(\Pi^4)} = \frac{1}{16} C_{(\chi^2)} \left( \frac{ C_{(\chi\gamma)} }{ 2 q b_2 C_{(\chi^2)} } + \frac{ 7 b_2^{\prime2} }{ 8 b_2^2 } - \frac{ b_2'' }{ b_2 } \right)^{-1},
\end{equation}
and all the other conditions coming from four momenta are solved.  Turning to the third order, the condition from $\pi\bp$ is solved by,
\begin{equation}
\begin{split}
    C_{(\Pi^3)} & = \frac{1}{12} \left\{ 
    C_{(\Pi\chi)} \left[ \frac{1}{qb_2} 
        \left( \frac{ 3 C_{(\chi\gamma)} }{ 2 C_{(\chi^2)} } - \frac{ 2 C_{(\Pi\gamma)} }{ C_{(\Pi\chi)} } \right) + \frac{ 7 b_2^{\prime2} }{ 8 b_2^2 } - \frac{ b_2'' }{ b_2 }
    \right]
\right. \\ &  \left.
        - \sqrt{q} C_{(\chi^2)} \left( 4 g_3' - \frac{3g_3b_2'}{b_2} \right) 
    \right\}
    \left( \frac{ C_{(\chi\gamma)} }{ 2 q b_2 C_{(\chi^2)} } + \frac{ 7 b_2^{\prime2} }{ 8 b_2^2 } - \frac{ b_2'' }{ b_2 } \right)^{-2},
\end{split}
\end{equation}
and the condition from $\pi\gamma$ is solved by,
\begin{equation}
    C_{(\Pi\gamma)} = \frac{ C_{(\Pi\chi)} C_{(\chi\gamma)} }{ 2 C_{(\chi^2)} },
\end{equation}
and the condition from $\pi^3$ is solved by,
\begin{equation}
    C_{(\Pi\chi)} = \frac{-1}{2} \sqrt{q} C_{(\chi^2)} \left( 4 g_3' - \frac{ 3 g_3 b_2' }{ b_2 } \right) \left( \frac{ C_{(\chi\gamma)} }{ 2 q b_2 C_{(\chi^2)} } + \frac{ 7 b_2^{\prime2} }{ 8 b_2^2 } - \frac{ b_2'' }{ b_2 } \right)^{-1},
\end{equation}
which completes all the terms from third order.  The only new condition coming from second order is solved by,
\begin{equation}
\begin{split}
    C_{(\Pi^2)} & = \left\{ 
    \quarter C_{(\Pi\chi)} \left[ \frac{1}{qb_2} 
        \left( \frac{ C_{(\chi\gamma)} }{ C_{(\chi^2)} } - \frac{ C_{(\gamma)} }{ C_{(\chi)} } \right) + \frac{ 7 b_2^{\prime2} }{ 8 b_2^2 } - \frac{ b_2'' }{ b_2 }
    \right]
\right. \\ &  \left.
        + \frac{1}{16} q C_{(\chi^2)} \left( 4 g_3' - \frac{3g_3b_2'}{b_2} \right)^2 
    \right\}
    \left( \frac{ C_{(\chi\gamma)} }{ 2 q b_2 C_{(\chi^2)} } + \frac{ 7 b_2^{\prime2} }{ 8 b_2^2 } - \frac{ b_2'' }{ b_2 } \right)^{-2},
\end{split}
\end{equation}
and the only new condition coming from first order is solved by,
\begin{equation}
\begin{split}
    C_{(\Pi)} & = \frac{-1}{4} \sqrt{q} C_{(\chi)} \left( 4 g_3' - \frac{ 3 g_3 b_2' }{ b_2 } \right)
    \left\{ \frac{1}{qb_2} \left( \frac{ C_{(\chi\gamma)} }{ C_{(\chi^2)} } - \frac{ C_{(\gamma)} }{ C_{(\chi)} } \right) + \frac{ 7 b_2^{\prime2} }{ 8 b_2^2 } - \frac{ b_2'' }{ b_2 }  \right\}
\\ & 
    \times \left( \frac{ C_{(\chi\gamma)} }{ 2 q b_2 C_{(\chi^2)} } + \frac{ 7 b_2^{\prime2} }{ 8 b_2^2 } - \frac{ b_2'' }{ b_2 } \right)^{-2},
\end{split}
\end{equation}
and from the zeroth order,
\begin{equation}
    C_{(\chi\gamma)} = \frac{ 2 C_{(\gamma)} C_{(\chi^2)} }{ C_{(\chi)} },
\end{equation}
%
%

When all of these terms are combined, I find the solution for the fourth order constraint,
\begin{equation}
    C = C_\0 
    + C_{(\chi)} \left( \chi_3 + \frac{ \Pi \left( \Pi - \Xi \right) }{ 4 \Omega } + \frac{ C_{(\gamma)} }{ C_{(\chi)} } \gamma \right)
    + C_{(\chi^2)} \left( \chi_3 + \frac{ \Pi \left( \Pi - \Xi \right) }{ 4 \Omega } + \frac{ C_{(\gamma)} }{ C_{(\chi)} } \gamma \right)^2,
        \label{eq:allst_constraint_4th}
\end{equation}
where
\begin{equation}
\begin{aligned}
    \Omega & =
    \frac{ C_{(\gamma)} }{ b_2 q C_{(\chi)} }
    + \frac{ 7 b_2^{\prime 2} }{ 8 b_2^2 }
    - \frac{ b_2'' }{ b _2 },
        &
    \Xi & = \sqrt{q} \left( 4 g_3' - \frac{ 3 g_3 b_2' }{ b_2 } \right).
\end{aligned}%
\end{equation}

If this solution is generalised to all orders,
\begin{equation}
    C = C_4 \left( q, \psi, \chi_4 \right),
        \quad
    \chi_4 = \bp - \frac{1}{6} p^2 + \sqrt{q} p g_3 - q \left( b_2 R - b_2' \Delta \right)
    + \frac{ \Pi \left( \Pi - \Xi \right) }{ 4 \Omega } 
    + \frac{ C_{(\gamma)} }{ C_{(\chi)} } \gamma,
\end{equation}
one can check that it satisfies all the conditions from \eqref{eq:allst_d2theta}, \eqref{eq:allst_d2eta}, \eqref{eq:allst_deta} and \eqref{eq:allst_dtheta}.  It is possible that directly generalising from the fourth order constraint rather than continuing to work generally means that this is not the most general solution.  However, at least I now know a form of the constraint which \emph{can} solve all the conditions.

Now that I have a form for the general constraint, I seek to compare it to the low-curvature limit, when $C\to\chi_4C_\chi+C_\0$, and match terms with that found previously \eqref{eq:2ndst_effective} in \chapref{sec:2ndst} and \cite{cuttell2018}.  I find that,
\begin{equation}
\begin{aligned}
    b_2 & = \frac{ \sigma_\beta \omega_R^2 }{ 4 },
        &
    \sigma_\beta & := \sgn{\beta} = \sgn{\beta_\0}.
\\
    C_\chi & = \frac{ 2 \sigma_\beta }{ \omega_R } \sqrt{ \frac{ \left| \beta_\0 \right| }{ q } },
        &
    C_\gamma & = \sqrt{ q \left| \beta_\0 \right| } \left( \frac{ \omega_\psi }{ 2 } + \omega_R'' \right),
\end{aligned}
\end{equation}
For convenience, I redefine the function determining the asymmetry, ${g_3=\xi/2}$,
and I expand the constraint in terms of the weightless (or `de-densitised') scalar ${\R:=\chi_4/q}$.
This means that the general form of the deformed constraint is given by,
\begin{subequations}
\begin{gather}
    C = C \left( q, \psi, \R \right),
\quad
    \beta = \frac{ \sigma_\beta }{ q } \left( \partdif{ C }{ \R } \right)^2,
        \label{eq:allst_constraint-solution_differential} \\
\begin{gathered}
    \R := 
    \frac{ 2 \sigma_\beta }{ q \omega_R } \left( \bp - \frac{1}{6} p^2 \right)
    - \frac{ \omega_R }{ 2 } R
    + \omega_R' \Delta \psi
    + \left( \frac{\omega_\psi}{2} + \omega_R'' \right) \partial^a \psi \partial_a \psi,
\\
    + \frac{ \sigma_\beta \omega_R }{ \omega_\psi \omega_R + \frac{ 3 }{ 2 } \omega_R^{\prime2} }
    \left\{ \frac{ 1 }{ 2 q } \left( \pi - \frac{ \omega_R' }{ \omega_R } p \right)^2 
    + \frac{ \xi }{ \sqrt{q} } \left[       
        \frac{ \omega_\psi }{ \omega_R } p 
        + \frac{ 3 \omega_R' }{ 2 \omega_R } \pi 
        - \frac{ \xi' }{ \xi } \left( \pi - \frac{ \omega_R' }{ \omega_R } p \right)
    \right] \right\}.
    \label{eq:allst_constraint-solution_curvature}
\end{gathered}
\end{gather}%
    \label{eq:allst_constraint-solution}%
\end{subequations}
It is probably more appropriate to see the deformation function itself as the driver of deformations to the constraint, 
so I rearrange \eqref{eq:allst_constraint-solution_differential},
\begin{equation}
    \partdif{ C }{ \R } = \sqrt{ q \left| \beta \right| },
\end{equation}
which can be integrated to find,
\begin{equation}
    C = 
    \int_0^\R \sqrt{ q \left| \beta( q, \psi, r ) \right| } \mathrm{d} r 
    + C_\0 ( q, \psi ).
    \label{eq:allst_constraint-solution_integral}
\end{equation}
From either form of the general solution \eqref{eq:allst_constraint-solution_differential} or \eqref{eq:allst_constraint-solution_integral}, one can now understand the meaning of \eqref{eq:methodology_order_constraint}, which relates the order of the constraint and the deformation, ${2n_C-n_\beta=4}$.  The differential form \eqref{eq:allst_constraint-solution_differential} is like
${n_\beta=2\left(n_C-2\right)}$,
and the integral form \eqref{eq:allst_constraint-solution_integral} is like
${n_C=2+n_\beta/2}$.

From the integral form of the solution \eqref{eq:allst_constraint-solution_integral}, I can now check a few examples of what constraint corresponds to certain deformations.  Here are a few examples of easily integrable functions with the appropriate limit,
\begin{equation}
\begin{split}
    \beta & = \beta_\0 \left( 1 + \beta_2 \R \right)^n
\\
    \to
    C & = C_\0 + 
\left\{
    \begin{aligned}
    & \frac{ 2 \sqrt{ q \left| \beta_\0 \right| } }{ \left(  n + 2 \right) \beta_2 }
    \left\{ \sgn{ 1 + \beta_2 \R } \left| 1 + \beta_2 \R \right|^{ \frac{ n + 2 }{ 2 } } - 1 \right\},
    & n \neq -2,
\\
    &
    \frac{ \sqrt{ q \left| \beta_\0 \right| } }{ \beta_2 } \sgn{ 1 + \beta_2 \R } \log \left| 1 + \beta_2 \R \right|,
    & n = -2,
    \end{aligned}
\right.
\\ &
    \simeq C_\0 + \sqrt{ q \left| \beta_\0 \right| } 
    \left\{ \R + \frac{ n \beta_2 }{ 4 } \R^2 + \cdots \right\}.
\end{split}
    \label{eq:allst_deftocon_linear}
\end{equation}
\begin{equation}
\begin{split}
    \beta = \beta_\0 e^{\beta_2 \R}
    \; \to \;
    C &= C_\0 
    + \frac{ 2 \sqrt{ q \left| \beta_\0 \right| } }{ \beta_2 } 
    \left( e^{ \beta_2 \R / 2} - 1 \right)
\\ &
    \simeq
    C_\0 + \sqrt{ q \left| \beta_\0 \right| } \left( \R + \frac{ \beta_2 }{ 4 } \R^2 + \cdots \right),
\end{split}
    \label{eq:allst_deftocon_exp}
\end{equation}
\begin{equation}
\begin{split}
    \beta = \beta_\0 \sech^2 \left( \beta_2 \R \right)
    \; \to \;
    C &= C_\0 
    + \frac{ \sqrt{ q \left| \beta_\0 \right| } }{ \beta_2 } \gud \left( \beta_2 \R \right),
\\ &
    \simeq
    C_\0 + \sqrt{ q \left| \beta_\0 \right| } \left( \R - \frac{ \beta_2^2 }{ 6 } \R^3 + \cdots \right),
\end{split}
    \label{eq:allst_deftocon_sech2}
\end{equation}
where $\gud(x):=\int_0^x\mathrm{d}t\,\sech(t)$ is the Gudermannian function.
Most other deformation functions would need to be integrated numerically to find the constraint.
As can be seen from the small $\R$ expansions, it would be possible to constrain $\beta_\0$ and $\beta_2$ phenomenologically but the asymptotic behaviour of $\beta$ would be difficult to determine.

The simplest constraint that can be expressed as a polynomial of $\R$ that contains higher orders than the classical solution is given by,
\begin{equation}
    \beta = \beta_\0 \left( 1 + \beta_2 \R \right)^2
    \to
    C = C_\0 + \sqrt{ q \left| \beta_\0 \right| } \left( \R + \frac{ \beta_2 }{ 2 } \R^2 \right),
        \label{eq:allst_deftocon_simple}
\end{equation}
which is equivalent to the fourth order constraint found in \eqref{eq:allst_constraint_4th}.


\section{Looking back at the constraint algebra}
\label{sec:allst_lookingback}

For this deformed constraint to mean anything, it must not reduce to the undeformed constraint through a simple transformation.  If I write the constraint as a function of the undeformed vacuum constraint $\bar{C}=\sqrt{q}\,\R$, I see that the deformation in the constraint algebra can be absorbed by a redefinition of the lapse functions,
\begin{subequations}
\begin{align}
    \{ C [ N ], C [ M ] \}
    & = \int \mathrm{d} x \mathrm{d} y N(x) M(y) \{ C (x), C (y) \},
\\ &
    = \int \mathrm{d} x \mathrm{d} y 
    \left( N \partdif{ C }{ \bar{C} } \right)_x
    \left( M \partdif{ C }{ \bar{C} } \right)_y \{ \bar{C} (x), \bar{C} (y) \},
\\ &
    = \int \mathrm{d} x \mathrm{d} y
    \left( \sigma_{\partial{}C} \bar{N} \right)_x \left( \sigma_{\partial{}C} \bar{M} \right)_y \{ \bar{C} (x), \bar{C} (y) \},
\\ &
    = \{ \bar{C} [ \sigma_{\partial{}C} \bar{N} ], \bar{C} [ \sigma_{\partial{}C} \bar{M} ] \},
\end{align}%
\end{subequations}
where 
$\bar{N}:=N\left|\partial{C}/\partial\bar{C}\right|$, 
$\bar{M}:=M\left|\partial{C}/\partial\bar{C}\right|$
and 
$\sigma_{\partial{}C}:=\sgn{\partial{C}/\partial{\bar{C}}}$, 
because the lapse functions should remain positive.
The other side of the equality,
\begin{equation}
\begin{split}
    D_a [ \beta q^{ab} ( N \partial_b M - \partial_b N M ) ],
    & =
    \int \mathrm{d} x D_a \beta q^{ab} ( N \partial_b M - \partial_b N M )
\\ &
    = \int \mathrm{d} x D_a \sigma_\beta \left( \partdif{ C }{ \bar{C} } \right)^2 ( N \partial_b M - \partial_b N M )
\\ &
    = \int \mathrm{d} x D_a \sigma_\beta ( \bar{N} \partial_b \bar{M} - \partial_b \bar{N} \bar{M} ),
\\ &
    = D_a [ \sigma_\beta q^{ab} ( \bar{N} \partial_b \bar{M} - \partial_b \bar{N} \bar{M} ) ],
\end{split}
\end{equation}
which I can combine to show the that the following two equations are equivalent,
\begin{subequations}
\begin{align}
    \{ C [ N ], C [ M ] \}
    & =
    D_a [ \beta q^{ab} ( N \partial_b M - \partial_b N M ) ],
\\
    \{ \bar{C} [ \sigma_{\partial{}C} \bar{N} ], \bar{C} [ \sigma_{\partial{}C} \bar{M} ] \}
    & =
    D_a [ \sigma_\beta q^{ab} ( \bar{N} \partial_b \bar{M} - \partial_b \bar{N} \bar{M} ) ].
\end{align}
\end{subequations}
The two $\sigma_{\partial{}C}$ on the left side should cancel out, but they are included here to show the limit to the redefinition of the lapse functions.  
While it may seem like I have regained the undeformed constraint algebra up to the sign $\sigma_\beta$ with a simple transformation, it shouldn't be taken to mean that this is actually the algebra of constraints.  That is, the above equation doesn't ensure that $\bar{C}\approx0$ instead of $C\approx0$ when on-shell.  The surfaces in phase space described by $\bar{C}=0$ and $C=0$ are different in general.


\section{Cosmology}
\label{sec:allst_cosmo}

I restrict to an isotropic and homogeneous space to find the background cosmological dynamics, following the definitions in \secref{sec:methodology_cosmo}.  Writing the constraint as ${C=C(a,\psi,\R)}$ where $\R=\R(a,\psi,\bar{p},\pi)$, the equations of motion are given by,
\begin{equation}
\begin{aligned}
    \frac{ \dot{a} }{ N } 
    &= \frac{ 1 }{ 6 a } \partdif{ \R }{ \bar{p} } \partdif{ C }{ \R },
&
    \frac{ \dot{\bar{p}} }{ N } 
    & = \frac{ - 1 }{ 6 a } \left( 
        \partdif{ C }{ a }
        + \partdif{ \R }{ a } \partdif{ C }{ \R }
    \right),
\\
    \frac{ \dot{\psi} }{ N }
    & = \partdif{ \R }{ \pi } \partdif{ C }{ \R },
&
    \frac{ \dot{\pi} }{ N }
    & = - \partdif{ C }{ \psi }
        - \partdif{ \R }{ \psi } \partdif{ C }{ \R },
\end{aligned}
\end{equation}
into which I can substitute 
$\displaystyle{\partdif{C}{\R}=a^3\sqrt{\left|\beta\right|}}$.
When I assume minimal coupling ($\omega_R'=0$, $\omega_\psi'=0$) and time-symmetry ($\xi=0$), the equations of motion become,
\begin{equation}
\begin{gathered}
    \R \to
    \frac{ - 3 \sigma_\beta \bar{p}^2 }{ \omega_R a^2 }
    - \frac{ 3 k \omega_R }{ a^2 }
    + \frac{ \sigma_\beta \pi^2 }{ 2 \omega_\psi a^6 },
\\
    \frac{ \dot{a} }{ N } 
    = \frac{ - \sigma_\beta \bar{p} }{ \omega_R } \sqrt{ \left| \beta \right| },
\quad
    \frac{ \dot{\psi} }{ N }
    = \frac{ \sigma_\beta \pi }{ \omega_\psi a^3 } \sqrt{ \left| \beta \right| },
\quad
    \frac{ \dot{\pi} }{ N }
    = - \partdif{ C }{ \psi },
\\
    \frac{ \dot{\bar{p}} }{ N } 
    = \frac{ - 1 }{ 6 a } \partdif{ C }{ a }
    - a \sqrt{ \left| \beta \right| } \left(
        \frac{ \sigma_\beta \bar{p}^2 }{ \omega_R a^2 }
        + \frac{ k \omega_R }{ a^2 }
        - \frac{ \sigma_\beta \pi^2 }{ 2 \omega_\psi a^6}
    \right) .
\end{gathered}
    \label{eq:allst_flrw_eom}
\end{equation}
To find the Friedmann equation, find the equation for $\mathcal{H}^2/N^2$, and substitute in for $\R$,
\begin{equation}
    \frac{ \mathcal{H}^2 }{ N^2 } 
    = \left| \beta \right| \frac{ \bar{p}^2 }{ \omega_R^2 }
    = \beta \left( \frac{ - \R }{ 3 \omega_R } - \frac{ k }{ a^2 } + \frac{ \sigma_\beta \pi^2 }{ 6 \omega_R \omega_\psi a^6 } \right),
        \label{eq:allst_friedmann_basic}
\end{equation}
and when the constraint is solved, $C\approx0$, then $\R$ can be found in terms of $C_\0$ .

\subsection{Cosmology with a perfect fluid}
\label{sec:allst_cosmo_fluid}

\begin{figure}[t]
\begin{center}
	{\subfigure[$\beta=\beta_\0\left(1+\beta_2\R\right)^n$]{
		\label{fig:allst_friedmann_lin}
		\includegraphics[width=0.44\textwidth]{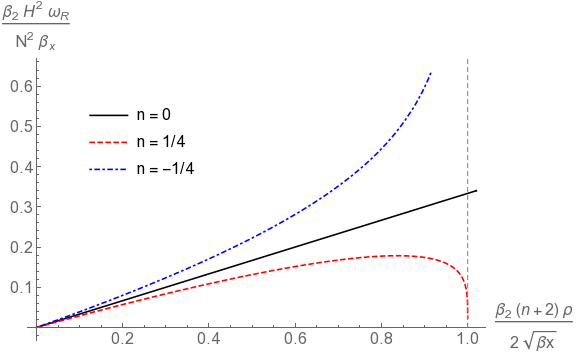}}}
	{\subfigure[$\beta\sim\exp{\R}$]{
		\label{fig:allst_friedmann_exp}
		\includegraphics[width=0.44\textwidth]{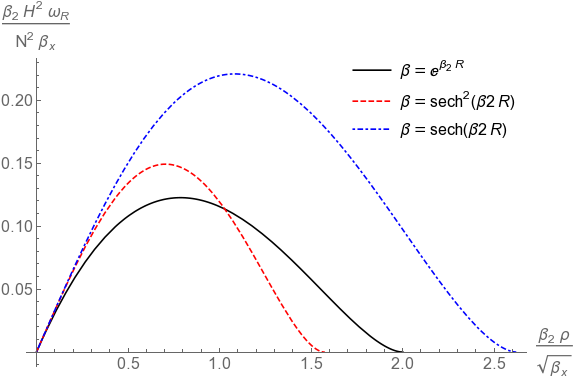}}}
\end{center}
    \caption[The Friedmann equation for various deformation functions $\beta(\R)$]
    {Behaviour of the Friedmann equation for various deformation functions $\beta(\R)$ when $k=0$.}
    \label{fig:allst_friedmann}
\end{figure}

I here find the deformed Friedmann equations for various forms of the deformation.  For simplicity, I ignore the scalar field and include a perfect fluid $C_\0=a^3\rho(a)$.
From the deformation function $\beta=\beta_\0\left(1+\beta_2\R\right)^n$, solving the constraint \eqref{eq:allst_deftocon_linear} gives
\begin{equation}
    \R = 
    \left\{
    \begin{aligned}
    & \frac{ \sigma_2 }{ \beta_2 } \left\{ 
        \sigma_2 - \frac{ \left( n + 2 \right) \sigma_2 \beta_2 \rho }{ 2 \sqrt{ \left| \beta_\0 \right| } } 
    \right\}^{ \frac{ n + 2 }{ 2 } } 
    - \frac{ 1 }{ \beta_2 },
    & n \neq -2,
\\ &
    \frac{ \sigma_2 }{ \beta_2 } \exp \left( \frac{ - \sigma_2 \beta_2 \rho }{ \sqrt{ \left| \beta_\0 \right| } } \right)
    - \frac{ 1 }{ \beta_2 },
    & n = -2,
    \end{aligned}
    \right.
        \label{eq:allst_R_lin}
\end{equation}
where $\sigma_\0:=\sgn{\beta_\0}$ and $\sigma_2:=\sgn{1+\beta_2\R}$.  When I simplify by assuming $\sigma_2=1$, the Friedmann equation is given by,
\begin{equation}
    \frac{ \mathcal{H}^2 }{ N^2 } = \left\{
    \begin{aligned}
    & \left( \frac{ \beta_\0 }{ 3 \omega_R \beta_2 } \left[
        1 - \left( 
            1 - \frac{ \rho }{ \rho_c (n) } 
        \right)^{\frac{2}{n+2}}
    \right] - \frac{ k \beta_\0 }{ a^2 } \right) \left( 
        1 - \frac{ \rho }{ \rho_c (n) } 
    \right)^{\frac{2n}{n+2}},
    & n \neq -2,
\\&
    \left( \frac{ \beta_\0 }{ 3 \omega_R \beta_2 } \left[
        1 - \exp{ \left( \frac{ - \beta_2 \rho }{ \sqrt{ \left| \beta_\0 \right| } } \right) }
    \right] - \frac{ k \beta_\0 }{ a^2 } \right)
    \exp{ \left( \frac{ 2 \beta_2 \rho }{ \sqrt{ \left| \beta_\0 \right| } } \right) }
    & n = -2,
    \end{aligned}
    \right.
        \label{eq:allst_friedmann_lin}
\end{equation}
where 
$\displaystyle{\rho_c(n) = \frac{ 2 \sqrt{ \left| \beta_\0 \right| } }{ \beta_2 \left( n + 2 \right) } }$.
To see the behaviour of the modified Friedmann equation for different values of $n$, look at \figref{fig:allst_friedmann_lin}.  For ${n>0}$, the Hubble rate vanishes as the universe approaches the critical energy density, this indicates that a collapsing universe reaches a turning point at which point the repulsive effect causes a bounce.  For ${0>n>-2}$, there appears a sudden singularity in $\mathcal{H}$ at finite $\rho$ (therefore finite $a$).
In the ${\rho\to\infty}$ limit, ${\mathcal{H}^2\sim{}e^{2\rho}}$ when ${n=-2}$ and ${\mathcal{H}^2\sim\rho^{\frac{2n}{n+2}}}$ when ${n<-2}$.

The singularities for $0>n>-2$ appear to be similar to sudden future singularities characterised in \cite{Cattoen:2005dx, FernandezJambrina:2006hj}.  However, the singularities here might instead be called sudden `past' singularities as they happen when $a$ is small (but non-zero) and $\rho$ is large.  Moreover, they happen for any perfect fluid with $w>-1$, i.e. including matter and radiation.

For the deformation function 
${\beta=\beta_\0\exp{\left(\beta_2\R\right)}}$ 
from \eqref{eq:allst_deftocon_exp}, solving the constraint gives,
\begin{equation}
    \R = \frac{ 2 }{ \beta_2 } \log \left(
        1 - \frac{ \beta_2 \rho }{ 2 \sqrt{ \left| \beta_\0 \right| } }
    \right),
        \label{eq:allst_R_exp}
\end{equation}
and the Friedmann equation is given by,
\begin{equation}
    \frac{ \mathcal{H}^2 }{ N^2 } = \left\{
        \frac{ - 2 \beta_\0 }{ 3 \omega_R \beta_2 } \log{ \left(
            1 - \frac{ \beta_2 \rho }{ 2 \sqrt{ \left| \beta_\0 \right| } }
        \right) } - \frac{ k \beta_2 }{ a^2 }
    \right\} \left(
        1 - \frac{ \beta_2 \rho }{ 2 \sqrt{ \left| \beta_\0 \right| } }
    \right)^2.
        \label{eq:allst_friedmann_exp}
\end{equation}
and a critical density appears for 
$\displaystyle{\rho\to\frac{2\sqrt{\left|\beta_\0\right|}}{\beta_2}}$.

For the deformation function
${\beta=\beta_\0\sech^2\left(\beta_2\R\right)}$ 
from \eqref{eq:allst_deftocon_sech2}, solving the constraint gives,
\begin{equation}
    \R = \frac{ - 1 }{ \beta_2 } \gud^{-1} \left( 
        \frac{ \beta_2 \rho }{ \sqrt{ \left| \beta_\0 \right| } } 
    \right).
        \label{eq:allst_R_sech2}
\end{equation}
Substituting this back into the deformation function gives,
\begin{equation}
    \beta = \beta_\0 \cos^2{ \left( 
        \frac{ \beta_2 \rho }{ \sqrt{ \left| \beta_\0 \right| } } 
    \right) }
        \label{eq:allst_def_sech2}
\end{equation}
and the Friedmann equation is given by
\begin{equation}
    \frac{ \mathcal{H}^2 }{ N^2 } = \left\{
        \frac{ \beta_\0 }{ 3 \omega_R \beta_2 } \gud^{-1} \left( 
            \frac{ \beta_2 \rho }{ \sqrt{ \left| \beta_\0 \right| } } 
        \right) - \frac{ k \beta_\0 }{ a^2 }
    \right\} 
    \cos^2{ \left( 
        \frac{ \beta_2 \rho }{ \sqrt{ \left| \beta_\0 \right| } } 
    \right) }.
        \label{eq:allst_friedmann_sech2}
\end{equation}
where there is a critical density%
\footnote{where $\pi_{\circ}\approx 3.14$.},
$\displaystyle{\rho\to\frac{\pi_\circ\sqrt{\left|\beta_\0\right|}}{2\beta_2}}$.  

These exponential-type deformation functions that I consider all predict a upper limit on energy density.  To illustrate this, I plot the modified Friedmann equations for these functions in \figref{fig:allst_friedmann_exp}.


\subsection{Cosmology with a minimally coupled scalar field}
\label{sec:allst_cosmo_scalar}

Since the metric and scalar kinetic terms must combine into one quantity, $\R$, a deformation function should not affect the relative structure between fields.  To illustrate this, take a free scalar field (without a potential) which is minimally coupled to gravity, and assume no perfect fluid component.  This means that the generalised potential term $C_\0$ will vanish, in which case solving the constraint, ${C\approx0}$, merely implies $\R=0$.  Consequently, since the deformation function $\beta$ is a function of $\R$, the only deformation remaining will be the zeroth order term ${\beta=\beta_\0\left(q,\psi\right)}$.  Combining the equations of motion \eqref{eq:allst_flrw_eom} allows me to find the Friedmann equation,
\begin{equation}
    \frac{ \mathcal{H}^2 }{ N^2 }
    = \frac{ \omega_\psi \dot{\psi}^2 }{ 6 \omega_R N^2 }
    - \frac{ k \beta_\0 }{ a^2 },
\end{equation}
that is, the minimally-deformed case.
For $\beta\neq\beta_\0$, it is required that $\R$ must not vanish, which itself requires that $C_\0$ must be non-zero.  Therefore, for the dynamics to depend on a deformation which is a function of curvature, there must be a non-zero potential term which acts as a background against which the fields are deformed.


\subsection{Deformation correspondence}
\label{sec:allst_deformation}

As discussed in the perturbative action chapter \ref{sec:pert}, the form of the deformation used in the literature which includes holonomy effects is given by the cosine of the extrinsic curvature \cite{Cailleteau2012a, Mielczarek:2012pf, Cailleteau2013}.  Of particular importance to this is that the deformation vanishes and changes sign for high values of extrinsic curvature.  Since the extrinsic curvature is proportional to the Hubble expansion rate, write the deformation \eqref{eq:pert_cosmo_beta_barrau_specific} here as,
\begin{equation}
    \beta = \beta_\0 \cos \left( \beta_k \mathcal{H} \right).
        \label{eq:allst_cosmo_beta_barrau}
\end{equation}
I wish to find $C(\R)$ and $\beta(\R)$ associated with this deformation of form $\beta(\K)$. To do so, I need to find the relationship between the Hubble parameter ${\mathcal{H}=\dot{a}/a}$ and the momentum $\bar{p}$, and thereby infer the form of $\beta(\R)$.  Then, using \eqref{eq:allst_constraint-solution_integral} I can find the constraint $C(\R)$.
So, using the equations of motion \eqref{eq:allst_flrw_eom}, I find 
\begin{equation}
    h = r \sqrt{\left|\cos{h}\right|},
\quad
    \mathrm{where, }
\quad
    h : = \beta_k \mathcal{H},
\quad
    r : = - \frac{ N \sigma_\beta \bar{p} }{ \omega_R a } \beta_k \sqrt{ \left| \beta_\0 \right| } ,
    \label{eq:allst_cosine_implicit}
\end{equation}
this is an implicit equation which cannot be solved analytically for $h(r)$, and so must be solved numerically.

\begin{figure}[t]
	\begin{center}
	{\subfigure[$\beta\left(h\right)$]{
		\label{fig:allst_cosine_def-h}
		\includegraphics[width=0.40\textwidth]{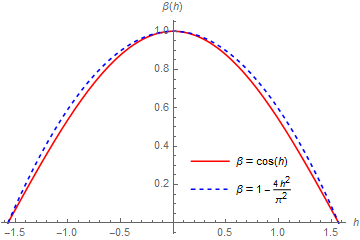}}}
	{\subfigure[$h(r)$]{
	    \label{fig:allst_cosine_def_matching}
		\includegraphics[width=0.40\textwidth]{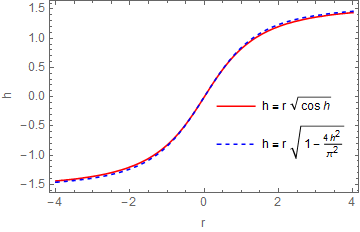}}}
	\\
	{\subfigure[$\beta\left(r\right)$]{
		\label{fig:allst_cosine_def-r}
		\includegraphics[width=0.40\textwidth]{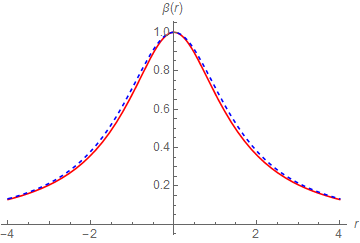}}}
	{\subfigure[$C_k\left(r\right)$]{
		\label{fig:allst_cosine_def_ck}
		\includegraphics[width=0.40\textwidth]{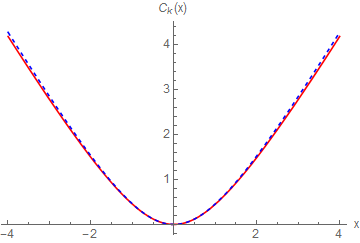}}}
	\end{center}
	\caption[Matching the Hubble expansion and canonical momentum for a cosine deformation]{
	Plot showing the process of starting from a deformation $\beta(h)$ \subref{fig:allst_cosine_def-h}, transforming $h(r)$ \subref{fig:allst_cosine_def_matching}, finding the new form of the deformation $\beta(r)$ \subref{fig:allst_cosine_def-r}, and finding the kinetic part of the constraint $C_k(r)$ \subref{fig:allst_cosine_def_ck}.  I include the function ${\beta=1-4\pi_\circ^{-2}h^{2}}$ (blue dashed line) because it has the same limits as ${\beta=\cos{h}}$ (red solid line) for the region ${|h|\leq\pi_{\circ}/2}$ but the transformation can be done analytically
	}
    	\label{fig:allst_cosine_def}
\end{figure}

For the general relation ${h=r\sqrt{|\beta(h)|}}$, there are similar $\beta$ functions which can be transformed analytically.  One example
is ${\beta(h)=1-4\pi_\circ^{-2}h^2}$,
which also has the same limits of ${\beta(0)=1}$ and ${\beta(h\to\pm\pi_{\circ}/2)=0}$,
and can be transformed to find
${\beta(r)=\left(1+4\pi_\circ^{-2}r^{2}\right)^{-1}}$.
In \figref{fig:allst_cosine_def}, I plot $\beta(h)$ and $h(r)$ in the region ${\left|h\right|\leq\pi_{\circ}/2}$.  After making the transformation, I find $\beta(r)$.
Note that, unlike for $h$, $\beta$ does not vanish for finite $r$.  So it seems that a deformation which vanishes for finite extrinsic curvature does not necessarily vanish for finite intrinsic curvature or metric momenta (at least not in the isotropic and homogeneous case).
In this respect, it matches the dynamics found for exponential-form deformations in \figref{fig:allst_friedmann}.

\begin{figure}[t]
	\begin{center}
	{\subfigure[$\beta\left(h\right)$]{
		\label{fig:allst_cosine_def2-h}
		\includegraphics[width=0.40\textwidth]{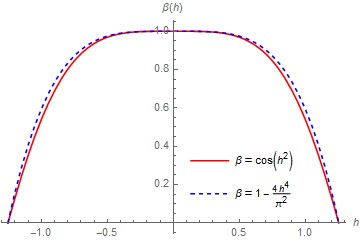}}}
	%
	{\subfigure[$\beta\left(r\right)$]{
		\label{fig:allst_cosine_def2-r}
		\includegraphics[width=0.40\textwidth]{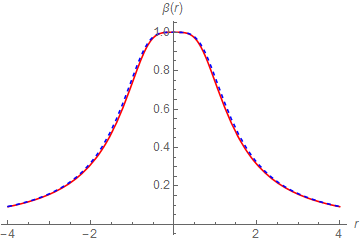}}}
	%
	%
	\end{center}
	\caption[]{
	Plot showing transformations for the deformations given by ${\beta(h)=\cos{h^2}}$ (red solid line) and ${\beta(h)=1-4\pi_{\circ}^{-2}h^4}$ (blue dashed line).
	}
    	\label{fig:allst_cosine_def2}
\end{figure}

Returning to the solution for the constraint, \eqref{eq:allst_constraint-solution_integral}, reducing it to depending on only $a$ and $\bar{p}$ gives
\begin{equation}
    C = \frac{ - 6 a }{ \omega_R } \int_0^{\bar{p}} 
    \mathrm{d} p' \,
    \sigma_\beta p'
    \sqrt{ \left| \beta \left( a, p' \right) \right| } 
    + C_\0 \left( a \right),
        \label{eq:allst-constraint-solution-cosmo}
\end{equation}  
and transforming from $\bar{p}$ to $r$ as defined in \eqref{eq:allst_cosine_implicit}, while making the assumptions $\sigma_\beta=1$, $N=1$, $\beta_\0=1$, and $\beta_k\sim\mathrm{constant}$, this becomes
\begin{equation}
    C = \frac{ - 6 \omega_R a^3 }{ \beta_k^2 } C_k (r) + C_\0 (a),
\quad
    C_k (r) := \int_0^{r} \mathrm{d} r' \; r' \sqrt{ \beta ( r' ) } .
\end{equation}
I numerically integrate the solution for $\beta(r)$ found for when $\beta=\cos(h)$.  I plot the function $C_k(r)$ in \figref{fig:allst_cosine_def_ck}. 

If instead of the extrinsic curvature itself, the deformation is a cosine of the standard extrinsic curvature contraction, $\beta=\cos{\beta_k\K}\sim\cos{h^2}$, it still cannot be transformed analytically.  However, it does match the function $\beta(h)=1-4\pi_{\circ}^{-2}h^4$ well, as I have plotted in \figref{fig:allst_cosine_def2}.  However, numerically finding the constraint for these two deformations, then considering the low $\R$ limit, I see that $C\sim\R^2+C_\0$.  Therefore, this deformation can be ruled out if $C\sim\R+C_\0$ is known to be the low curvature limit of the Hamiltonian constraint.

Considering the function $\beta(h)=1-4\pi_{\circ}^{-2}h^2$ in \figref{fig:allst_cosine_def}, transforming from $h$ to $\K$ and from $r$ to $\R$ to $R$, we can see the correspondence between different limits of the deformation function,
\begin{equation}
    \beta \left( \K, 0 \right) = 1 - \beta_2 \K,
\quad \to \quad
    \beta \left( 0, R \right) = \frac{ 1 }{ 1 + \beta_2 R }.
\end{equation}
This is what I found in \chapref{sec:allact}, where the general form of this particular deformation is actually the product of these two limits.  However, for non-linear deformation functions, $\beta(\K,R)$ cannot be determined so easily from $\beta(\K,0)$ and $\beta(0,R)$. 
That being said, given $\beta(\R)$, the dependence on $\K$ could be found by simply solving and evolving the equations of motion.



\section{Discussion}
\label{sec:allst_discussion}

In this chapter, I have found the general form that a deformed constraint can take for non-minimally coupled scalar-tensor variables.  The momenta and spatial derivatives for all fields must maintain the same relative structure in how they appear compared to the minimally-deformed constraint.  This means that the constraint is a function of the fields and the general kinetic term $\R$.  The freedom within this kinetic term comes down to the coupling functions.
While a lapse function transformation can apparently take the constraint algebra back to the undeformed form, this seems to be merely a cosmetic change as it does not in fact alter the Hamiltonian constraint itself.

I have shown how to obtain the cosmological equations of motion, and given a few simple examples of how they are modified.  For some deformation functions, a upper bound on energy density appears, which probably generates a cosmological bounce.  For other deformation functions, a sudden singularity in the expansion appears when the deformation diverges for high densities.  I have shown that deformations to the field dynamics requires a background general potential against which the deformation must be contrasted.

Using the cosmological equations of motion, I made contact with the holonomy-generated deformation which is a cosine of the extrinsic curvature.  Through this, I have demonstrated how the relationship of momenta and extrinsic curvature becomes non-linear with a non-trivial deformation.  It seems that when the deformation produces an upper bound on extrinsic curvature, there does not seem to be an upper bound on intrinsic curvature or momenta.

\chapter{Deformed gravitational action to all orders}
\label{sec:allact}

As shown in \secref{sec:methodology_order}, the deformed action must be calculated either perturbatively, as has been done in \chapref{sec:pert}, or completely generally.  It appears that this is because it does not permit a closed polynomial solution when the deformation depends on curvature.  In this chapter I attempt this general calculation.  This material has been subsequently published in ref.~\cite{cuttellconstraint}.

Take the equations \eqref{eq:pert_d2theta} and \eqref{eq:pert_dtheta}, which solve the distribution equation for the gravitational action when I expand it in terms of the variables $(q,v,w,R)$, and see what can be deduced about the action when it is treated non-perturbatively.

Start with the equation for $\partial^aF\partial^b\theta_{ab}$ where $F\in\{v,w,R\}$, \eqref{eq:pert_dtheta_8}, this can be rewritten as
\begin{equation}
    0 = \beta \left( \partdif{ L }{ w } \right)^2 
    \partdif{}{ F } \log \left\{ \beta \left( \partdif{ L }{ w } \right)^2 \right\},
\end{equation}
which implies that
\begin{equation}
    \beta \left( \partdif{ L }{ w } \right)^2 = \lambda_1(q),
\end{equation}
and so I can solve up to a sign, $\displaystyle\sigma_L:=\mathrm{sgn}\left(\partdif{L}{w}\right)$,
\begin{equation}
    \partdif{ L }{ w } = \sigma_L \sqrt{ \left| \frac{ \lambda_1 }{ \beta } \right| }. 
    \label{eq:allact_dLdw}
\end{equation}
Then, from $Q^{abcd}\partial_{cd}\theta_{ab}$, \eqref{eq:pert_d2theta_2}, I find
\begin{equation}
    \partdif{ L }{ R } = 4 \beta \partdif{ L }{ w } 
    = 4 \sigma_L \sigma_\beta \sqrt{ \left| \lambda_1 \beta \right| },
    \label{eq:allact_dLdR}
\end{equation}
where $\sigma_\beta:=\sgn{\beta(q,v,w,R)}$.
If I then compare the second derivative of the action, 
$\displaystyle{\partdif{^2L}{w\partial{R}}}$, 
using both equations, I find a nonlinear partial differential equation for the deformation function,
\begin{equation}
    0 = \partdif{ \beta }{ R } + 4 \beta \partdif{ \beta }{ w },
    \label{eq:allact_def_pde}
\end{equation}
which is the same form as Burgers' equation for a fluid with vanishing viscosity \cite{smoller}.  However, before I attempt to interpret this, I will find further restrictions on the action and deformation.

I now seek to find how the trace of the metric's normal derivative, $v$, appears.
Take the condition for $v_\T^{ab}\partial^2\theta_{ab}$, \eqref{eq:pert_d2theta_4}
\begin{equation}
    0 = \frac{v}{3} \partdif{^2 L }{ R \partial w } - \beta \partdif{^2 L }{ v \partial w }
    = \frac{ \sigma_L }{ 2 } \sqrt{ \left| \frac{ \lambda_1 }{ \beta } \right| } \left( 
        \frac{ 4 v }{ 3 } \partdif{ \beta }{ w } + \partdif{ \beta }{ v } 
    \right)
\end{equation}
which I can solve to find that $\beta=\beta\left(q,\bar{w},R\right)$, where $\bar{w}=w-2v^2/3$. So in the deformation, the trace $v$ must always be paired with the traceless tensor squared $w$ like this.  I can see that this is related to the standard extrinsic curvature contraction by $\bar{w}=-4\K$.
To find how the trace appears in the action, I look at the condition from $q^{ab}\partial^2\theta_{ab}$, \eqref{eq:pert_d2theta_1},
\begin{equation}
    0 = \partdif{ L }{ R } - \frac{ 2 v }{ 3 } \partdif{^2 L }{ v \partial R }
    + 2 \beta \left( \partdif{^2 L }{ v^2 } - \frac{2}{3} \partdif{ L }{ w } \right)
\end{equation}
inputting my solutions so far, I can solve for the second derivative with respect to the trace,
\begin{equation}
    \partdif{^2 L }{ v^2 } = \frac{ - 4 \sigma_L }{ 3 } \sqrt{ \left| \frac{ \lambda_1 }{ \beta } \right| } \left(
        1 - \frac{ v }{ 2 } \partdif{ \beta }{ v }
    \right).
\end{equation}
I integrate over $v$ to find the first derivative,
\begin{equation}
    \partdif{ L }{ v } = \frac{ - 4 v \sigma_L}{ 3 } \sqrt{ \left| \frac{ \lambda_1 }{ \beta } \right| } + \xi_1 ( q, w, R )
    = \frac{ - 4 v }{ 3 } \partdif{ L }{ w } + \xi_1 ( q, w, R ).
        \label{eq:allact_dLdv}
\end{equation}
To make sure that the solutions \eqref{eq:allact_dLdw}, \eqref{eq:allact_dLdR} and \eqref{eq:allact_dLdv} match for the second derivatives 
$\displaystyle\partdif{^2L}{v\partial{R}}$ and 
$\displaystyle\partdif{^2L}{v\partial{w}}$, 
I find that $\xi_1=\xi_1(q)$.
Therefore, from this I can see that the action should have the metric normal derivatives appear in the combined form $\bar{w}$ apart from a single linear term $L\supset{}v\xi_1(q)$.

I now just have to see what conditions there are on how the metric determinant appears in the action.
First I have the condition from $X^a\partial^b\theta_{ab}$, \eqref{eq:pert_dtheta_1}, 
\begin{equation}
\begin{split}
    0 & = \partdif{ L }{ R } - 4 \left( \partial_q \beta + 2 \beta \partial_q \right) \partdif{ L }{ w },
\\ &
    = 4 \sigma_L \sigma_\beta \sqrt{ \left| \lambda_1 \beta \right| } \left( 1 - \frac{ \partial_q \lambda_1 }{ \lambda_1 } \right),
\\ &
    \therefore \lambda_1 (q) = q \lambda_2 ,
\end{split}
\end{equation}
and second I have the condition from $v_\T^{ab}X^c\partial_c\theta_{ab}$, \eqref{eq:pert_dtheta_3}, 
\begin{equation}
\begin{split}
    0 & = \frac{v}{3} \left( 4 \partial_q - 1 \right) \partdif{^2 L }{ w \partial R }
    + \partdif{ \beta }{ w } \left( 1 - 2 \partial_q \right) \partdif{ L }{ v }
    + \left( \beta - 2 \partial_q \beta - 4 \beta \partial_q \right) \partdif{^2 L }{ v \partial w },
\\ &
    = \partdif{ \beta }{ w } \left( \xi_1 - 2 \partial_q \xi_1 \right),
    \quad
    \therefore \xi_1 (q) = \xi_2 \sqrt{q},
\end{split}
\end{equation}
and both these results show that my action will indeed have the correct density weight when $\beta\to1$, that is $L\propto\sqrt{q}$.

All the remaining conditions from the distribution equation that have not been explicitly referenced are solved by what I have found so far, so to make progress I must now attempt to consolidate my equations to find an explicit form for the action.
If I integrate \eqref{eq:allact_dLdw}, I find
\begin{equation}
    L = \sigma_L \sqrt{ \left| q \lambda_2 \right| } 
    \int_0^{\bar{w}} 
    \frac{ \mathrm{d} x }{ \sqrt{ \left| \beta ( q, x, R ) \right| } }
    + f_1 ( q, v, R ),
        \label{eq:allact_first_action}
\end{equation}
and then if I match the derivative of this with respect to $v$ with \eqref{eq:allact_dLdv}, I find the $v$ part of the second term,
\begin{equation}
    f_1 ( q , v , R ) = v \xi_2 \sqrt{q} + f_2 ( q , R ).
\end{equation}
If I then match the derivative of \eqref{eq:allact_first_action} with respect to $R$ with \eqref{eq:allact_dLdR}, I see that
\begin{equation}
    \partdif{ L }{ R }
    =
    4 \sigma_L \sigma_\beta \sqrt{ \left| q \lambda_2 \beta \right| } = \partdif{ f_2 }{ R }
    - \frac{\sigma_L}{2} \sqrt{ \left| q \lambda_2 \right| } \int_0^{\bar{w}} \frac{ \sigma_\beta \; \mathrm{d} x }{ \left| \beta(q,x,R) \right|^{3/2} }
    \partdif{}{ R } \beta ( q, x, R )
\end{equation}
and using \eqref{eq:allact_def_pde} to change the derivative of $\beta$,
\begin{equation}
     4 \sigma_L \sigma_\beta \sqrt{ \left| q \lambda_2 \beta \right| }
     = \partdif{ f_2 }{ R }
    + 2 \sigma_L \sqrt{ \left| q \lambda_2 \right| } \int_0^{\bar{w}} \frac{ \mathrm{d} x }{ \sqrt{ \left| \beta ( q, x, R ) \right| } }
    \partdif{}{x} \beta (q, x, R ),
\end{equation}
and so I can change the integration variable,
\begin{equation}
    4 \sigma_L \sigma_\beta \sqrt{ \left| q \lambda_2 \beta \right| }
    = \partdif{ f_2 }{ R }
    + 2 \sigma_L \sqrt{ \left| q \lambda_2 \right| } 
    \int_{\beta(q,0,R)}^{\beta(q,\bar{w},R)} 
    \frac{ \mathrm{d} b }{ \sqrt{|b|} },
\end{equation}
the upper integration limit cancels with the left hand side of the equality, and therefore
\begin{equation}
    \partdif{ f_2 }{ R } = 4 \sigma_L \sgn{ \beta ( q, 0, R ) } \sqrt{ \left| q \lambda_2 \beta ( q, 0, R ) \right| }.
\end{equation}
Then integrating this over $R$,
\begin{equation}
    f_2 ( q, R ) = 
    4 \sigma_L \sqrt{ \left| q \lambda_2 \right| } \int_0^R \sgn{ \beta ( q, 0, r ) } \sqrt{ \left| \beta ( q, 0, r ) \right| } \mathrm{d} r + f_3 ( q ),
\end{equation}
which means that finally I have my solution for the general action,
\begin{equation}
\begin{split}
    L & = \sigma_L \sqrt{ \left| q \lambda_2 \right| } \left(
        \int_0^{\bar{w}} \frac{ \mathrm{d} x }{ \sqrt{ \left| \beta ( q, x, R ) \right| } }
        + 4 \int_0^R \sgn{ \beta ( q, 0, r ) } \sqrt{ \left| \beta ( q, 0, r ) \right| } \mathrm{d} r
    \right)
\\ &
    + v \xi_2 \sqrt{q}
    + f_3 (q).
\end{split}
\end{equation}
Now, I test this with a zeroth order deformation so I can match terms with my previous results. Using $\beta=\beta_\0(q)$,
\begin{equation}
    L = \sigma_L \sqrt{ \left| q \lambda_2 \right| } \left( 
        \frac{ \bar{w} }{ \sqrt{ \left| \beta_\0 \right| } }
        + 4 R \sgn{ \beta_\0 } \sqrt{ \left| \beta_\0 \right| }
    \right)
    + v \xi_2 \sqrt{q} + f_3 (q),
\end{equation}
comparing this to \eqref{eq:pert_action_sol} and using $\bar{w}=-4\K$ leads to
\begin{equation}
    \sigma_L = \sigma_\beta,
\quad
    \sqrt{ \left| \lambda_2 \right| } = \frac{ \omega }{ 8 },
\quad
    \xi_2 = \xi,
\quad
    f_3 = - \sqrt{q} V(q),
\end{equation}
and therefore, the full solution is given by,
\begin{equation}
\begin{split}
    L & = \frac{ \omega \sigma_\beta \sqrt{ q } }{ 2 } \left( 
        \int_0^R \sgn{ \beta ( q, 0, r ) } \sqrt{ \left| \beta ( q, 0, r ) \right| } \mathrm{d} r
        - \int_0^{\K} \frac{ \mathrm{d} k }{ \sqrt{ \left| \beta ( q, k, R ) \right| } }
    \right) 
\\ &
    + \sqrt{q} \left( v \xi - V (q) \right),
        \label{eq:allact_action}
\end{split}
\end{equation}
and the deformation function must satisfy the non-linear partial differential equation,
\begin{equation}
    0 = \partdif{ \beta }{ R } - \beta \partdif{ \beta }{ \K }.
        \label{eq:allact_def_pde2}
\end{equation}

By performing a Legendre transform, I can see that the Hamiltonian constraint associated with this action is given by,
\begin{equation}
\begin{split}
    C & = \frac{ \omega \sigma_\beta \sqrt{q} }{ 2 }
    \left\{ 
        \int_0^{\K} \frac{ \mathrm{d} k }{ \sqrt{ \left| \beta ( q, k, R ) \right| } }
        - \frac{ 2 \K }{ \sqrt{ \left| \beta ( q, \K, R ) \right| } }
\right. \\ & \left.
        - \int_0^R \sgn{ \beta ( q, 0, r ) } \sqrt{ \left| \beta ( q, 0, r ) \right| } \mathrm{d} r
    \right\}
    + \sqrt{q}\, V,
        \label{eq:allact_constraint}
\end{split}
\end{equation}

\section{Solving for the deformation}
\label{sec:allact_def}

The nonlinear partial differential equation for the deformation function is an unexpected result, and invites a comparison to a very different area of physics.
I can compare it to Burgers' equation for nonlinear diffusion, \cite{smoller},
\begin{equation}
    \partdif{ u }{ t } + u \partdif{ u }{ x } = \eta \partdif{^2 u }{ x^2 },
        \label{eq:burgers}
\end{equation}
(where $u$ is a density function), and see that the deformation equation is very similar to the limit of vanishing viscosity $\eta\to0$.
This equation is not trivial to solve because it can develop discontinuities where the equation breaks down, termed `shock waves'.
Returning to my own equation \eqref{eq:allact_def_pde2}, I analyse its characteristics. It implies that there are trajectories parameterised by $s$ given by
\begin{equation}
    \totdif{ q }{ s } = 0,
\quad
    \totdif{ R }{ s } = 1,
\quad
    \totdif{ \K }{ s } = - \beta \left( q, \K, R \right),
\end{equation}
along which $\beta$ is constant.  These trajectories have gradients given by,
\begin{equation}
    \totdif{ R }{ \K } = \frac{-1}{\beta\left(q,\K,R\right)}
\end{equation}
and because $\beta$ is constant along the trajectories, they are a straight line in the $(\K,R)$ plane.  I must have an `initial' condition in order to solve the equation, and because $R$ is here the analogue of $-t$ in \eqref{eq:burgers} I define the initial function when $R=0$, given by $\beta(q,\K,0)=:\alpha(q,\K)$.
Since there are trajectories along which $\beta$ is constant, I can use $\alpha$ to solve for $R(\K)$ along those curves, given an initial value $\K_0$,
\begin{equation}
    R = \frac{ \K_0 - \K }{ \alpha ( \K_0 ) }.
\end{equation}
Reorganising to get, $\K_0=\K+R\alpha(\K_0)$, and then substituting into $\beta$, this leads to the implicit relation,
\begin{equation}
    \beta ( q, \K, R ) = \alpha \left( q,
        \K + R \beta ( q, \K, R )
    \right).
\end{equation}
I invoke the implicit function theorem to calculate the derivatives of $\beta$,
\begin{equation}
    \partdif{ \beta }{ \K } = \frac{ \alpha' }{ 1 - R \alpha' },
\quad
    \partdif{ \beta }{ R } = \frac{ - \beta \alpha' }{ 1 - R \alpha' },
\end{equation}
which show that a discontinuity develops when $R\alpha'\to1$.  This is the point where the characteristic trajectories along which $\beta$ is constant converge to form a caustic. Beyond this point, $\beta$ seems to become a multi-valued function.

\begin{figure}[t]
	\begin{center}
	{\subfigure[$\beta$]{
		\label{fig:allact_def_num_tanhbeta}
		\includegraphics[width = 0.3\textwidth]{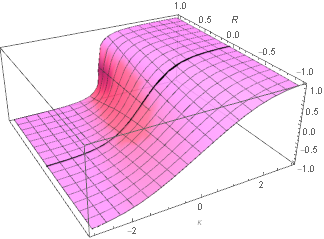}}}
	{\subfigure[$\partial\beta/\partial\K$]{
		\label{fig:allact_def_num_tanhdbetadk}
		\includegraphics[width = 0.3\textwidth]{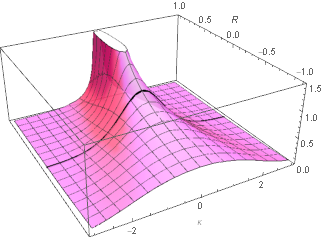}}}
	{\subfigure[$\partial\beta/\partial{R}$]{
		\label{fig:allact_def_num_tanhdbetadr}
		\includegraphics[width = 0.3\textwidth]{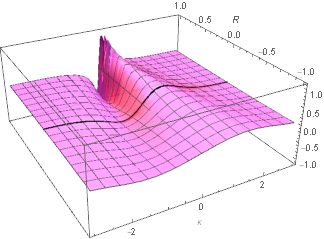}}}
	\end{center}
\caption[Numerically solved deformation with initial function \texorpdfstring{$\alpha=\tanh{\mathcal{\omega{K}}}$}{\pm tanh(K)}]
{Numerically solved deformation function for initial function $\alpha=\tanh{(\omega\K)}$.  The numerical evolution breaks for $R>\omega$ because a discontinuity has developed.  The initial function is indicated by the black line. The plots are in $\omega=1$ units.
}
	\label{fig:allact_def_num_tanh}
\end{figure}

An analytic solution to $\beta$ only exists when $\alpha$ is linear,
\begin{equation}
    \alpha = \alpha_1 (q) + \alpha_2 (q) \K,
\quad
    \beta = \frac{ \alpha_1 (q) + \alpha_2 (q) \K }{ 1 - \alpha_2 (q) R },
        \label{eq:allact_def_sol}
\end{equation}
and when $\alpha_2(q)$ is small, I can expand $\beta$ into a series,
\begin{equation}
    \beta \simeq \alpha_1 + \alpha_2 \left( \K + \alpha_1 R \right) \sum_{n=0}^{\infty} R^n \alpha_2^n,
        \label{eq:allact_def_sol_expand}
\end{equation}
and by comparing this to the perturbative deformation found previously, \eqref{eq:pert_def_sol}, I can see the correspondence $\alpha_1=\beta_\0$ and 
$\alpha_2=\varepsilon^2\beta_{(R)}/\beta_\0=\varepsilon^2\beta_2$.
For other initial functions, I must numerically solve the deformation.  As a test, in \figref{fig:allact_def_num_tanh}, I numerically solve for $\beta$ when $\alpha=\tanh{(\omega\K)}$.  I see that, as $R$ increases, the positive gradient in $\K$ intensifies to form a discontinuity, and softens as $R$ decreases.

\begin{figure}[t]
	\begin{center}
	{\subfigure[$\beta$]{
		\label{fig:allact_def_num_cosbeta}
		\includegraphics[width = 0.333\textwidth]{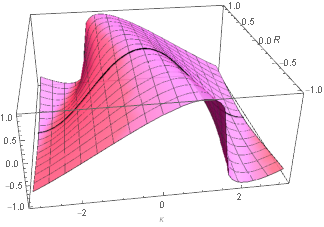}}}
	{\subfigure[$\partial\beta/\partial\K$]{
		\label{fig:allact_def_num_cosdbetadk}
		\includegraphics[width = 0.333\textwidth]{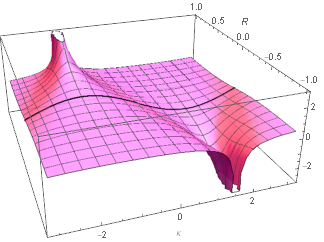}}}
\\
	{\subfigure[$\partial\beta/\partial{R}$]{
		\label{fig:allact_def_num_cosdbetadr}
		\includegraphics[width = 0.333\textwidth]{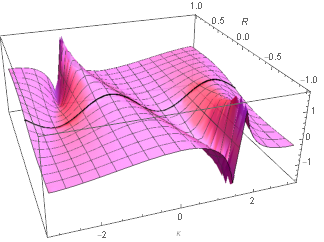}}}
	{\subfigure[$\beta$ cross sections]{
		\label{fig:allact_def_num_cosbetacross}
		\includegraphics[width = 0.333\textwidth]{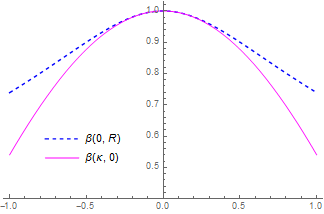}}}
	\end{center}
\caption[Numerically solved deformation with initial function \texorpdfstring{$\alpha=\cos{(\omega\K)}$}{cos(K)}]
{Numerically solved deformation function with an initial function $\alpha=\cos{(\omega\K)}$ and periodic boundary conditions.  The numerical evolution breaks for $|R|>\omega$ because discontinuities have developed.  The initial function is indicated by the black line.  The plots are in $\omega=1$ units.
}
	\label{fig:allact_def_num_cos}
\end{figure}

I have also numerically solved for the deformation when the initial function is given by $\alpha=\cos{(\omega\K)}$, shown in \figref{fig:allact_def_num_cos}.
This function is  motivated by loop quantum cosmology models with holonomy corrections \cite{Cailleteau2012a, Mielczarek:2012pf, Cailleteau2013}.  As with the $\tanh$ numerical solution in \figref{fig:allact_def_num_tanh}, I see the positive gradient intensify and the negative gradient soften.  I could not evolve the equations past the formation of the shock wave so I cannot say for certain whether a periodicity emerges in $R$, but I can compare the cross sections for $\beta$ in \figref{fig:allact_def_num_cosbetacross}.  

This cross section appears to match what was found in \secref{sec:allst_deformation} when I attempted to find the correspondence between $\beta(\K,0)$ and $\beta(\R)$.  It would seem that $\beta(0,R)$ should be a non-vanishing function of the shape as shown in \figref{fig:allst_cosine_def-r}.

When the inviscid Burgers' equation is being simulated in the context of fluid dynamics, a choice must be made on how to model the shock wave \cite{smoller}.  The direct continuation of the equation means that the density function $u$ becomes multi-valued, and the physical intepretation of it as a density breaks down.  The alternative is to propagate the shock wave as a singular object, which requires a modification to the equations.  

Considering my case of the deformation function, allowing a shock wave to propagate does not seem to make sense.  It might require being able to interpret $\beta$ as a density function and the space of $\left(\K,R\right)$ to be interpreted as a medium. 
Whether or not the shock wave remains singular or becomes multi-valued, the most probable interpretation is that it represents a disconnection between different branches of curvature configurations.  That is, for a universe to transition from one side of the discontinuity to the other may require taking an indirect path through the phase space.


\section{Linear deformation}
\label{sec:allact_linear}

If I take the analytic solution for the deformation function when its initial condition is linear \eqref{eq:allact_def_sol}, I can substitute it into the general form for the gravitational action \eqref{eq:allact_action}.  If I assume I am in a region where $1-\alpha_2R>0$, I get the solution,
\begin{equation}
\begin{split}
    L &= \frac{ \omega \sqrt{q} }{ \alpha_2 } \left\{
        \mathrm{sgn} \left( 1 + \frac{ \alpha_2 \K }{ \alpha_1 } \right) \sqrt{ \left| \alpha_1  \right| }
        - \sqrt{ \left| \alpha_1 \! + \! \alpha_2 \K \right| }
        \sqrt{ \left| 1 \! - \! \alpha_2 R \right| }
    \right\}
    + \sqrt{q} \left( v \xi \! - \! V \right),
        \label{eq:allact_action_linear}
\end{split}
\end{equation}
and expanding in series for small $\alpha_2$ when I am in a region where $|\alpha_1|\gg|\alpha_2\K|$,
\begin{equation}
    L = \frac{ \omega }{ 2 } \sqrt{ q \left| \alpha_1 \right| } \left(
        R - \frac{ \K }{ \alpha_1 }
        - \frac{ \alpha_2 }{ 4 } \left( R + \frac{ \K }{ \alpha_1 } \right)^2
        + \mathcal{O} \left( \alpha_2^3 \right)
    \right)
    + \sqrt{q} \left( v \xi - V \right),
        \label{eq:allact_action_linear_pert}
\end{equation}
which matches exactly the fourth order perturbative action I found previously \eqref{eq:pert_action_sol}.
The Hamiltonian constraint associated with the non-perturbative action can be found from \eqref{eq:allact_constraint}, and then I can solve for $\K$ when the constraint vanishes (as long as I specify that it must be finite in the limit $\alpha_2\to0$),
\begin{equation}
    \K = \left\{
        \frac{ 2 }{ \omega } \sgn{ \alpha_1 } \sqrt{ \left| \alpha_1 \right| } V \left(
            1 - \frac{ \alpha_2 V }{ 2 \omega \sqrt{ \left| \alpha_1 \right| } }
        \right)
        - \alpha_1 R
    \right\}
    \left( 1 - \frac{ \alpha_2 V }{ \omega \sqrt{ \left| \alpha_1 \right| } }
    \right)^{-2},
\end{equation}
and if I restrict to the FLRW metric and a perfect fluid as in \secref{sec:methodology_cosmo},
I find the modified Friedmann equation,
\begin{equation}
    \frac{ \mathcal{H}^2 }{ N^2 } = 
    \left\{
        \frac{ \sgn{ \alpha_1 } \sqrt{ \left| \alpha_1 \right| } }{ 3 \omega } \rho \left(
            1 - \frac{ \alpha_2 \rho }{ 2 \omega \sqrt{ \left| \alpha_1 \right| } }
        \right)
        - \frac{ \alpha_1 k }{ a^2 }
    \right\}
        \left( 1 - \frac{ \alpha_2 \rho }{ \omega \sqrt{ \left| \alpha_1 \right| } }
    \right)^{-2}.
    \label{eq:allact_friedmann_linear}
\end{equation}
There is a correction term similar to that found for the fourth order perturbative action which suggests there could be a bounce when $\rho\to2\omega\sqrt{\left|\alpha_1 \right|}/\alpha_2$.  However, there is also an additional factor which causes $\mathcal{H}$ to diverge when $\rho\to\omega\sqrt{\left|\alpha_1\right|}/\alpha_2$, which is before that potential bounce.

This is directly comparable to the modified Friedmann equation found for the deformation function $\beta(\R)=\beta_\0\left(1+\beta_2\R\right)^{-1}$, \eqref{eq:allst_friedmann_lin} investigated in \secref{sec:allst_cosmo_fluid}, with 
${\alpha_1=\beta_\0}$ and
${\alpha_2=\omega\beta_2/2}$.  
As is found here, those results suggested a sudden singularity where $\mathcal{H}$ diverges when $a$ and $\rho$ remain finite.

\section{Discussion}
\label{sec:allact_discussion}

I have found the general form of the deformed gravitation action when considering tensor combinations of derivatives up to second order.  The way in which the deformation, and thereby the action, depends on the extrinsic and intrinsic curvature was found to be highly non-linear.  Curiously, its form matches an equation found in fluid dynamics.  The meaning of this comparison is far from clear.

For different initial functions, I numerically solved for the deformation function until a discontinuity formed.  The meaning of this discontinuity is not clear, but might manifest as a barrier across which paths through phase space cannot cross.

\chapter{Conclusions}
\label{sec:conclusions}

I have attempted to thoroughly investigate the effects that a quantum-motivated deformation to the hypersurface deformation algebra of general relativity has in the semi-classical limit.  Starting from the algebra, I have shown how to regain a deformed gravitational action or a deformed scalar-tensor constraint.

Finding the minimally-deformed version of a non-minimally coupled scalar-tensor model, I was able to establish the classical low-curvature reference point.  I was able to show how the higher-order curvature terms arising from a deformation are qualitatively different from conventional higher-order terms which can absorbed by a non-minimally coupled scalar field.  I also investigated some of the interesting effects which non-minimal coupling has on cosmology.

As a first step towards including higher-order curvature terms coming from a deformation, I derived the fourth order gravitational action perturbatively.  The nearest order corrections demonstrate a change in the relative structure between time and space since the higher order curvature terms appear with a different sign.  I investigated the cosmological implications of the higher order terms, albeit while using the assumption that the action found perturbatively could be extended beyond the perturbative regime.

In attempting to find the deformed scalar-tensor constraint to any order, I was able to show how the momenta and spatial derivatives maintain the same relative kinetic structure.  Interestingly, the way the scalar field and gravitational kinetic terms combine must also be unchanged.  That is to say that higher order gravitational terms are necessarily accompanied by higher order scalar terms of the same form.  The main consequence of this seems to be that a potential term (in a general sense) must be present for a deformation of the kinetic terms to affect the dynamics.
By testing different deformation functions, I was able to show what kinds of cosmological effects should be expected.  Interestingly, the deformations which cause a big bounce seem to be required to vanish, but are not required to change sign.

For the final chapter, I derived the general deformed gravitational action.  The way the deformation function is differently affected by extrinsic and intrinsic curvature (or, equivalently, by time and space derivatives) was found to be similar to a differential equation which usually appears in fluid mechanics.  Discontinuities in the deformation function seem to be inevitable, but the interpretation of what they mean is not clear.
By checking the nearest order perturbative corrections, I was able to validate the perturbative action derived in an earlier chapter.

One of the original motivations of this study was to provide insight into the problem of incorporating spatial derivatives, local degrees of freedom and matter fields into models of loop quantum cosmology which deform space-time covariance.  From my results, it would seem that the problem comes from considering the kinetic terms as separable, or as differently deformed.
The kinetic term, when constructed with canonical variables, cannot have its internal structure deformed beyond a sign.  The deformation can only be a function of the combined term, which means that matter field derivatives deform the space-time covariance in a similar way to curvature.
This may strike at the heart of the way the loop quantisation project, which attempts to first find a quantum theory of gravity, typically adds in matter as an afterthought.

That being said, there are important caveats to this work which must be kept in mind.  The fact that I used metric variables rather than the preferred connection or loop variables might limit the applicability of my results when comparing to the motivating theory.  Moreover, the deformation of the constraint algebra is only predicted for real values of $\BI$.
I also only considered combinations of derivatives or momenta that were a maximum of two orders, when higher order combinations and higher order derivatives are likely to appear in true quantum corrections.

As said in the introduction, \ref{sec:intro}, there are potentially wider implications for this study.
The deformation can lead to a modified dispersion relation, possibly indicating a variable speed of light or an invariant energy scale.  
It might be related to non-classical geometric qualities such a non-commutativity or scale-dependent dimensionality.
In the literature, it is indicated that the deformation function may change sign, implying a transition from a Lorentzian to a Euclidean geometry at high densities.  In such a way, it might be a potential mechanism for the Hartle-Hawking no-boundary proposal.


\appendix 



\chapter{Decomposing the curvature}
\label{sec:curvature}

In our calculations, we need to decompose the three dimensional Riemann curvature frequently, so we collect the relevant identities in this appendix.

The Riemann tensor is defined as the commutator of two covariant derivatives of a vector
\begin{equation}
    \nabla_c \nabla_d A^a - \nabla_d \nabla_c A^a 
    = R^a_{\;\;bcd} A^b,
\end{equation}
and can be given in terms of the Christoffel symbols,
\begin{equation}
    R^a_{\;\;bcd} = 
    \partial_c \Gamma^a_{db} - \partial_d \Gamma^a_{cb}
    + \Gamma^a_{ce} \Gamma^e_{db} - \Gamma^a_{de} \Gamma^e_{cb},
\end{equation}
which are given by 
\begin{equation}
    \Gamma^a_{bc} = q^{ad} \partial_{(b} q_{c)d} - \half \partial^a q_{bc},
\end{equation}

The variation of the Riemann tensor is given by the Palatini equation,
\begin{equation}
	\delta R^a_{\;\,bcd} = \nabla_c \delta \Gamma^a_{db} - \nabla_d \delta \Gamma^a_{cb},
	    \label{eq:var_riemann_1}
\end{equation}
where the variation of the connection is
\begin{equation}
	\delta \Gamma^a_{bc} = 
	q^{ad} \nabla_{(b} \delta q_{c)d}
	- \half \nabla^a \delta q_{bc},
	    \label{eq:var_connection}
\end{equation}
from which we can calculate,
\begin{equation}
	\delta R^a_{\;\,bcd} = 
	\Theta^{a \;\;\;\;\; ef}_{\;\,bcd} \delta q_{ef}
	+ \Phi^{a \;\;\;\;\; efgh}_{\;\,bcd} \nabla_{ef} \delta q_{gh}
	    \label{eq:var_riemann_2}
\end{equation}
where we've defined the useful tensors,
\begin{subequations}
\begin{align}
	\Theta^{a \;\;\;\;\; ef}_{\;\;bcd} & =
	\frac{-1}{2} \left( q^{a(e} R^{f)}_{\;\;\;\;bcd} + \delta^{(e}_b R^{f)a}_{\;\;\;\;\;\;\;cd} \right),
	    \label{eq:var_coeff_Theta} \\
	\Phi^{a \;\;\;\;\; efgh}_{\;\;bcd} & = \half \left( q^{a(e} \delta^{f)}_d \delta^{gh}_{bc} + q^{a(g} \delta^{h)}_d \delta^{ef}_{bc} - q^{a(e} \delta^{f)}_c \delta^{gh}_{bd} - q^{a(g} \delta^{h)}_c \delta^{ef}_{bd} \right),
	    \label{eq:var_coeff_Phi}
\end{align}%
    \label{eq:var_coeff}%
\end{subequations}
but contracted versions of these are more useful,
\begin{subequations}
\begin{align}
	\Theta^{cd}_{ab}
	& := \delta^{ef}_{ab} \Theta^{g \;\;\;\;\; cd}_{\;\;egf} = \half \left( Q^{cdef} R_{e(ab)f} + \delta^{(c}_{(a} R^{d)}_{b)}	\right),
	\quad q_{cd} \Theta^{cd}_{ab} = 0,
	\quad q^{ab} \Theta^{cd}_{ab} = 0,
	    \label{eq:var_coeff_contract_Theta} \\
	\Phi_{ab}^{cdef}
	& := \delta^{gh}_{ab} \Phi^{i \;\;\;\;\; cdef}_{\;\;gih} = \half \left( q^{c(e} \delta^{f)d}_{ab} + q^{d(e} \delta^{f)c}_{ab} - q^{cd} \delta^{ef}_{ab} - q^{ef} \delta^{cd}_{ab} \right),
	    \label{eq:var_coeff_contract_Phi} \\
	\Phi^{abcd}
	& := q^{ef} \Phi^{abcd}_{ef} = Q^{abcd} - q^{ab} q^{cd}.
	    \label{eq:var_coeff_dblcontract_Phi}
\end{align}%
    \label{eq:var_coeff_contract}%
\end{subequations}
To decompose the Riemann tensor in terms of partial derivatives, use this formula for decomposing the second covariant derivative of the variation of the metric,
\begin{equation}
\begin{split}
    \nabla_d \nabla_c \delta q_{ab}
    & = \partial_d \partial_c \delta q_{ab} + \partial_g \delta q_{ef} \left( - \Gamma^g_{dc} \delta^{ef}_{ab} - 4 \delta^{(e}_{(a} \Gamma^{f)}_{b)(c} \delta^g_{d)} \right)
        \\
    &  + \delta q_{ef} \left( - 2 \partial_d \Gamma^{(e}_{c(a} \delta^{f)}_{b)} + 2 \Gamma^g_{dc} \Gamma^{(e}_{g(a} \delta^{f)}_{b)} + 2 \Gamma^g_{d(a} \delta^{(e}_{b)} \Gamma^{f)}_{cg} + 2 \Gamma^{(e}_{d(a} \Gamma^{f)}_{b)c} \right).
\end{split}
    \label{eq:var_metric}
\end{equation}

The two equations we need most are the derivative of the Ricci scalar with respect to the first and second spatial derivative of the metric, and we can find these from combining the above equations,
\begin{subequations}
\begin{align}
    \partdif{ R }{ \left( \partial_d \partial_c q_{ab} \right) }
    & =
    \partdif{ \left( \nabla_h \nabla_g q_{ef} \right) }{ \left( \partial_d \partial_c q_{ab} \right)}
    \partdif{ R }{ \left( \nabla_h \nabla_g q_{ef} \right) }
    = \delta_h^d \delta_g^c \delta_{ef}^{ab} \Phi^{efgh} = \Phi^{abcd}
\nonumber \\
    \therefore \partdif{ R }{ q_{ab,cd} } & = 
    \Phi^{abcd} = Q^{abcd} - q^{ab} q^{cd},
\\
    \partdif{ R }{ \left( \partial_c q_{ab} \right) }
    & =
    \partdif{ \left( \nabla_h \nabla_g q_{ef} \right) }{ \left( \partial_c q_{ab} \right)}
    \partdif{ R }{ \left( \nabla_h \nabla_g q_{ef} \right) }
    = \left( 
        - \Gamma^c_{gh} \delta^{ab}_{ef}
        - 4 \delta^{(a}_{(e} \Gamma^{b)}_{f)(g} \delta^c_{h)} 
    \right) \Phi^{efgh},
\nonumber \\
\begin{split}
    \therefore \partdif{ R }{ q_{ab,c} } & =
        \frac{3}{2} Q^{abde} \partial^c q_{de} 
        - Q^{edc(a} \partial^{b)} q_{de}
\\ &
        + q^{ab} Y^c
        - 2 q^{c(b} Y^{a)}
        - \half q^{ab} X^c
        + q^{c(b} X^{a)},
\end{split}
\end{align}
\end{subequations}
where 
$ X_a : = q^{bc} \partial_a q_{bc} $
and
$ Y_a : = q^{bc} \partial_{(c} q_{b)a} = \partial^b q_{ba} $.

\chapter{The general diffeomorphism constraint}
\label{sec:diff}

I start from the assumption that the equal-time slices of our foliation are internally diffeomorphism covariant.  That is to say that spatial transformations and distortions are not deformed by the deformation of the constraint algebra.  As such, the Hamiltonian constraint is susceptible to deformation and the diffeomorphism constraint is not.  Therefore I need to consider what form the diffeomorphism constraint has.
In the hyperspace deformation algebra \eqref{eq:con-alg}, the diffeomorphism constraint forms a closed sub-algebra,
\begin{equation}
    \{ D_a [N^a], D_b [ M^b ] \} = D_a [ \mathcal{L}_M N^a].
\end{equation}
This equation shows that the diffeomorphism constraint is the generator of spatial diffeomorphisms (hence the name),
\begin{equation}
    \{ F, D_a [ N^a ] \} = \mathcal{L}_N F,
    \label{eq:diff_translation}
\end{equation}
for any phase space function $F$.  Using this relation, I can determine the unique form of the constraint for any field content.

For these calculations, I must include the concept of a tensor density, which does not transform under a change of coordinates as a tensor does.  A tensor density of weight $w_\Psi\in\mathbb{R}$ transforms under the change $x^a\to{}x^{\prime{}a^{\prime}}$,
\begin{equation}
    \Psi^{ \prime\, b^{\prime}_1 \ldots b^{\prime}_i }_{ \;\, a^{\prime}_1 \ldots a^{\prime}_j }
    = \left| \det \left( \partdif{ x^c }{ x^{ \prime \, c^{\prime} } } \right) \right|^{w_\Psi}
    \Psi^{ b_1 \ldots b_i }_{ a_1 \ldots a_j }
    \partdif{ x^{ \prime \, b^{\prime}_1 } }{ x^{b_1} }
    \cdots
    \partdif{ x^{ \prime \, b^{\prime}_i } }{ x^{b_i} }
    \partdif{ x^{a_1} }{ x^{ \prime \, a^{\prime}_1 } }
    \cdots
    \partdif{ x^{a_j} }{ x^{ \prime \, a^{\prime}_j } },
\end{equation}
and one can `de-densitise' to find a tensor%
\footnote{a tensor is a tensor density of weight zero, which I sometimes also call weightless.  If something is called a tensor density without any reference to its weight, it is probably of weight one.}
by multiplying it by $q^{-w_\Psi/2}$, because $\sqrt{q}$ is a scalar density of weight one
\cite[p.\~276]{bojowald2010canonical}.
The integration measure ${\mathrm{d}^3x}$ has a weight of ${-1}$, so for an integral to be appropriately tensorial, the integrand must have a weight of ${+1}$, e.g. ${\int\mathrm{d}^3x\sqrt{q}}$.  
Since making a Legendre transformation requires using the term ${\int\mathrm{d}^3x\,\dot{\psi}\,\pi}$ for a conjugate pair $\left(\psi,\pi\right)$, when the variable $\psi$ is of weight $w_\psi$, the momentum $\pi$ is of weight ${1-w_{\psi}}$.

\section{Diffeomorphism constraint for a scalar field}
\label{sec:diff_scalar}

I consider a scalar field $\left(\psi,\pi\right)$.
Take \eqref{eq:diff_translation} with $F=\psi$,
\begin{subequations}
\begin{align}
\begin{split}
    \{ \psi (x), D_a [ N^a ] \}
    & =
    \int \mathrm{d}^3 y N^a (y) \funcdif{ D_a (y) }{ \pi (x) },
\\ &
    = N^a \partdif{ D_a }{ \pi } 
    - \partial_b \left( N^a \partdif{ D_a }{ \pi_{,b} } \right)
    + \partial_{bc} \left( N^a \partdif{ D_a }{ \pi_{,bc} } \right)
    + \ldots
\\ & 
    = N^a \left\{ 
        \partdif{ D_a }{ \pi } 
        - \partial_b \left( \partdif{ D_a }{ \pi_{,b} } \right) 
        + \partial_{bc} \left( \partdif{ D_a }{ \pi_{,bc} } \right)
    \right\}
\\ &
    + \partial_b N^a \left\{ 
        - \partdif{ D_a }{ \pi_{,b} }
        + 2 \partial_c \left( \partdif{ D_a }{ \pi_{,bc} } \right)
    \right\}
    + \partial_{bc} N^a \left( \partdif{ D_a }{ \pi_{,bc} } \right)
    + \ldots,
\end{split}
\\
    \mathcal{L}_N \psi & = N^a \partial_a \psi,
\end{align}%
\end{subequations}
comparing these two equations, one can easily see that 
\begin{equation}
    \partdif{ D_a }{ \pi } = \partial_a \psi,
\quad
    \partdif{ D_a }{ \pi_{,b} } = 0,
\quad
    \partdif{ D_a }{ \pi_{,bc} } = 0.
\end{equation}
Checking what result I get for $F=\pi$ merely produces the same equations
and therefore the diffeomorphism constraint for a scalar field is given by,
\begin{equation}
    D_a = \pi \partial_a \psi .
        \label{eq:diff_scalar}
\end{equation}
I considered up to second order spatial derivatives here as a demonstration, but no diffeomorphism constraint goes beyond first order, so I will not bother with them for further equations below.

\section{Diffeomorphism constraint for a vector}
\label{sec:diff_vector}

I consider a weightless contravariant vector $\left(A^a,P_b\right)$.
Take \eqref{eq:diff_translation} with $F=A^a$, 
\begin{subequations}
\begin{align}
\begin{split}
    \{ A^a (x), D_b [ N^b ] \}
    & = \int \mathrm{d}^3 y N^b (y) \funcdif{ D_b (y) }{ P_a (x) },
\\ &
    = N^b \partdif{ D_b }{ P_a } 
    - \partial_c \left( N^b \partdif{ D_b }{ P_{a,c} } \right)
    + \ldots
\\ & 
    = N^b \left\{ 
        \partdif{ D_b }{ P_a } 
        - \partial_c \left( \partdif{ D_b }{ P_{a,c} } \right) 
    \right\}
    + \partial_c N^b \left(
        - \partdif{ D_b }{ P_{a,c} } 
    \right) + \ldots
\end{split}\\
    \mathcal{L}_N A^a &
    = N^b \partial_b A^a - A^b \partial_{b} N^a,
\end{align}%
\end{subequations}
looking at the derivative of $N^a$, I can see that 
$\partdif{ D_b }{ P_{a,c} } = \delta^a_b A^c$, 
and substituting this back into the equation I find,
$\partdif{ D_b }{ P_a } = \delta_b^a  \partial_c A^c + \partial_b A^a$.  
If I check with $F=P_a$ I find the same equations, leading us to the diffeomorphism constraint 
\begin{equation}
    D_a = P_b \partial_a A^b + \partial_b \left( P_a A^a \right).
        \label{eq:diff_vector}
\end{equation}

\section{Diffeomorphism constraint for a tensor}
\label{sec:diff_tensor}

I consider a rank-2 tensor defined on a three dimensional spatial manifold 
$\left( q_{ab}, p^{cd} \right)$.  
I use the example of the metric, but our result is general.
Test \eqref{eq:diff_translation} using $F=q_{ab}$, 
\begin{subequations}
\begin{align}
\begin{split}
    \{ q_{ab} (x), D_c [ N^c ] \}
    & = \int \mathrm{d}^3 y N^c (y) \funcdif{ D_c (y) }{ p^{ab} (x) },
\\ &
    = N^c \partdif{ D_c }{ p^{ab} } 
    - \partial_d \left( N^c \partdif{ D_c }{ p^{ab}_{,d} } \right)
    + \ldots
\\ & 
    = N^c \left\{ 
        \partdif{ D_c }{ p^{ab} } 
        - \partial_d \left( \partdif{ D_c }{ p^{ab}_{,d} } \right) 
    \right\}
    + \partial_d N^c \left(
        - \partdif{ D_c }{ p^{ab}_{,d} } 
    \right) + \ldots
\end{split}\\
    \mathcal{L}_N q_{ab} &
    = N^c \partial_c q_{ab} + 2 q_{c(b} \partial_{a)} N^c,
\end{align}%
\end{subequations}
looking at the derivative of $N^a$, I can see that 
$\partdif{ D_c }{ p^{ab}_{,d} } = - 2 q_{c(b} \delta_{a)}^d$, and substituting this back into the equation I find,
$\partdif{ D_c }{ p^{ab} } = \partial_c q_{ab} - 2 \partial_{(a} q_{b)c}$.  If I check with $F=p^{ab}$ I find the same equations, leading us to the diffeomorphism constraint 
\begin{equation}
    D_a = p^{bc} \partial_a q_{bc} - 2 \partial_{(c} \left( q_{b)a} p^{bc} \right),
    \label{eq:diff_tensor}
\end{equation}
and for the specific example of the metric, this reduces to 
\begin{equation}
    D_a = - 2 q_{ab} \nabla_c p^{bc}.
    \label{eq:diff_metric}
\end{equation}

\section{Diffeomorphism constraint for a tensor density}
\label{sec:diff_general}

For the general case of a tensor density with $n$ covariant indices, $m$ contravariant indices and weight $w_\Psi$,
$ \left(
\Psi_{ a_1 \cdots a_n}^{ b_1 \cdots b_m},
\Pi^{ c_1 \cdots c_n}_{ d_1 \cdots d_m}
\right)$
where the canonical momentum has weight $1-w_\Psi$,
the associated diffeomorphism constraint is given by,
\begin{equation}
\begin{split}
    D_a & = 
    \Pi^{ b_1 \cdots b_n}_{ c_1 \cdots c_m} \partial_a 
    \Psi_{ b_1 \cdots b_n}^{ c_1 \cdots c_m}
    - w_\Psi \, \partial_a \left( \Pi^{ b_1 \cdots b_n}_{ c_1 \cdots c_m}
    \Psi_{ b_1 \cdots b_n}^{ c_1 \cdots c_m} \right)
\\ &
    - n \, \partial_{(b_1} \left( 
        \Psi_{ b_2 \cdots b_n) a }^{ c_1 \cdots c_m } 
        \Pi^{ b_1 \cdots b_n}_{ c_1 \cdots c_m} 
    \right)
    + m \, \partial_{(c_1} \left( 
        \Pi_{ c_2 \cdots c_m) a }^{ b_1 \cdots b_n } 
        \Psi_{ b_1 \cdots b_n}^{ c_1 \cdots c_m}
    \right).
\end{split}
    \label{eq:diff_general}
\end{equation}

\chapter{Fourth order perturbative gravitational action: Extras}
\label{sec:pert_extras}

For convenience, I use the definitions,
\begin{equation}
    X_a = q^{bc} \partial_a q_{bc},
\quad
    Y_a = q^{bc} \partial_c q_{ba} = \partial^b q_{ab},
\quad
    Z_a = v_\T^{bc} \partial_a q_{bc},
\quad
    W_a = v_\T^{bc} \partial_c q_{ba}.
        \label{eq:pert_tensor_combinations}
\end{equation}
Evaluating each term in the $\partial_{cd}\theta_{ab}$ bracket of \eqref{eq:pert_dist-eqn-sol}, by substituting in the variables 
\begin{equation}
    q := \det{q_{ab}},
\quad
    v := q^{ab} v_{ab},
\quad
    w := v_{ab}v^{ab} - \third v^2,
\quad
    R := q^{bc} R^a_{\;\;bac}
\end{equation}
and using the equations derived for decomposing $R$ in \appref{sec:curvature},
\begin{subequations}
\begin{align}
    \partdif{ L }{ q_{ab,cd} }
    & = \left( Q^{abcd} - q^{ab} q^{cd} \right) \partdif{ L }{ R },
\\ 
    v_{ef} \partdif{^2 L }{ q_{ef,cd} \partial v_{ab} } 
    & =
    \left( v_\T^{cd} - \frac{2}{3} v q^{cd} \right) \left( 
        q^{ab} \partdif{^2 L }{ v \partial R } 
        + 2 v_\T^{ab} \partdif{^2 L }{ w \partial R } 
    \right),
\\
\begin{split}
    \partdif{^2 L }{ v_{ab} \partial v_{cd} }
    & =
    q^{ab} q^{cd} \left( \partdif{^2 L }{ v^2 } - \frac{2}{3} \partdif{ L }{ w } \right)
    + 2 Q^{abcd} \partdif{ L }{ w }
\\ & 
    + 2 \left( q^{ab} v_\T^{cd} + v_\T^{ab} q^{cd} \right) \partdif{^2 L }{ v \partial w }
    + 4 v_\T^{ab} v_\T^{cd} \partdif{^2 L }{ w^2 }.
\end{split}
\end{align}%
    \label{eq:pert_d2theta_components}%
\end{subequations}

Evaluating each term in the $\partial_{c}\theta_{ab}$ bracket of \eqref{eq:pert_dist-eqn-sol},
\begin{subequations}
\begin{equation}
\begin{split}
    \partdif{ L }{ q_{ab,c} } & = 
    \partdif{ L }{ R } \left(
        \frac{3}{2} Q^{abde} \partial^c q_{de} 
        - q^{c(d} q^{e)(a} \partial^{b)} q_{de}
\right. \\ & \left.
        + q^{ab} Y^c
        - \half q^{ab} X^c
        - 2 q^{c(a} Y^{b)}
        + q^{c(a} X^{b)}
    \right),
\end{split}
\end{equation}
\begin{equation}
\begin{split}
    v_{ef} \partdif{^2 L }{ q_{ef,c} \partial v_{ab} } &=
    \left( 
        \frac{3}{2} Z^c - W^c - 2 v_\T^{cd} Y_d + v_\T^{cd} X_d + \frac{v}{3} X^c
    \right)
\\ & \times
    \left( 
        q^{ab} \partdif{^2 L }{ v \partial R } 
        + 2 v_\T^{ab} \partdif{^2 L }{ w \partial R } 
    \right),
\end{split}
\end{equation}
\begin{equation}
\begin{gathered}
    v_{ef} \partial_d \left( \partdif{^2 L }{ q_{ef,cd} \partial v_{ab} } \right)  =
    \left( v_\T^{cd} - \frac{2v}{3} q^{cd} \right) \left\{
        \left( q^{ab} \partial_d - Q^{abef} \partial_d q_{ef} \right) \partdif{^2 L }{ v \partial R }
\right. \\  \left.
        + 2 \left( v_\T^{ab} \partial_d + Q^{abef} \partial_d v^\T_{ef} - 2 v_\T^{e(a} q^{b)f} \partial_d q_{ef} \right) \partdif{^2 L }{ w \partial R }
    \right\} 
\\ 
    + \left( Z^c - W^c + \frac{v}{3} X^c + \frac{v}{3} Y^c - v_\T^{cd} Y_d \right) \left( q^{ab} \partdif{^2 L }{ v \partial R } + 2 v_\T^{ab} \partdif{^2 L }{ w \partial R } \right),
\end{gathered}
\end{equation}
\begin{equation}
\begin{split}
    \Gamma^c_{de} \partdif{^2 L }{ v_{ab} \partial v_{de} } & =
    \left( Y^c - \half X^c \right) \left\{ q^{ab} \left( \partdif{^2 L }{ v^2 } - \frac{2}{3} \partdif{ L }{ w } \right) + 2 v_\T^{ab} \partdif{^2 L }{ v \partial w } \right\}
\\ & 
    + \left( 2 q^{cd} q^{e(a} \partial^{b)} q_{de} - Q^{abde} \partial^c q_{de} \right) \partdif{ L }{ w }
\\ &
    + \left( 2 W^c - Z^c \right) \left( q^{ab} \partdif{^2 L }{ v \partial w } + 2 v_\T^{ab} \partdif{^2 L }{ w^2 } \right),
\end{split}
\end{equation}
\begin{equation}
\begin{split}
    \Gamma^c_{de} \partdif{ \beta }{ v_{ab} } \partdif{ L }{ v_{cd} } & =
    \left( q^{ab} \partdif{ \beta }{ v } + 2 v_\T^{ab} \partdif{ \beta }{ w } \right) \left\{
        \left( Y^c - \half X^c \right) \partdif{ L }{ v }
        + \left( 2 W^c - Z^c \right) \partdif{ L }{ w }
    \right\},
\end{split}
\end{equation}
\begin{equation}
\begin{split}
    \partial_d \beta \partdif{^2 L }{ v_{ab} \partial v_{cd} } & =
    \partial^c \beta \left\{ 
        q^{ab} \left( \partdif{^2 L }{ v^2 } - \frac{2}{3} \partdif{ L }{ w } \right) + 2 v_\T^{ab} \partdif{^2 L }{ v \partial w }
    \right\}
    + 2 q^{c(a} \partial^{b)} \beta \partdif{ L }{ w }
\\ &
    + 2 v_\T^{cd} \partial_d \beta \left( q^{ab} \partdif{^2 L }{ v \partial w } + 2 v_\T^{ab} \partdif{^2 L }{ w^2 } \right),
\end{split}
\end{equation}
\begin{equation}
\begin{split}
    \partial_d \left( \partdif{^2 L }{ v_{ab} \partial v_{cd} } \right) & =
    \left( q^{ab} \partial^c - q^{ab} Y^c - Q^{abef} \partial^c q_{ef} \right) \left( \partdif{^2 L }{ v^2 } - \frac{2}{3} \partdif{ L }{ w } \right)
\\ &
    + 2 \left( q^{c(a} \partial^{b)} - q^{c(a} Y^{b)} - q^{c(e} q^{f)(a} \partial^{b)} q_{ef} \right) \partdif{ L }{ w }
\\ &
    + 2 \left\{ q^{ab} \left( v_\T^{cd} \partial_d - v_\T^{cd} Y_d - W^c + q^{cd} \partial^e v^\T_{de} \right)
    +v_\T^{ab} \partial^c - v_\T^{ab} Y^c
\right. \\ & \left.
    + Q^{abef} \left( \partial^c v^\T_{ef} - v_\T^{cd} \partial_d q_{ef} \right) - 2 v_\T^{e(a} q^{b)f} \partial^c q_{ef}
    \right\} \partdif{^2 L }{ v \partial w }
\\ &
    + 4 \left\{ v_\T^{ab} \left( v_\T^{cd} \partial_d - W^c - v_\T^{cd} Y_d + q^{cd} \partial^e v^\T_{de} \right)
\right. \\ & \left.
    + Q^{abef} v_\T^{cd} \partial_d v^\T_{ef} - 2 v_\T^{e(a} q^{b)f} v_\T^{cd} \partial_d q_{ef}
    \right\} \partdif{^2 L }{ w^2 },
\end{split}
\end{equation}
\begin{equation}
\begin{split}
    \partdif{ \beta }{ v_{ab} } \partial_d \left( \partdif{ L }{ v_{cd} } \right) & = 
    \left( q^{ab} \partdif{ \beta }{ v } + 2 v_\T^{ab} \partdif{ \beta }{ w } \right) 
\\ &
    \times \left\{ 
        \left( \partial^c - Y^c \right) \partdif{ L }{ v }
        + 2 \left( v_\T^{cd} \partial_d + q^{cd} \partial^e v^\T_{de} - v_\T^{cd} Y_d - W^c \right) \partdif{ L }{ w }
    \right\}.
\end{split}
\end{equation}%
    \label{eq:pert_dtheta_components}%
\end{subequations}

\chapter{Deformed scalar-tensor constraint to all orders: Extras}
\label{sec:allst_extras}

Use the following definitions for convenience,
\begin{equation}
    X_a = q^{bc} \partial_a q_{bc},
\quad
    Y_a = q^{bc} \partial_c q_{ba} = \partial^b q_{ab},
\quad
    Z_a = p_\T^{bc} \partial_a q_{bc},
\quad
    W_a = p_\T^{bc} \partial_c q_{ba}.
        \label{eq:allst_tensor_combinations}
\end{equation}
Evaluating each term in the $\partial_{cd}\theta^{ab}$ bracket of \eqref{eq:allst_dist-eqn-sol-metric},
\begin{subequations}
\begin{equation}
\begin{split}
    \partdif{ C }{ q_{ef,cd} } \partdif{^2 C }{ p^{ab} \partial p^{cd} }
    & = 2 \delta^{cd}_{ab} \partdif{ C }{ R } \partdif{ C }{ \bp }
    - 2 q_{ab} q^{cd} \partdif{ C }{ R } \left( \partdif{^2 C }{ p^2 } + \third \partdif{ C }{ \bp } \right)
\\ &
    + 2 \left( q_{ab} p_\T^{cd} - 2 p^\T_{ab} q^{cd} \right) \partdif{ C }{ R } \partdif{^2 C }{ p \partial \bp }
    + 4 p^\T_{ab} p_\T^{cd} \partdif{ C }{ R } \partdif{^2 C }{ \bp^2 }
\end{split}
\end{equation}
\begin{equation}
    - \partdif{ C }{ p^{ef } } \partdif{^2 C }{ q_{ef,cd} \partial p^{ab} }
    = 2 \left( 
        q^{cd} \partdif{ C }{ p } - p_\T^{cd} \partdif{ C }{ \bp } 
    \right) \left( 
        q_{ab} \partdif{^2 C }{ p \partial R }
    + 2 p^\T_{ab} \partdif{^2 C }{ \bp \partial R } 
    \right)
\end{equation}
\begin{equation}
    \partdif{ C }{ \psi_{,cd} }
    \partdif{^2 C }{ p^{ab} \partial \pi }
    = q^{cd} \partdif{ C }{ \Delta } \left(
        q_{ab} \partdif{^2 C }{ p \partial \pi }
        + 2 p^\T_{ab} \partdif{^2 C }{ \bp \partial \pi }
    \right)
\end{equation}
\begin{equation}
    - \partdif{ C }{ \pi } \partdif{^2 C }{ \psi_{,cd}} \partial p^{ab}
    = - \partdif{ C }{ \pi } q^{cd} \left(
        q_{ab} \partdif{^2 C }{ p \partial \Delta } + 2 p^\T_{ab} \partdif{^2 C }{ \bp \partial \Delta }
    \right)
\end{equation}
\begin{equation}
    - \beta \partdif{ D^c }{ p^{ab}_{,d} }
    = 2 \beta \delta_{ab}^{cd},
\end{equation}
    \label{eq:allst_extras_d2theta}%
\end{subequations}

Evaluating each term in the $\partial_c\theta^{ab}$ bracket of \eqref{eq:allst_dist-eqn-sol-metric},
\begin{subequations}
\begin{equation}
\begin{gathered}
    \partdif{ C }{ q_{ef,c} } \partdif{^2 C }{ p^{ab} \partial p^{ef} } = 
    \partdif{ C }{ \Delta }
    \left\{ q_{ab} \left[ 
        \partial^c \psi \left( \half \partdif{^2 C }{ p^2 } + \frac{2}{3} \partdif{ C }{ \bp } \right)
        - 2 p_\T^{cd} \partial_d \psi \partdif{^2 C }{ p \partial \bp }
    \right]
\right. \\ \left.
    - 2 \delta^c_{(a} \partial_{b)} \psi \partdif{ C }{ \bp }
    + p^\T_{ab} \left[
        \partial^c \psi \partdif{^2 C }{ p \partial \bp }
        - 4 p_\T^{cd} \partial_d \psi \partdif{^2 C }{ \bp^2 }
    \right]
    \right\}
\\
   + \partdif{ C }{ R } \left\{
    \partdif{ C }{ \bp } \left[
        3 \partial^c q_{ab} 
        - 2 q^{cd} \partial_{(a} q_{b)d} 
        + 2 q_{ab} Y^c
        - q_{ab} X^c
        - 4 \delta^c_{(a} Y_{b)}
        + 2 \delta^c_{(a} X_{b)}
    \right]
\right. \\ \left.
    + q_{ab} X^c \left( \partdif{^2 C }{ p^2 } - \frac{2}{3} \partdif{ C }{ \bp } \right)
    + 2 p^\T_{ab} X^c \partdif{^2 C }{ p \partial \bp }
\right. \\ \left.
    + \left( 3 Z^c - 2 W^c - 4 p_\T^{cd} Y_d + 2 p_\T^{cd} X_d \right) \left(
        q_{ab} \partdif{^2 C }{ p \partial \bp }
        + 2 p^\T_{ab} \partdif{^2 C }{ \bp^2 } 
    \right)
    \right\},
\end{gathered}
\end{equation}
\begin{equation}
\begin{gathered}
    \partdif{^2 C }{ q_{ef,cd} } \partial_d \left( \partdif{^2 C }{ p^{ab} \partial p^{ef} } \right) =
    \partdif{ C }{ R } \left\{ 
        \left[ 
            q_{ab} \left( Y^c - X^c - 2 \partial^c \right)
            - 2 \partial^c q_{ab}
        \right] \left( \partdif{^2 C }{ p^2 } \! - \! \frac{2}{3} \partdif{ C }{ \bp } \right)
\right. \\ \left.
    + 2 \left( 
        \delta^c_{(a} \partial_{b)}
        - q_{ab} \partial^c
        + q^{cd} \partial_{(a} q_{b)d}
        + \delta^c_{(a} Y_{b)}
        - 2 \partial^c q_{ab}
    \right) \partdif{ C }{ \bp }
    + 2 \left[ 
        q_{ab} \left( 
            p_\T^{cd} \partial_d 
            + \partial_d p_\T^{cd}
\right. \right. \right. \\ \left. \left. \left.
            + W^c - Z^c
            + p_\T^{cd} Y_d
        \right)
        + p_\T^{cd} \partial_d q_{ab}
        + p^\T_{ab} \left(
            Y^c - X^c - 2 \partial^c
        \right)
        - 2 Q_{abde} \partial^c p_\T^{de}
\right. \right. \\ \left. \left.
        - 4 \partial^c q_{d(a} p_{\;b)}^{\T\,d}
    \right] \partdif{^2 C }{ p \partial \bp }
    + 4 \left[ 
        Q_{abef} p_\T^{cd} \partial_d p_\T^{ef}
        + 2 p_\T^{cd} \partial_d q_{e(a} p_{\;b)}^{\T\,e}
\right. \right. \\ \left. \left.
        + p^\T_{ab} \left( 
            \partial_d p_\T^{cd}
            + W^c - Z^c + p_\T^{cd} Y_d + p_\T^{cd} \partial_d
        \right)
    \right] \partdif{^2 C }{ \bp^2 }
    \right\},
\end{gathered}
\end{equation}
\begin{equation}
\begin{gathered}
    \partdif{ C }{ p^{ef} } \partdif{^2 C }{ q_{ef,c} \partial p^{ab} } =
    \left\{ \partdif{ C }{ p } \left[
        X^c \partdif{}{ R } 
        + \half \partial^c \psi \partdif{}{ \Delta }
    \right]
\right. \\ \left.
    + \partdif{ C }{ \bp } \left[ 
        \left( 3 Z^c - 2 W^c - 4 p_\T^{cd} Y_d \right) \partdif{}{ R }
        - 2 p_\T^{cd} \partial_d \psi \partdif{}{ \Delta }
    \right]
    \right\} \left( q_{ab} \partdif{ C }{ p } + 2 p^\T_{ab} \partdif{ C }{ \bp } \right)
\\
    + 2 p_\T^{cd} \partdif{ C }{ \bp } \left\{
        q_{ab} \partial_d \partdif{^2 C }{ p \partial R }
        + 2 p^\T_{ab} \partial_d \partdif{^2 C }{ \bp \partial R }
    \right\},
\end{gathered}
\end{equation}
\begin{equation}
\begin{gathered}
    \partdif{ C }{ p^{ef} } \partial_d \left( \partdif{^2 C }{ q_{ef,cd} \partial p^{ab} } \right) =
    \partdif{ C }{ p } \left\{ 
        \left[
            q_{ab} \left( X^c + Y^c - 2 \partial^c \right)
            - 2 \partial^c q_{ab}
        \right] \partdif{^2 C }{ p \partial R }
\right. \\ \left.
        + 2 \left[
            p^\T_{ab} \left( X^c + Y^c - 2 \partial^c \right)
            - 2 Q_{abef} \partial^c p_\T^{ef}
            - 4 \partial^c q_{d(a} p^{\T\,d}_{\;b)}
        \right] \partdif{^2 C }{ \bp \partial R }
    \right\}
\\
    + 2 \partdif{ C }{ \bp } \left\{ 
        \left[
            q_{ab} \left( Z^c \! - \! W^c \! - \! p_\T^{cd} Y_d \! + \! p_\T^{cd} \partial_d \right)
            + p_\T^{cd} \partial_d q_{ab}
        \right] \partdif{^2 C }{ p \partial R }
\right. \\ \left.
        + 2 \left[
            p^\T_{ab} \left( 
                Z^c \! - \! W^c \! - \! p_\T^{cd} Y_d \! + \! p_\T^{cd} \partial_d
            \right)
            + Q_{abef} p_\T^{cd} \partial_d p_\T^{ef}
            + 2 p_\T^{cd} \partial_d q_{e(a} p_{\;b)}^{\T\,e}
        \right] \partdif{^2 C }{ \bp \partial R }
    \right\},
\end{gathered}
\end{equation}
\begin{equation}
    \partdif{ C }{ \psi_{,c} } \partdif{^2 C }{ \pi \partial p^{ab} } =
    \left\{ 
        2 \partial^c \psi \partdif{ C }{ \gamma }
        + \left( \half X^c - Y^c \right) \partdif{ C }{ \Delta }
    \right\} \left( 
        q_{ab} \partdif{ C }{ \pi \partial p }
        + 2 p^\T_{ab} \partdif{ C }{ \pi \partial \bp }
    \right),
\end{equation}
\begin{equation}
\begin{split}
    \partdif{ C }{ \psi_{,cd} } \partial_d \left( \partdif{^2 C }{ \pi \partial p^{ab} } \right)
    & =
    \partdif{ C }{ \Delta } \left\{ \left( 
            \partial^c q_{ab} + q_{ab} \partial^c 
        \right) \partdif{^2 C }{ \pi \partial p }
\right. \\ & \left.
        + 2 \left(
            Q_{abef} \partial^c p_\T^{ef} 
            + 2 \partial^c q_{d(a} p_{\;a)}^{\T\,d}
            + p^\T_{ab} \partial^c
        \right) \partdif{^2 C }{ \pi \partial \bp }
    \right\},
\end{split}
\end{equation}
\begin{equation}
\begin{gathered}
    \partdif{ C }{ \pi } \partial_d \left( \partdif{^2 C }{ \psi_{,cd} \partial p^{ab} } \right)
    =
    \partdif{ C }{ \pi } \left\{
        \left( 
            q_{ab} \partial^c + \partial^c q_{ab} - q_{ab} Y^c
        \right) \partdif{^2 C }{ p \partial \Delta }
\right. \\ \left.
        + 2 \left(
            Q_{abef} \partial^c p_\T^{ef}
            + 2 \partial^c q_{d(a} p_{\;b)}^{\T\,d}
            + p^\T_{ab} \partial^c
            - p^\T_{ab} Y^c
        \right) \partdif{^2 C }{ \bp \partial \Delta }
    \right\},
\end{gathered}
\end{equation}
\begin{equation}
\begin{split}
    \partdif{ ( \beta D^c ) }{ p^{ab} }
    & = 
    \beta \left( \partial^c q_{ab} - 2 q^{cd} \partial_{(a} q_{b)d} \right)
    + \left( 
        q_{ab} \partdif{ \beta }{ p }
        + 2 p^\T_{ab} \partdif{ \beta }{ \bp }
    \right)
\\ & \times
    \left(
        \pi \partial^c \psi
        - 2 \partial_d p_\T^{cd}
        - \frac{2}{3} \partial^c p
        - 2 W^c + Z^c
        + \third p X^c
    \right),
\end{split}
\end{equation}
\begin{equation}
    \partial_d \left( \beta \partdif{ D^c }{ p^{ab}_{,d} } \right)
    = - 2 \delta^c_{(a} \partial_{b)} \beta,
\end{equation}
    \label{eq:allst_extras_dtheta}%
\end{subequations}

Evaluating each term in the $\partial_{cd}\eta^{ab}$ bracket of \eqref{eq:allst_dist-eqn-sol-scalar},
\begin{subequations}
\begin{equation}
    \partdif{ C }{ q_{cd,ab} } \partdif{^2 C }{ \pi \partial p^{cd} } =
    \partdif{ C }{ R } \left(
        - 2 q^{ab} \partdif{^2 C }{ \pi \partial p }
        + 2 p_\T^{ab} \partdif{^2 C }{ \pi \partial \bp }
    \right),
\end{equation}
\begin{equation}
    \partdif{ C }{ p^{cd} } \partdif{^2 C }{ q_{cd,ab} \partial \pi } =
    \partdif{^2 C }{ R \partial \pi } \left(
        - 2 q^{ab} \partdif{ C }{ p }
        + 2 p_\T^{ab} \partdif{ C }{ \bp }
    \right),
\end{equation}
\begin{equation}
    \partdif{ C }{ \psi_{,ab} } \partdif{^2 C }{ \pi^2 }
    - \partdif{ C }{ \pi } \partdif{^2 C }{ \psi_{,ab} \partial \pi }
    =
    q^{ab} \left( 
        \partdif{ C }{ \Delta } \partdif{^2 C }{ \pi^2 }
        - \partdif{ C }{ \pi } \partdif{^2 C }{ \Delta \partial \pi }
    \right),
\end{equation}
    \label{eq:allst_extras_d2eta}%
\end{subequations}

Evaluating each term in the $\partial_c\eta^{ab}$ bracket of \eqref{eq:allst_dist-eqn-sol-scalar},
\begin{subequations}
\begin{equation}
\begin{split}
    \partdif{ C }{ q_{cd,a} } \partdif{^2 C }{ \pi \partial p^{cd} } & =
    \partdif{ C }{ R } \left\{ 
        X^a \partdif{^2 C }{ \pi \partial p }
        + \left( 
            3 Z^a - 2 W^a - 4 p_\T^{ab} Y_b + 2 p_\T^{ab} \partial_b
        \right) \partdif{^2 C }{ \pi \partial \bp }
    \right\}
\\ &
    + \partdif{ C }{ \Delta } \left\{
        \half \partial^a \psi \partdif{^2 C }{ \pi \partial p }
        - 2 p_\T^{ab} \partial_b \psi \partdif{^2 C }{ \pi \partial \bp }
    \right\},
\end{split}
\end{equation}
\begin{equation}
\begin{split}
    \partdif{ C }{ q_{cd,ab} } \partial_b \left( \partdif{^2 C }{ \pi \partial p^{cd} } \right) & =
    \partdif{ C }{ R } \left\{ 
        \left( Y^a - X^a - 2 \partial^a 
        \right) \partdif{^2 C }{ \pi \partial p }
\right. \\ & \left.
        + 2 \left( 
            \partial_b p_\T^{ab} + W^a - Z^a + p_\T^{ab} Y_b + p_\T^{ab} \partial_b
        \right) \partdif{^2 C }{ \pi \partial \bp }
    \right\},
\end{split}
\end{equation}
\begin{equation}
\begin{gathered}
    \partdif{ C }{ p^{cd } } \partdif{^2 C }{ q_{cd,a} \partial \pi } =
    \partdif{ C }{ p } \left\{ 
        X^a \partdif{^2 C }{ R \partial \pi }
        + \half \partial^a \psi \partdif{^2 C }{ \Delta \partial \pi }
    \right\}
\\
    + \partdif{ C }{ \bp } \left\{
        \left( 
            3 Z^a - 2 W^a - 4 p_\T^{ab} Y_b + 2 p_\T^{ab} X_b
        \right) \partdif{^2 C }{ R \partial \pi }
        - 2 p_\T^{ab} \partial_b \psi \partdif{^2 C }{ \Delta \partial \pi }
    \right\},
\end{gathered}
\end{equation}
\begin{equation}
\begin{split}
    \partdif{ C }{ p^{cd} } \partial_b \left( \partdif{^2 C }{ q_{cd,ab} \partial \pi } \right) & =
    \left\{ 
        \partdif{ C }{ p } \left( X^a + Y^a - 2 \partial^a \right)
\right. \\ & \left.
        + 2 \partdif{ C }{ \bp } \left( Z^a - W^a - p_\T^{ab} Y_b + p_\T^{ab} \partial_b \right)
    \right\} \partdif{^2 C }{ R \partial \pi },
\end{split}
\end{equation}
\begin{equation}
\begin{split}
    \partdif{ C }{ \psi_{,a} } \partdif{^2 C }{ \pi^2 }
    + \partdif{ C }{ \pi } \partdif{^2 C }{ \psi_{,a} \partial \pi }
    & = 
    2 \partial^a \psi \left(
        \partdif{ C }{ \gamma } \partdif{^2 C }{ \pi^2 }
        + \partdif{ C }{ \pi } \partdif{^2 C }{ \pi \partial \gamma }
    \right)
\\ &
    + \left( \half X^a + Y^a \right) \left(
        \partdif{ C }{ \Delta } \partdif{^2 C }{ \pi^2 }
        + \partdif{ C }{ \pi } \partdif{^2 C }{ \pi \partial \Delta }
    \right),
\end{split}
\end{equation}
\begin{equation}
    \partdif{ C }{ \psi_{,ab} } \partial_b \left( \partdif{^2 C }{ \pi^2 } \right)
    - \partdif{ C }{ \pi } \partial_b \left( \partdif{^2 C }{ \psi_{,ab} \partial \pi } \right)
    = \partdif{ C }{ \Delta } \partial^a \left( \partdif{^2 C }{ \pi^2 } \right)
    + \partdif{ C }{ \pi } \left( Y^a - \partial^a \right) \partdif{^2 C }{ \pi \partial \Delta },
\end{equation}
\begin{equation}
    \partdif{ ( \beta D^a ) }{ \pi } =
    \partdif{ \beta }{ \pi } \left( 
        - 2 \partial_b p_\T^{ab} 
        - \frac{2}{3} \partial^a p
        + 2 W^a - Z^a - \third p X^a
    \right)
    + \partial^a \psi \left( \beta + \pi \partdif{ \beta }{ \pi } \right).
\end{equation}
    \label{eq:allst_extras_deta}%
\end{subequations}


\label{Bibliography}
\bibliographystyle{./utphys}  
\bibliography{bibliography}  


\end{document}